\documentclass[12pt,a4paper]{iopart}

\usepackage[english]{babel}

\usepackage{iopams}
\usepackage{verbatim}
\usepackage{color}

\usepackage{graphicx}

\usepackage{color}

\usepackage[colorlinks,citecolor=blue,linkcolor=blue,urlcolor=blue]{hyperref}

\def\iotabar{\lower3pt\hbox{$\mathchar'26$}\mkern-7mu\iota}
\newcommand {\aplt}{\ {\raise-.5ex\hbox{$\buildrel<\over\sim$}}\ }
\newcommand{\dd}{\mbox{d}}

\newcommand{\spe}{{\sigma}}
\newcommand{\lw}{{\rm lw}}
\newcommand{\sw}{{\rm sw}}
\newcommand{\eq}[1]{(\ref{#1})}
\newcommand{\bun}{\hat{\mathbf{b}}}
\newcommand{\eun}{\hat{\mathbf{e}}}

\newcommand{\phiwig}{\widetilde{\phi}}

\newcommand{\boldr}{\mathbf{r}}
\newcommand{\bv}{\mathbf{v}}
\newcommand{\bA}{\mathbf{A}}
\newcommand{\bR}{\mathbf{R}}

\newcommand{\bJ}{\mathbf{J}}

\newcommand{\bB}{\mathbf{B}}

\newcommand{\bZ}{\mathbf{Z}}
\newcommand{\bK}{\mathbf{K}}

\newcommand{\bX}{\mathbf{X}}
\newcommand{\bV}{\mathbf{V}}

\newcommand{\matrixtop}[1]{\buildrel\leftrightarrow\over{#1}}
\newcommand{\matI}{\matrixtop{\mathbf{I}}}
\newcommand{\matW}{\matrixtop{\mathbf{W}}}
\newcommand{\matM}{\matrixtop{\mathbf{M}}}

\newcommand{\matPi}{\matrixtop{\mathbf{\Pi}}}

\newcommand{\dotcross}{ \raise 0.65ex\hbox{${\scriptstyle {{_{\displaystyle \cdot}}\atop\times}}$} }
\newcommand{\crossdot}{ \raise 0.5ex\hbox{${\scriptstyle {{_\times}\atop{\displaystyle \cdot}}}$} }
\newcommand{\rhobf}{\mbox{\boldmath$\rho$}}

\newcommand{\kappabf}{\mbox{\boldmath$\kappa$}}

\newcommand{\Gammabf}{\mbox{\boldmath$\Gamma$}}
\newcommand{\cT}{{\cal T}}
\newcommand{\modTinv}{{\mathbb T_{\spe,0}}}
\newcommand{\modTinvprime}{{\mathbb T_{\spe',0}}}

\newcommand{\potH}{{p}}
\newcommand{\potL}{{q}}
\newcommand{\bw}{\mathbf{w}}

\newcommand{\sumsig}{ \raise -1.3ex\hbox{${{\displaystyle \sum}\atop{\scriptstyle \sigma}}$} }
\newcounter{appnumb}

\begin{document}

\title[Stellarators close to quasisymmetry]{Stellarators close to
  quasisymmetry}


\author{Iv\'an Calvo$^{1}$}
\vspace{-0.2cm}
\eads{\mailto{ivan.calvo@ciemat.es}}
\vspace{-0.5cm}
\author{Felix I Parra$^{2}$}
\vspace{-0.2cm}
\eads{\mailto{fparra@mit.edu}}
\vspace{-0.5cm}
\author{Jos\'e Luis Velasco$^{1}$}
\vspace{-0.2cm}
\eads{\mailto{joseluis.velasco@ciemat.es}}
\vspace{-0.5cm}
\author{J Arturo Alonso$^{1}$}
\vspace{-0.2cm}
\eads{\mailto{arturo.alonso@ciemat.es}}

\vspace{0.3cm}

\address{$^1$Laboratorio Nacional de Fusi\'on, Asociaci\'on
EURATOM-CIEMAT, 28040 Madrid, Spain}
\address{$^2$Plasma Science and Fusion Center, MIT, Cambridge, MA 02139, USA}


\begin{abstract}
  Rotation is favorable for confinement, but a stellarator can rotate
  at high speeds if and only if it is sufficiently close to
  quasisymmetry. This article investigates how close it needs to
  be. For a magnetic field $\bB = \bB_0 + \alpha \bB_1$, where $\bB_0$
  is quasisymmetric, $\alpha\bB_1$ is a deviation from quasisymmetry,
  and $\alpha\ll 1$, the stellarator can rotate at high velocities if
  $\alpha < \epsilon^{1/2}$, with $\epsilon$ the ion Larmor radius
  over the characteristic variation length of $\bB_0$. The cases in
  which this result may break down are discussed. If the
  stellarator is sufficiently quasisymmetric in the above sense, the
  rotation profile, and equivalently, the long-wavelength radial
  electric field, are not set neoclassically; instead, they can be
  affected by turbulent transport. Their computation requires the
  $O(\epsilon^2)$ pieces of both the turbulent and the long-wavelength
  components of the distribution function. This article contains the
  first step towards a formulation to calculate the rotation profile
  by providing the equations determining the long-wavelength
  components of the $O(\epsilon^2)$ pieces.
\end{abstract}

\vspace{-0.3cm}




\maketitle

\section{Introduction}
\label{sec:Introduction}

In a stellarator, quasisymmetry and the capability to sustain large
rotation are equivalent~\cite{Helander08,Simakov11}. This means that
only when the stellarator is quasisymmetric, there exists a component
of the flow (the component in the symmetry direction) on the flux
surface that is neoclassically undamped.  For design purposes, large
rotation only requires that the stellarator be sufficiently close to
quasisymmetry, as is physically reasonable. This is important because,
on the one hand, large rotation is beneficial for magnetohydrodynamic
modes and confinement~\cite{Mantica09}, and therefore desirable; on
the other hand, Garren and Boozer proved~\cite{Garren1991} that except
for the axisymmetric case, quasisymmetric toroidal magnetic fields do
not exist\footnote{Garren and Boozer proved that
  quasisymmetry is possible on one flux surface but not throughout the
  entire plasma volume.}. Then, it is natural to ask what is the
maximum size of a deviation from quasisymmetry that still allows sonic
rotation speeds. Let us write the magnetic field as $\bB = \bB_0 +
\alpha \bB_1$, where $\bB_0$ is quasisymmetric, $\alpha\ll 1$, and
$\alpha \bB_1$ represents a deviation from
quasisymmetry. Reference~\cite{Garren1991} proved that $\alpha = 0$
cannot be achieved. In this paper, and in the framework of gyrokinetic
theory, we find that in general a stellarator plasma can rotate fast
if
\begin{equation}\label{eq:ourScaling_intro}
\alpha < \epsilon^{1/2}.
\end{equation}
Here, $\epsilon := \rho_i/L$, where $\rho_i$ is a typical ion Larmor
radius and $L$ is the characteristic variation length of $\bB_0$. This
scaling was already obtained in a fluid description of a highly
collisional plasma in reference \cite{Simakov2009}. Advancing results
of Section \ref{sec:deviationQS}, we point out that
\eq{eq:ourScaling_intro} is the correct criterion to evaluate
closeness to quasisymmetry as long as the helicity of $\bB_1$ is small
enough.

Whether the deviation from quasisymmetry satisfies
\eq{eq:ourScaling_intro} or not has consequences for the
theoretical computation of the rotation profile or, equivalently, the
long-wavelength radial electric field. If \eq{eq:ourScaling_intro} is
not satisfied, the problem is neoclassical. If
\eq{eq:ourScaling_intro} holds, the effect of microturbulent
fluctuations has to be taken into account. Let us explain the problem
in detail.

Only in recent years has the problem of the determination of the
long-wavelength radial electric field in a turbulent tokamak in the
low flow ordering (i.e. with subsonic speeds) been
understood~\cite{parra08,parra09b,parra10c}.  First, it has been shown
that computing the long-wavelength radial electric field is equivalent
to the calculation of the toroidal momentum transport or to the
determination of the toroidal rotation profile. The momentum flux in
tokamaks is mostly due to microturbulence, well described by
gyrokinetic theory.  It requires gyrokinetic equations
that are accurate to $O(\epsilon^2)$, because the expressions for
obtaining the toroidal rotation profile involve the $O(\epsilon^2)$
pieces of the distribution function (the turbulent and long-wavelength
components) and some $O(\epsilon^2)$ pieces of the electrostatic
potential (the complete turbulent component and the part
corresponding to the long-wavelength poloidal electric field). An
introductory overview of the problem is given in
\cite{ParraBarnesCalvoCatto2012}. The derivation of the electrostatic
gyrokinetic equations to $O(\epsilon^2)$ in general magnetic geometry
has been carried out in reference \cite{ParraCalvo2011}.  A complete
set of equations to find the tokamak long-wavelength radial electric
field has been given~\cite{parra10a,parra10b,parra11d} that is valid
when $B/B_p\gg 1$, where $B$ is the modulus of the magnetic field and
$B_p$ is the modulus of the poloidal magnetic field.  The general
proof that the turbulent tokamak is intrinsically ambipolar and the
subset of long-wavelength equations that will eventually be needed to
correctly compute toroidal momentum transport in a tokamak, without
relying on the assumption $B/B_p\gg 1$, have been given in references
\cite{parra09b} and \cite{CalvoParra2012}.

The origin of the difficulty is the intrinsic ambipolarity of the
tokamak, even in the presence of turbulence; that is, the fact that to
lowest order the flux-surface averaged radial flux of charge is
identically zero, whatever the value of the long-wavelength radial
electric field. This is the property that allows the tokamak to
rotate. Intrinsic ambipolarity in the tokamak implies that the flow in
the toroidal direction is neoclassically undamped, that the plasma can
rotate freely, and that the long-wavelength radial electric field is
not determined by the lowest-order calculation in $\epsilon$. The
theoretical calculation of the long-wavelength radial electric field
in a generic stellarator is comparatively much easier because it is
determined by neoclassical physics due to the lack of intrinsic
ambipolarity: parallel viscosity damps the two components of the flow
on the flux surface, setting the long-wavelength radial electric
field.  This is true even for some types of stellarators optimized
with respect to neoclassical transport. For example, collisionless
particle trajectories in omnigeneous~\cite{Cary97, Cary97b,
  Landreman12} and isodynamic stellarators~\cite{Helander09} have
vanishing average radial velocity but the radial particle fluxes need
not be intrinsically ambipolar, and the radial electric field is
computed from the standard neoclassical ambipolarity condition. The
most constrained stellarator design concept is
quasisymmetry~\cite{Boozer83,Nuehrenberg88}. Quasisymmetric
stellarators are omnigenous and isodynamic, but the converse is not
true. Since a stellarator is quasisymmetric if and only if it is
intrinsically ambipolar~\cite{Helander08,Simakov11}, it is expected
that the problem of solving for the radial electric field in these
devices be similar to the one in the tokamak. The same can be said for
the computation of the rotation profile along the direction of
symmetry of the quasisymmetric stellarator.

Although the general picture is analogous in quasisymmetric
stellarators and in tokamaks, the former exhibit very interesting
subtleties~\cite{Helander07, Sugama11}. In particular, a remark in
connection with the striking result of \cite{Sugama11} is in order. In
this reference it was found that stellarators, even if quasisymmetric,
cannot rotate to speeds that are strictly sonic; only the axisymmetric
tokamak can. However, the result is not as negative for rotation in
stellarators as this statement might suggest. Let us denote by $M =
V_i/v_{ti}$ the Mach number, where $V_i$ is the plasma rotation speed
and $v_{ti}$ is the ion thermal speed. Typically, there is much room
between $\epsilon v_{ti}$ and $v_{ti}$. If $\epsilon v_{ti} \ll M
v_{ti} \ll v_{ti}$, then a subsidiary expansion in the Mach number is
possible and the mathematical obstructions for rotation found in
\cite{Sugama11} disappear. It is in this setting that our work should
be understood.

The problem of writing the equations that would determine the
long-wavelength radial electric field, or the rotation profile, in a
quasisymmetric stellarator (actually, in a stellarator satisfiying
equation \eq{eq:ourScaling_intro}) has never been considered. We take
the first step towards a formulation that unifies the
non-quasisymmetric and quasisymmetric cases. Such a formulation has to
be addressed in the framework of gyrokinetic theory and implies
obtaining equations for the $O(\epsilon^2)$ pieces of both the
turbulent and long-wavelength components of the distribution function,
for the $O(\epsilon^2)$ pieces of the turbulent electrostatic
potential, and for the $O(\epsilon^2)$ pieces of the long-wavelength
poloidal and toroidal components of the electric field. In this paper
we formulate the equations for the long-wavelength components. The
long-wavelength Fokker-Planck and quasineutrality equations to
$O(\epsilon^2)$ for an arbitrary stellarator are derived, and explicit
expressions are given that could eventually be implemented in computer
codes.  Obviously, if the stellarator is non-quasisymmetric it is not
necessary to perform the calculation to $O(\epsilon^2)$. The proof of
intrinsic ambipolarity in \cite{CalvoParra2012} fails for
non-quasisymmetric stellarators, and the equations presented here give
the neoclassical radial electric field. In the quasisymmetric case,
the $O(\epsilon^2)$ accuracy to which the gyrokinetic Fokker-Planck
and quasineutrality equations are derived in this paper is necessary;
physically this implies, in particular, that both neoclassical and
turbulent effects have to be included. Progress in computational
stellarator optimization and design techniques~\cite{Reiman99} as well
as in computer simulation of the gyrokinetic equations~\cite{dimits96,
  dorland00, dannert05, candy03, chen03, peeters09, kornilov04,
  xanthopoulos07, baumgaertel11, watanabe11} make the topics of this
paper relevant to present-day fusion research.

The rest of the paper is organized as follows. In Section
\ref{sec:secondordergyrokinetics} we introduce the basic assumptions,
mathematical tools and equations of gyrokinetics, and set the
notation.  We also recall the results from \cite{ParraCalvo2011} that
are needed in subsequent sections. Section \ref{sec:FokkerPlancklong}
is devoted to taking the long-wavelength limit of the gyrokinetic
Fokker-Planck equation to $O(\epsilon^2)$ by assuming only the
existence of nested flux surfaces; that is, in a general
stellarator. In Section \ref{sec:longwavePoisson} we do the same for
the quasineutrality equation. Although it was not emphasized there,
many expressions in reference~\cite{CalvoParra2012} are valid for
general magnetic geometry, not only for tokamaks. Hence, much of the
effort involved in the derivation of the gyrokinetic Fokker-Planck and
quasineutrality equations to $O(\epsilon^2)$ has already been
done. Accordingly, sections \ref{sec:FokkerPlancklong} and
\ref{sec:longwavePoisson} will be shorter than the corresponding ones
in~\cite{CalvoParra2012}. Section
\ref{sec:indeterminacyRadialElectricField} shows why the
long-wavelength radial electric field is not determined, in any
stellarator, by using exclusively the quasineutrality equation to
second order. To find the long-wavelength radial electric field, we
need the solvability conditions of the Fokker-Planck equation to order
$\epsilon^j$, discussed in Section \ref{sec:transportEqs}. For $j=1$
it turns out that the solvability conditions are trivially satisfied,
whereas for $j=2$ they yield transport equations for density and
energy, that we give explicitly. Imposing compatibility of the density
transport equation with the lowest-order quasineutrality equation
leads to the well-known neoclassical ambipolarity condition, i.e. the
vanishing of the lowest-order flux-surface averaged radial
current. Section \ref{sec:ambipolarityConditionInformation} discusses
in detail when the neoclassical ambipolarity condition is not
automatically satisfied, giving the long-wavelength radial electric
field. In other words, we ask when the system is intrinsically
ambipolar, which is equivalent to being quasisymmetric.  In Section
\ref{sec:deviationQS} we discuss the violation of ambipolarity due to
small deviations from quasisymmetry, determining when a stellarator
can be considered quasisymmetric in practice. The conclusions are
contained in Section \ref{sec:conclusions}.

\section{Electrostatic gyrokinetics}
\label{sec:secondordergyrokinetics}

The kinetic description of a plasma in a time-independent magnetic
field requires the Fokker-Planck equation for each species $\spe$,
\begin{eqnarray}\label{eq:FPinitial}
\fl\partial_t f_\spe + \bv\cdot\nabla_\boldr f_\spe
+\frac{Z_\spe e}{m_\spe}\left(-\nabla_\boldr\varphi +
c^{-1}\, \bv\times\bB\right)\cdot\nabla_\bv f_\spe =\nonumber\\[5pt]
\fl\hspace{1cm}
\sum_{\spe'}C_{\spe \spe'}[f_\spe,f_{\spe'}](\boldr,\bv),
\end{eqnarray}
and Poisson's equation,
\begin{eqnarray}\label{eq:Poisson}
\nabla^2_\boldr \varphi (\boldr,t)= -4\pi e\sum_\spe Z_\spe\int \,
f_\spe(\boldr,\bv,t)\dd ^3 v.
\end{eqnarray}
Here, $c$ is the speed of light, $e$ the charge of the proton,
$\varphi(\boldr, t)$ the electrostatic potential,
$\bB(\boldr)=\nabla_\boldr\times\bA(\boldr)$ a time-independent
magnetic field, $\bA(\boldr)$ the magnetic vector potential,
$f_\spe(\boldr,\bv,t)$ the phase-space probability distribution, and
$Z_\spe e$ and $m_\spe$ the charge and the mass of species
$\spe$. The Fokker-Planck collision operator between species $\spe$
and $\spe'$ is
\begin{eqnarray}\label{eq:collisionoperator}
\fl
C_{\spe \spe'}[f_\spe,f_{\spe'}](\boldr,\bv) =
\nonumber\\[5pt]
\hspace{1cm}
\fl
\frac{\gamma_{\spe\spe'}}{m_\sigma}
\nabla_\bv\cdot
\int
\matW(\bv-\bv')
\cdot
\Bigg(
\frac{1}{m_\spe}f_{\spe'}(\boldr,\bv',t)\nabla_\bv f_\spe(\boldr,\bv,t)
\nonumber\\[5pt]
\hspace{1cm}
\fl
-
\frac{1}{m_{\spe'}}
f_{\spe}(\boldr,\bv,t)\nabla_{\bv'} f_{\spe'}(\boldr,\bv',t)
\Bigg)
\dd^3v',
\end{eqnarray}
where
\begin{equation}
\gamma_{\spe\spe'}:= 2\pi Z_\spe^2 Z_{\spe'}^2 e^4 \ln\Lambda,
\end{equation}
\begin{equation}\label{eq:defmatW}
{\bf \matW}({\bold w})
 := \frac{|{\bold w}|^2{\matI}-{\bold w}{\bold w}}{|{\bold w}|^3},
\end{equation}
$\ln\Lambda$ is the Coulomb logarithm, and $\matI$ is the identity
matrix. Some important properties of this operator are collected in
\ref{sec:collisionoperator}. A direct check shows that the
Fokker-Planck equation can also be written as
\begin{eqnarray}
\partial_t f_\spe + \{f_\spe,H_\spe\}_\bX =
\sum_{\spe'}C_{\spe \spe'}[f_\spe,f_{\spe'}](\bX),
\end{eqnarray}
where we designate by $\bX\equiv (\boldr,\bv)$ a set of euclidean
coordinates in phase-space,
\begin{eqnarray}
H_\spe(\boldr,\bv,t) = \frac{1}{2}m_\spe \bv^2 + Z_\spe
e\varphi(\boldr,t)
\end{eqnarray}
is the Hamiltonian of species $\spe$, and the Poisson bracket of two
functions on phase space, $g_1(\boldr,\bv)$ and $g_2(\boldr,\bv)$, is
\begin{eqnarray}
\{g_1,g_2\}_\bX &= \frac{1}{m_\spe}\left(\nabla_\boldr g_1\cdot\nabla_\bv g_2 -
\nabla_\bv g_1\cdot\nabla_\boldr g_2\right)\nonumber\\[5pt]
 &+ \frac{Z_\spe e}{m_\spe^2c} \bB\cdot(\nabla_\bv g_1\times\nabla_\bv g_2).
\end{eqnarray}

\subsection{Gyrokinetic ordering}
\label{sec:orderingANDseparationOFscales}

Relevant frequencies in magnetically confined fusion plasmas are
usually much smaller than the gyrofrequency, i.e. the frequency that
corresponds to the approximately circular motion of a charged particle
around a magnetic field line. Gyrokinetic theory~\cite{catto78,
  krommes2012} is the reduced kinetic theory obtained by averaging out
the gyrofrequency time scale. This is possible due to the assumed
smallness of the gyroradius, as explained in what follows.

Denote by $L\sim|\nabla_\boldr\ln|\bB||^{-1}$ the typical length of
variation of the magnetic field and by $B_0$ a typical value of the
magnetic field strength. The sound speed is defined as
$c_s:=\sqrt{T_{e0}/m_i}$, where $T_{e0}$ is a typical electron
temperature and $m_i$ is the mass of the singly-charged dominant ion
species. Gyrokinetics is formulated as an asymptotic expansion in the
small parameter $\epsilon_s=\rho_s/L$, where $\rho_s=c_s/\Omega_i$ is
a characteristic sound gyroradius, and $\Omega_i={eB_0/(m_ic)}$ is a
characteristic ion gyrofrequency. Even if formally the only expansion
parameter is $\epsilon_s$, many expressions are more conveniently
written in terms of the species-dependent parameter $\epsilon_\spe
=\rho_\spe/L$, where $\rho_\spe=v_{t\spe}/\Omega_\spe$ is the
characteristic gyroradius of species $\spe$, $v_{t\spe} =\sqrt{T_{e
    0}/m_\spe}$ is the thermal velocity, and $\Omega_\spe=Z_\spe e
B_0/(m_\spe c)$ is the characteristic gyrofrequency. Observe that the
relation between $\epsilon_\spe$ and $\epsilon_s$ is
$\epsilon_s=\lambda_\spe\epsilon_\spe$, with
\begin{equation}
\lambda_\spe = \frac{\rho_s}{\rho_\spe} = Z_\spe
\sqrt{\frac{m_i}{m_\spe}}\, .
\end{equation}
In this paper we assume that $T_{e 0}$ is the typical temperature of
all species, which is justified when the time between collisions is
shorter than the transport time scale. It is not difficult to relax
this hypothesis. We also stress that our ordering is maximal, in the
sense that typical expansions, like those in the mass ratio, can be
obtained by performing subsidiary expansions in parameters such as
$\lambda_\spe$.

It has been experimentally observed that the characteristic
correlation length of the turbulence is of the order of the gyroradius
and scales with it, and that the size of the turbulent fluctuations
scales with the ion gyroradius~\cite{mckee01}. Also, the
characteristic length of the turbulent eddies and the size of the
fluctuations are related to each other by the background gradient. An
eddy of length $\ell_\bot \sim \rho_s$ mixes the plasma contained
within it. In the presence of a gradient this eddy will lead to
fluctuations on top of the background density of order $\delta n_e
\sim \ell_\bot |\nabla n_e| \sim \epsilon_s n_e \ll n_e$. These facts
suggest the ordering that we set below.

We start by introducing a transport average, that acting on a function
$g(\boldr,t)$ gives the component corresponding to long wavelengths
and small frequencies.  Let $\{\psi,\Theta,\zeta\}$ be a set of flux
coordinates, where $\psi$ is the flux surface label, $\Theta$ is the
poloidal angle, and $\zeta$ is the toroidal angle. This averaging
operation is
\begin{equation}
\fl\langle g \rangle_{\rm{T}} 
= \frac{1}{\Delta t \Delta \psi \Delta\Theta\Delta\zeta}
\int_{\Delta t}\dd t
\int_{\Delta \psi}\dd \psi
\int_{\Delta \Theta}\dd \Theta \int_{\Delta \zeta}\dd \zeta\, g,
\end{equation}
where $\epsilon_s\ll \Delta\psi/\psi \ll 1$, $\epsilon_s\ll
\Delta\Theta \ll 1$, $\epsilon_s\ll \Delta\zeta \ll 1$, and
$L/c_s\ll\Delta t\ll \tau_E$. Here $\tau_E := \epsilon_s^{-2} L/c_s$
is the transport time scale. For any function $g(\boldr,t)$, we define
\begin{eqnarray}
g^\lw &:= \left\langle g
\right\rangle_{\rm{T}}\nonumber\\[5pt]
g^\sw &:= g - g^\lw,
\end{eqnarray}
which satisfies the following obvious properties:
\begin{eqnarray}
&&\left[g^\lw\right]^\lw = g^\lw,\nonumber\\[5pt]
&&\left[g^\sw\right]^\lw = 0,\nonumber\\[5pt]
&&\left[g h\right]^\lw = g^\lw h^\lw + \left[g^\sw h^\sw\right]^\lw, 
\end{eqnarray}
for any two functions $g(\boldr,t)$ and $h(\boldr,t)$.

We decompose the fields of our theory using the transport average:
\begin{eqnarray}
f_\spe& = f_\spe^\lw + f_\spe^\sw,\nonumber\\
\varphi& = \varphi^\lw
+ \varphi^\sw.
\end{eqnarray}
The length and time scales of $f_\spe^\sw$ and $\varphi^\sw$
correspond to the turbulence, whereas the scales of $f_\spe^\lw$ and
$\varphi^\lw$ are those of the slowly varying profiles. Now, we are
ready to start enumerating the ordering assumptions. The
long-wavelength component of the distribution function is assumed to
be larger than the short-wavelength piece by a factor of
$\epsilon_s^{-1} \gg 1$; the long-wavelength piece of the potential is
itself comparable to the kinetic energy of the particles and its
short-wavelength component is also small in $\epsilon_s$, i.e.
\begin{eqnarray} \label{orderings}
\frac{v_{t\spe}^3 f_\spe^\sw}{n_{e0}}\sim \frac{Z_\spe e
\varphi^\sw}{m_\spe v_{t\spe}^2} \sim \epsilon_s,
\nonumber\\[5pt]
\frac{v_{t\spe}^3 f_\spe^\lw}{n_{e0}}\sim
\frac{Z_\spe e \varphi^\lw}{m_\spe v_{t\spe}^2}
\sim 1.
\end{eqnarray}
Here, $n_{e0}$ is a typical electron density.

We turn to the size of the space and time
derivatives of the long- and short-wavelength components of our
fields. The long-wavelength components $f_\spe^\lw$ and $\varphi^\lw$ are
characterized by large spatial scales, of the order of the macroscopic
scale $L$, and long time scales, of the order of the transport time
scale, $\tau_E$, i.e.
\begin{eqnarray}
\nabla_\boldr \ln f_\spe^\lw, \ \nabla_\boldr
\ln \varphi^\lw
\sim 1/L, \nonumber\\[5pt]
\partial_t \ln f_\spe^\lw ,\
\partial_t \ln \varphi^\lw \sim\epsilon_s^2 c_s/L.
\end{eqnarray}
The short-wavelength components $f_\spe^\sw$ and $\varphi^\sw$ have
perpendicular wavelengths of the order of the sound gyroradius, and
short time scales of the order of the turbulence correlation time. The
parallel correlation length of the short-wavelength component is much
longer than its characteristic perpendicular
wavelength, and it is comparable to the size of the machine. That is,
$f_\spe^\sw$ and $\varphi^\sw$ are characterized by
\begin{eqnarray}
\bun\cdot\nabla_\boldr \ln f_\spe^\sw,\ \bun\cdot\nabla_\boldr
\ln \varphi^\sw \sim 1/L,
\nonumber\\[5pt]
\nabla_{\boldr_\perp} \ln f_\spe^\sw,\ \nabla_{\boldr_\perp}
\ln \varphi^\sw \sim 1/\rho_s,
\nonumber\\[5pt]
\partial_t \ln f_\spe^\sw, \ \partial_t 
\ln \varphi^\sw \sim
c_s/L,
\end{eqnarray}
with $\bun :=\bB/B$.

The magnetic field only contains long-wavelength
components,
\begin{eqnarray}
\nabla_\boldr \ln |\bB| \sim 1/L.
\end{eqnarray}

Finally, we order the collision frequency by the relation
\begin{equation}
  \epsilon_s \ll \nu_{*\spe
    \spe^\prime}\ll \frac{1}{\epsilon_s}\, ,
\end{equation}
where we are defining the collisionality as
\begin{equation}\label{eq:def_collisionality}
\nu_{*\spe \spe^\prime}:=L\nu_{\spe \spe^\prime}/v_{t\spe}
\end{equation}
 and the collision frequency as
\begin{equation}
\nu_{\spe \spe^\prime}:= \frac{4\sqrt{2\pi}}{3}
\frac{Z_\spe^2 Z_{\spe'}^2 n_{e0} e^4 }{m_\spe^{1/2} T_{\spe}^{3/2}}\ln\Lambda,
\end{equation}
which coincides with Braginskii's definition~\cite{Braginskii65} for
$\spe=e$ and $\spe'=i$.

The ordering introduced above leads to consistent equations to each
order in $\epsilon_s$ that capture the physics of microturbulence in
strongly magnetized plasmas. This helps not only from a conceptual
perspective but also, and above all, from the simulation point of
view. The removal of the gyromotion time scale saves enormous
computational time, allowing routine simulation of magnetized plasma
turbulence~\cite{dorland00, dannert05, candy03, peeters09,
  barnes11}. After introducing suitably normalized variables in
subsection \ref{sec:dimensionlessvariables}, we explain how to carry
out the gyrokinetic expansion in subsection
\ref{sec:gyrokinCoorTransf}.

\subsection{Dimensionless variables}
\label{sec:dimensionlessvariables}

The expansion of the equations to high order in $\epsilon_s$ is easier
if we work in non-dimensionalized variables adapted to the gyrokinetic
ordering~\cite{ParraCalvo2011}. We employ the species-independent
normalization
\begin{eqnarray} \label{norm_spindep}
\underline{t} = \frac{c_s t}{L}, \ \underline{\boldr} =
\frac{\boldr}{L}, \ \underline{\bA} = \frac{\bA}{B_0 L}, \
\underline{\varphi} = \frac{e \varphi}{\epsilon_s T_{e0}}, \
\nonumber\\[5pt]
\underline{H_\spe} = \frac{H_\spe}{T_{e0}}, \
\underline{n_{\sigma}}=\frac{n_\sigma}{n_{e0}}, \
\underline{T_{\sigma}}=\frac{T_\sigma}{T_{e0}},
\end{eqnarray}
\noindent for time, space, vector potential, electrostatic
potential, Hamiltonian, particle density, and temperature; and the
species-dependent normalization
\begin{equation} \label{norm_spdep}
\underline{\bv_\spe} = \frac{\bv_\spe}{v_{t\spe}}, \
\underline{f_\spe} = \frac{v_{t\spe}^3}{n_{e0}} f_\spe,
\end{equation}
\noindent for velocities and distribution functions.

In dimensionless variables, the Fokker-Planck
equation~\eq{eq:FPinitial} becomes
\begin{eqnarray}\label{eq:FPnon-dim}
\partial_{\underline{t}}\, \underline{f_\spe} +
\tau_\spe
\left\{\underline{f_\spe},\underline{H_\spe}
\right\}_{\underline{\bX}} = \tau_\spe
\sum_{\spe'}\underline{C_{\spe \spe'}}
[\underline{f_\spe},\underline{f_{\spe'}}]
(\underline{\boldr},\underline{\bv}),
\end{eqnarray}
where
\begin{equation}
\tau_\spe = \frac{v_{t\spe}}{c_s} =
\sqrt{\frac{m_i}{m_\spe }}\, ,
\end{equation}
and the Poisson bracket of two functions
$g_1(\underline{\boldr},\underline{\bv})$,
$g_2(\underline{\boldr},\underline{\bv})$ (we no longer write the
subindex $\spe$ in $\bv_\spe$) is defined by
\begin{eqnarray}
\{g_1,g_2\}_{\underline{\bX}} &=
\left(\nabla_{\underline{\boldr}}
        g_1\cdot\nabla_{\underline{\bv}} g_2 - \nabla_{\underline{\bv}}
        g_1\cdot\nabla_{\underline{\boldr}} g_2\right) 
\nonumber\\[5pt]
&+
        \frac{1}{\epsilon_\spe}
        \underline{\bB}\cdot(\nabla_{\underline{\bv}}
        g_1\times\nabla_{\underline{\bv}} g_2).
\end{eqnarray}
Here $\underline{\bX}\equiv (\underline{\boldr},\underline{\bv})$ are
the dimensionless cartesian coordinates. The normalized collision
operator is
\begin{eqnarray}\label{eq:collisionoperatornondim}
\fl
\underline{C_{\spe \spe'}}
[\underline{f_\spe},\underline{f_{\spe'}}]
(\underline{\boldr},\underline{\bv}) =\nonumber\\[5pt]
\fl
\hspace{1cm}\underline{\gamma_{\spe\spe'}} \nabla_{\underline{\bv}}\cdot
\int \matW\left(\tau_\spe\underline{\bv} -
\tau_{\spe^\prime}\underline{\bv'}\right) \cdot
\Bigg( \tau_\spe\underline{f_{\spe'}}(\underline{\boldr},\underline{\bv'},
\underline{t})
\nabla_{\underline{\bv}}
\underline{f_\spe}(\underline{\boldr},\underline{\bv},\underline{t})
\nonumber\\[5pt]
\hspace{1cm}
\fl
-
\tau_{\spe'}
\underline{f_{\spe}}(\underline{\boldr},\underline{\bv},\underline{t})
\nabla_{\underline{\bv'}} \underline{f_{\spe'}}
(\underline{\boldr},\underline{\bv'},\underline{t}) \Bigg)
\dd^3\underline{v'},\nonumber\\
\end{eqnarray}
with
\begin{equation}
\underline{\gamma_{\spe\spe'}}:= \frac{2 \pi
Z_\spe^2 Z_{\spe'}^2 n_{e0} e^4 L}{T_{e0}^2}\ln\Lambda.
\end{equation}
Observe that $\underline{\gamma_{\spe\spe'}}$ is the usual
collisionality parameter $\nu_{*\spe \spe^\prime}$ in
\eq{eq:def_collisionality} up to a factor of order unity.

As for Poisson's equation~\eq{eq:Poisson},
\begin{eqnarray}\label{eq:Poissonnondim}
 \frac{\epsilon_s \lambda_{De}^2}{L^2} \nabla_{\underline{\boldr}}^2\, 
\underline{\varphi} (\underline{\boldr}, \underline{t}) = -\sum_\spe Z_\spe
\int
\underline{f_\spe}(\underline{\boldr},
\underline{\bv}, \underline{t}) \dd^3 \underline{v},
\end{eqnarray}
where
\begin{equation}
\lambda_{De} = \sqrt{\frac{T_{e0}}{4\pi e^2 n_{e0}}}
\end{equation}
is the electron Debye length. We assume that the Debye
length is sufficiently small that we can neglect
the left-hand side of \eq{eq:Poissonnondim}, so quasineutrality
\begin{eqnarray}\label{eq:Quasineutralitynondim}
 \sum_\spe Z_\spe
\int \underline{f_\spe}(\underline{\boldr}, \underline{\bv},
\underline{t}) \dd^3 \underline{v}=0
\end{eqnarray}
 holds. 

In dimensionless variables the short-wavelength
electrostatic potential and distribution functions satisfy
\begin{eqnarray}\label{eq:ordering_sw_dimensionless}
\underline{\varphi}^\sw(\underline{\boldr}, \underline{t})\sim 1,
\nonumber\\[5pt]
\underline{f_\spe^\sw} (\underline{\boldr},\underline{\bv},
\underline{t})
\sim \epsilon_s,\nonumber\\[5pt]
{\bun}(\underline{\boldr}) \cdot \nabla_{\underline{\boldr}}\,
\underline{\varphi}^\sw (\underline{\boldr}, \underline{t}) \sim 1,
\nonumber\\[5pt]
{\bun}(\underline{\boldr}) \cdot \nabla_{\underline{\boldr}}\,
\underline{f_\spe^\sw} (\underline{\boldr},\underline{\bv},
\underline{t}) \sim \epsilon_s,
\nonumber\\[5pt]
\nabla_{\underline{\boldr}_{\perp}}\, \underline{\varphi}^\sw
 (\underline{\boldr},
\underline{t})  \sim 1/\epsilon_s,\nonumber\\[5pt]
\nabla_{\underline{\boldr}_{\perp}}\, \underline{f_\spe^\sw}
 (\underline{\boldr},\underline{\bv},
\underline{t})  \sim 1,
\nonumber\\[5pt] \partial_{\underline{t}} \underline{\varphi}^\sw (\underline{\boldr}, \underline{t}) \sim 1,
\nonumber\\[5pt] \partial_{\underline{t}} \underline{f_\spe^\sw} (\underline{\boldr},\underline{\bv},
\underline{t}) \sim \epsilon_s.
\end{eqnarray}
The normalized functions $\underline{\varphi}^\sw$ and
$\underline{f_\spe}^\sw$ are of different size due to our choice of
dimensionless variables (see \eq{norm_spindep} and \eq{norm_spdep}).

As for the long-wavelength
components,
\begin{eqnarray}\label{eq:ordering_lw_dimensionless}
  \underline{\varphi}^\lw(\underline{\boldr},
  \underline{t})\sim 1/\epsilon_s,\nonumber\\[5pt]
  \underline{f_\spe^\lw}(\underline{\boldr},\underline{\bv},
  \underline{t})\sim 1,\nonumber\\[5pt]
  \nabla_{\underline{\boldr}}\, \underline{\varphi}^\lw
  (\underline{\boldr},
  \underline{t})  \sim 1/\epsilon_s,\nonumber\\[5pt]
  \nabla_{\underline{\boldr}}\, \underline{f_\spe^\lw}
  (\underline{\boldr},
  \underline{t})  \sim 1,
  \nonumber\\[5pt] \partial_{\underline{t}} \underline{\varphi}^\lw (\underline{\boldr}, \underline{t}) \sim \epsilon_s,
  \nonumber\\[5pt] \partial_{\underline{t}} \underline{f_\spe^\lw}(\underline{\boldr},\underline{\bv},
  \underline{t}) \sim \epsilon_s^2.
\end{eqnarray}

The following notation is adopted when we expand
$\underline{\varphi}^\lw(\underline{\boldr},
\underline{t})$ in powers of $\epsilon_s$:
\begin{eqnarray}\label{eq:potentiallw1}
\fl
\underline{\varphi}^\lw(\underline{\boldr},
\underline{t}) :=
\frac{1}{\epsilon_s}\underline{\varphi_0}(\underline{\boldr},
\underline{t}) +
\underline{\varphi_1}^\lw(\underline{\boldr},
\underline{t}) + \epsilon_s
\underline{\varphi_2}^\lw(\underline{\boldr},
\underline{t})
\nonumber\\[5pt]
\fl\hspace{1cm} +
O(\epsilon_s^2).
\end{eqnarray}
Similarly,
\begin{eqnarray}
\fl
\underline{\varphi}^\sw(\underline{\boldr},
\underline{t}) :=
\underline{\varphi_1}^\sw(\underline{\boldr},
\underline{t}) + \epsilon_s
\underline{\varphi_2}^\sw(\underline{\boldr},
\underline{t}) +
O(\epsilon_s^2).
\end{eqnarray}

From now on we do not underline variables but assume that we are
working with the dimensionless ones. Only in Section
\ref{sec:deviationQS} we go back to dimensionful variables.

\subsection{Gyrokinetic coordinate transformation}
\label{sec:gyrokinCoorTransf}

The goal of the gyrokinetic expansion is to use the ordering
assumptions in subsection \ref{sec:orderingANDseparationOFscales} to
remove from the kinetic equation the degree of freedom associated to
the gyromotion, order by order. Grossly speaking, there are two
typical approaches to this objective. Iterative methods act by
directly averaging the equations of motion~\cite{frieman82, lee83,
  wwlee83, bernstein85, parra08} to each order in the expansion
parameter. Hamiltonian and Lagrangian methods rely on the machinery of
analytical mechanics and the averaging procedure is implemented by
requiring that certain geometrical objects be gyrophase
independent~\cite{dubin83, hahm88, brizard07,ParraCalvo2011}. Both
approaches try to find new coordinates on phase space in which the
slow and fast degrees of freedom are decoupled. Such coordinates, that
are not uniquely defined, are called gyrokinetic coordinates.

The complete calculation of the gyrokinetic system of equations to
second order is given in reference \cite{ParraCalvo2011} in the
phase-space Lagrangian formalism. The latter was applied to the
problem of guiding-center motion by Littlejohn~\cite{littlejohn83} and
has been used extensively in modern formulations of
gyrokinetics~\cite{brizard07}. In reference \cite{ParraCalvo2011} a
change of variables is performed in \eq{eq:FPnon-dim} and
\eq{eq:Quasineutralitynondim}, that decouples the fast degree of
freedom, the gyrophase, from the slow ones in the absence of
collisions. This decoupling is achieved by eliminating the dependence
of the phase-space Lagrangian on the gyrophase order by order in
$\epsilon_\spe$. Let us denote the transformation from the new
phase-space coordinates $\bZ\equiv\{\bR,u,\mu,\theta\}$ to the
euclidean ones $\bX\equiv\{\boldr,\bv\}$ by $\cT_\spe$,
\begin{equation}
(\boldr,\bv) = \cT_\spe(\bR,u,\mu,\theta,t).
\end{equation}
The transformation is, in general, explicitly time-dependent and is
expressed as a power series in $\epsilon_\spe$.  Here, $\bR$, $u$,
$\mu$, and $\theta$ are deformations of the guiding-center position,
parallel velocity, magnetic moment, and gyrophase,
respectively. Namely,
\begin{eqnarray}\label{eq:TinvZerothOrder}
\bR = \boldr
 -\epsilon_\spe\frac{1}{B(\boldr)}\bun(\boldr)\times\bv
+ O(\epsilon_\spe^2)
,\nonumber\\[5pt]
u = \bv\cdot\bun(\boldr) + O(\epsilon_\spe)
,\nonumber\\[5pt]
\mu = \frac{1}{2 B(\boldr)}
\left(\bv - \bv\cdot\bun(\boldr)\bun(\boldr)\right)^2
+ O(\epsilon_\spe)
,\nonumber\\[5pt]
\theta = \arctan
\left(\frac{\bv\cdot\eun_2(\boldr)}{\bv\cdot\eun_1(\boldr)}
\right)
+ O(\epsilon_\spe),
\end{eqnarray}
where the unit vectors $\eun_1 (\boldr)$ and $\eun_2 (\boldr)$ are
orthogonal to each other and to $\bun (\boldr)$, and satisfy $\eun_1
(\boldr) \times \eun_2 (\boldr) = \bun (\boldr)$ at every location
$\boldr$.

We want to write the Fokker-Planck equation in gyrokinetic
coordinates. Denote by $\cT^*_\spe$ the pull-back transformation
induced by $\cT_\spe$. Acting on a function $g(\bX, t)$, $\cT_\spe^* g
(\bZ,t)$ is simply the function $g$ written in coordinates $\bZ$,
i.e.
\begin{equation}
\cT_\spe^* g (\bZ,t) = g(\cT_\spe(\bZ,t), t).
\end{equation}
Now, defining $ F_\spe:=\cT_\spe^* f_\spe $, we transform
\eq{eq:FPnon-dim} and get
\begin{eqnarray}\label{eq:FokkerPlancknondimgyro}
\fl\partial_{t} F_\spe + \tau_\spe
\left\{F_\spe,\overline{H}_\spe \right\}_\bZ =
\nonumber
\\[5pt]
\fl\hspace{1cm}
\tau_\spe \sum_{\spe'}\cT^{*}_{\spe} C_{\spe
\spe'} [\cT^{-1*}_{\spe} F_\spe,\cT^{-1*}_{\spe'}F_{\spe'}]
(\bZ, t),
\end{eqnarray}
where $\cT^{-1*}_{\spe}$ is the pull-back transformation that
corresponds to $\cT_\spe^{-1}$, i.e. $\cT_\spe^{-1*} F_\spe (\bX,t) =
F_\spe(\cT_\spe^{-1}(\bX,t),t)$, and the Poisson bracket in the new
coordinates is expressed as
\begin{eqnarray} \label{eq:poissonbracket}
\fl
\{G_1,G_2\}_\bZ &=
\frac{1}{\epsilon_\spe}\left(\partial_\mu G_1
\partial_\theta G_2
 - \partial_\theta
G_1\partial_\mu G_2
\right) 
\nonumber
\\[5pt]
\fl&
+
\frac{1}{B_{||,\spe}^*}{\bf B}_\spe^* \cdot
\left(\nabla^*_{\bR}G_1\partial_u G_2
-
\partial_u G_1
\nabla_{\bR}^*G_2\right) \nonumber
\\[5pt]
\fl&
+ \frac{\epsilon_\spe }{B_{||}^*} \nabla_{\bR}^*G_1 \cdot (
\bun \times \nabla_{\bR}^*G_2 ),
\end{eqnarray}
with
\begin{eqnarray} \label{Bstar}
\fl
\bB_\spe^* (\bR, u, \mu) := \bB(\bR) + \epsilon_\spe u \nabla_\bR \times
\bun (\bR)
\nonumber \\[5pt]
\fl\hspace{1cm} - \epsilon^2_\spe \mu \nabla_\bR \times \bK (\bR),
\end{eqnarray}
\begin{eqnarray} \label{Bstar_par}
\fl B^*_{||,\spe} (\bR, u, \mu) := \bB_\spe^* (\bR, u, \mu) \cdot \bun (\bR)
\nonumber \\[5pt] = B(\bR) + \epsilon_\spe u \bun(\bR) \cdot \nabla_\bR
\times \bun (\bR)
\nonumber \\[5pt]
 - \epsilon^2_\spe \mu \bun(\bR) \cdot \nabla_\bR
\times \bK (\bR),
\end{eqnarray}
\begin{equation}
\nabla_{\bR}^* := \nabla_{\bR} - {\bf
K}(\bR)\partial_\theta,
\end{equation}
and
\begin{equation}\label{eq:defvectorK}
\fl
\bK (\bR) = \frac{1}{2} \bun(\bR) \bun(\bR) \cdot \nabla_\bR
\times \bun(\bR) - \nabla_\bR \eun_2 (\bR) \cdot \eun_1 (\bR).
\end{equation}

The gyrokinetic transformation is not unique. To compare with standard
references, we have chosen $(\bR,u,\mu,\theta)$ so that the Poisson
bracket has the form \eq{eq:poissonbracket}, precisely the same
employed in \cite{brizard07}.

In gyrokinetic variables the quasineutrality equation reads
\begin{eqnarray}\label{eq:gyroQuasineutrality}
\fl\sum_\spe Z_\spe  \int |\det\left(J_{\spe}\right)|
F_\spe
\delta\Big(\pi^{\boldr}\Big(\cT_{\spe}(\bZ,
t)\Big)-\boldr\Big)\dd ^6Z = 0,
\end{eqnarray}
where $\pi^\boldr(\boldr,\bv):=\boldr$, and the Jacobian of the
transformation to $O(\epsilon_\spe ^2)$ is
\begin{eqnarray}\label{eq:Jacobian}
|\det(J_{\spe} )| = B_{||,\spe}^*.
\end{eqnarray}

We do not include in this paper all the details of the derivation of
the gyrokinetic Hamiltonian and change of coordinates, and we refer
the reader to \cite{ParraCalvo2011} and \cite{CalvoParra2012}. Using
the explicit expressions for $\overline{H}_\spe$ and for $\cT_\spe^*$
given in those references, we will compute in Section
\ref{sec:gyrokineticstosecondorder} the long-wavelength limit of the
Fokker-Planck equation \eq{eq:FokkerPlancknondimgyro} and the
quasineutrality equation \eq{eq:gyroQuasineutrality} up to second
order in the expansion parameter $\epsilon_s$. The manipulations are
very similar to those made in \cite{CalvoParra2012} for the tokamak,
so we will present the final equations and then turn
to discuss the topics that are specific to stellarators.

Finally, we recall that the gyrokinetic equations are written naturally
in terms of a function $\phi_\spe$ defined as
\begin{equation}
\phi_\spe(\bR,\mu,\theta,t) :=
\varphi(\bR+\epsilon_\spe\rhobf(\bR,\mu,\theta),t),
\end{equation}
with the gyroradius vector, $\rhobf$, given by
\begin{equation}\label{eq:defrho}
\rhobf ( \bR, \mu, \theta) = - \sqrt{\frac{2
\mu}{B(\bR)}} \left[ \sin \theta\, \eun_1 (\bR) - \cos
\theta\, \eun_2 (\bR) \right].
\end{equation}

The gyroaverage of a function $G(\bR,u,\mu,\theta)$ is defined by
\begin{equation}
\langle
G
\rangle(\bR,\mu,t) :=
\frac{1}{2\pi}\int_0^{2\pi}G(\bR,\mu,\theta,t)
\dd\theta.
\end{equation}

It is useful to
introduce the gyrophase-dependent piece of
$\phi$,
\begin{equation}
\tilde\phi_\spe(\bR,\mu,\theta,t) :=
\phi_\spe(\bR,\mu,\theta,t)
 - \langle\phi_\spe\rangle(\bR,\mu,t).
\end{equation}
From the ordering and scale separation assumptions on $\varphi$,
equations~\eq{eq:ordering_sw_dimensionless} and
\eq{eq:ordering_lw_dimensionless}, we obtain that the short-wavelength
component of $\phi_\spe$ is $O(1)$, i.e.
\begin{eqnarray}
\phi^\sw_\spe = \phi^\sw_{\spe 1} + O(\epsilon_s),\nonumber\\
\tilde\phi^\sw_\spe = \tilde\phi^\sw_{\spe 1} + O(\epsilon_s).
\end{eqnarray}
For the long wavelength piece $\phi^\lw_\spe$, we use that it is
possible to expand around $\boldr = \bR$. With the double-dot
convention for an arbitrary matrix $\matM$ defined by $\mathbf
u\mathbf v :\matM = \mathbf v \cdot\matM \cdot \mathbf u$, we write
\begin{eqnarray}\label{eq:potentiallw2}
\fl\langle\phi_\spe^\lw\rangle(\bR,\mu,t) =
\frac{1}{\epsilon_s}\varphi_0(\bR,t)
+\varphi_1^\lw(\bR,t)
\nonumber\\[5pt]
\fl
\hspace{0.5cm}
+\epsilon_s \Bigg( \frac{\mu}{2\lambda_\spe^2 B(\bR)}
 (\matI-\bun(\bR)\bun(\bR)):\nabla_{\bR}\nabla_{\bR}
\varphi_0(\bR,t)
\nonumber\\[5pt]
\fl
\hspace{0.5cm}
+\varphi_2^\lw(\bR,t) \Bigg)
 +O(\epsilon_s^2)
\end{eqnarray}
and
\begin{equation}\label{eq:potentiallw3}
\fl\tilde\phi_\spe^\lw(\bR,\mu,\theta,t) =
\frac{1}{\lambda_\spe} \rhobf(\bR,\mu,\theta)
\cdot\nabla_{\bR}\varphi_0(\bR,t) + O(\epsilon_s),
\end{equation}
giving $\tilde\phi^\lw_\spe = O(1)$. We have expanded up to first
order in $\epsilon_s$ in \eq{eq:potentiallw2} because it will be
needed later in this paper. This ordering, in which the gyroaveraged
long-wavelength component is large and the short-wavelength component
is small, is similar to the one used in references \cite{Dimits92} and
\cite{Dimits12}.

\section{Long-wavelength gyrokinetic equations up to 
second order}
\label{sec:gyrokineticstosecondorder}

As advanced in subsection \ref{sec:gyrokinCoorTransf}, once the
explicit expressions for the gyrokinetic transformation, $\cT_\spe^*$,
and Hamiltonian, $\overline{H}_\spe$, are known, the remaining task
consists of expanding equations \eq{eq:FokkerPlancknondimgyro} and
\eq{eq:gyroQuasineutrality} to $O(\epsilon_s^2)$ and taking the
long-wavelength limit. The basic ingredients, $\cT_\spe^*$ and
$\overline{H}_\spe$ to $O(\epsilon_s^2)$, have been computed for
general magnetic geometry in \cite{ParraCalvo2011}. These results were
employed for the first time in \cite{CalvoParra2012} to obtain the
long-wavelength limit of the set of gyrokinetic equations in a
tokamak. Here, we perform the same calculation for
stellarators. Although not emphasized or exploited in
\cite{CalvoParra2012}, a number of intermediate expressions in
\cite{CalvoParra2012} are valid for general toroidal systems. Hence,
when working out the long-wavelength gyrokinetic Fokker-Planck and
quasineutrality equations in the following subsections, we make use of
some results of \cite{CalvoParra2012} and explain in detail the
problems specific to stellarators. More specifically, in this section
we use Appendix A in \cite{CalvoParra2012}, expand
\eq{eq:FokkerPlancknondimgyro} and \eq{eq:gyroQuasineutrality} to
$O(\epsilon_s^2)$, and take the long-wavelength limit following
Sections 3 and 4 of \cite{CalvoParra2012}, but we do not assume that
the magnetic field is axisymmetric as in a tokamak.

\subsection{Fokker-Planck equation at long wavelengths}
\label{sec:FokkerPlancklong}

The objective in this subsection is to calculate the long-wavelength
limit of the gyrokinetic Fokker-Planck equation
\eq{eq:FokkerPlancknondimgyro} up to second order in $\epsilon_s$ for
an arbitrary stellarator. We will expand $F_\spe$ as
\begin{eqnarray}\label{eq:orderingF}
F_\spe&=\sum_{n=0}^\infty
\epsilon_\spe^n F_{\spe n} = 
\sum_{n=0}^\infty
\epsilon_\spe^n F^\lw_{\spe n}
+
\sum_{n=1}^\infty
\epsilon_\spe^n F^\sw_{\spe n}.
\end{eqnarray}
From the
ordering assumptions enumerated in Section
\ref{sec:secondordergyrokinetics} it follows that
\begin{eqnarray}
F_{\spe n}\sim 1,\  n\ge 0, \nonumber\\[5pt]
\bun (\bR) \cdot \nabla_{\bR} F_{\spe n}\sim 1,\  n\ge 0.
\end{eqnarray}

The long-wavelength component of every $F_{\spe n}$ must have
perpendicular derivatives of order unity and time derivatives of order
$\epsilon_s^2$ in normalized variables, i.e.
\begin{eqnarray}
\nabla_{\bR_\perp}F_{\spe n}^\lw\sim 1,\  n\ge 0,\nonumber\\[5pt]
\partial_t F_{\spe n}^\lw\sim \epsilon_s^2,\  n\ge 0.
\end{eqnarray}
Finally, the zeroth-order distribution function must have an
identically vanishing short-wavelength component. The perpendicular
gradient of the rest of the short-wavelength components is of order
$\epsilon_\spe^{-1}$ and the time derivative is of order unity,
\begin{eqnarray}
F_{\spe 0}^\sw\equiv 0, \nonumber\\[5pt]
 \nabla_{\bR_\perp}F_{\spe n}^\sw\sim\epsilon_\spe^{-1},\  n\ge 1,
\nonumber\\[5pt]
 \partial_t F_{\spe n}^\sw\sim 1,\  n\ge 1.
\end{eqnarray}

\subsubsection{Long-wavelength Fokker-Planck equation to 
$O(\epsilon_\spe^{-1})$.}
\label{sec:FokkerPlancklongminus1}

The coefficient of $\epsilon_\spe^{-1}$ in
\eq{eq:FokkerPlancknondimgyro} simply gives
\begin{eqnarray}
-\tau_\spe B\partial_\theta F_{\spe 0} = 0,
\end{eqnarray}
implying that $F_{\spe 0}$ is independent of $\theta$.

\subsubsection{Long-wavelength Fokker-Planck equation to 
$O(\epsilon_\spe^0)$.}
\label{sec:FokkerPlancklong0}

From the $O(\epsilon_\spe^0)$ terms in \eq{eq:FokkerPlancknondimgyro}
it is deduced, after a calculation identical to the one in
\cite{CalvoParra2012}, that the first-order distribution function is
gyrophase independent,
\begin{equation}
\fl\partial_\theta F_{\spe 1} = 0,
\end{equation}
that $F_{\spe 0}$ is a Maxwellian with zero mean flow,
\begin{eqnarray}\label{eq:F0}
&& \fl F_{\spe 0}(\bR,u,\mu,t)=
\frac{n_{\spe}}{(2\pi T_{\spe})^{3/2}}
\exp\left(-\frac{\mu B(\bR) + u^2/2}{T_{\spe}}\right),
\end{eqnarray}
and that $n_\spe\equiv n_\spe(\psi,t)$, $T_\spe \equiv
T_\spe(\psi,t)$, and $\varphi_0 \equiv \varphi_0(\psi,t) $ are flux
functions. The temperature has to be the same for all the species;
that is, $T_\spe = T_{\spe'}$ for every pair $\spe, \spe'$. If a
subsidiary expansion in the mass ratio $\sqrt{m_e/m_i}\ll 1$ is
performed, or equivalently, if $\tau_e\sim\lambda_e\gg 1$ is used, the
electron temperature can decouple from the ion temperature.

\subsubsection{Long-wavelength Fokker-Planck equation to $O(\epsilon_\spe)$.}
\label{sec:FokkerPlancklong1}

The equations presented in this subsection involve the collision
operator, which is typically written in coordinates $\bX\equiv
(\boldr,\bv)$. We avoid transforming the kernel that defines this
operator by transforming, instead, the gyrokinetic
distribution function $F_\spe(\bR,u,\mu,\theta,t)$ from gyrokinetic
coordinates $\bZ\equiv (\bR,u,\mu,\theta)$ to euclidean coordinates
$\bX\equiv (\boldr,\bv)$. We denote the coefficients of the expansion of
$\cT_\spe$ and its inverse $\cT_\spe^{-1}$ by
\begin{eqnarray}
\fl\bX = \cT_{\spe}(\bZ,t) 
= \cT_{\spe,0}(\bZ,t)  + \epsilon_\spe \cT_{\spe,1}(\bZ,t)  +
O(\epsilon_\spe^2),\\[5pt]
\fl\bZ = \cT_{\spe}^{-1}(\bX,t) = \cT_{\spe,0}^{-1}(\bX,t)
+ \epsilon_\spe \cT_{\spe,1}^{-1}(\bX,t)
\nonumber\\ 
 +
\epsilon_\spe^2 \cT_{\spe,2}^{-1}(\bX,t) 
+O(\epsilon_\spe^3).
\end{eqnarray}
In the present subsection we only need $\cT_{\spe,0}$, the transformation
$\cT_\spe$ for $\epsilon_\spe = 0$ (compare with
\eq{eq:TinvZerothOrder}),
\begin{eqnarray}\label{eq:transformation_oderzero}
\cT_{\spe,0}(\bR,u,\mu,\theta)=(\bR,u\bun(\bR)+\rhobf(\bR,\mu,\theta)\times
\bB(\bR)).
\end{eqnarray}
In subsequent subsections some pieces of $\cT_{\spe,1}$ and
$\cT_{\spe,2}$ are required.

From the Fokker-Planck equation to $O(\epsilon_\spe)$ one gets an
equation for the gyrophase-dependent piece of $F_{\spe 2}^\lw$,
\begin{eqnarray}
\fl  - B\partial_\theta (F_{\spe 2}^\lw -\langle F_{\spe 2}^\lw\rangle)
\nonumber\\[5pt]
\fl\hspace{1cm}
=
\sum_{\sigma'}\cT^{*}_{\spe,0}
C_{\spe\spe'}\Bigg[ \frac{1}{T_{\spe}} \Bigg( \bv\cdot{\mathbf
V}^p_\spe 
\nonumber\\[5pt]
\fl\hspace{1cm}
+ \Bigg( \frac{v^2}{2T_{\spe}}-\frac{5}{2}
\Bigg)\bv\cdot{\mathbf V}^T_{\spe}
\Bigg)\cT^{-1*}_{\spe,0}F_{\spe 0} ,\cT^{-1*}_{\spe',0}F_{\spe'
0} \Bigg]
\nonumber\\[5pt]
\fl\hspace{1cm} + \sum_{\sigma'}
\frac{\lambda_\spe}{\lambda_{\spe'}}\cT^{*}_{\spe,0}
C_{\spe\spe'} \Bigg[ \cT^{-1*}_{\spe,0}F_{\spe 0}, \frac{1}{
T_{\spe'}} \Bigg( \bv\cdot{\mathbf V}^p_{\spe'}
\nonumber\\[5pt]
\fl\hspace{1cm}
 + \left(
\frac{v^2}{2T_{\spe'}}-\frac{5}{2} \right)\bv\cdot{\mathbf
V}^T_{\spe'} \Bigg)\cT^{-1*}_{\spe',0}F_{\spe' 0}
\Bigg],
\end{eqnarray}
where the velocities ${\mathbf V}^p_{\spe}$ and
${\mathbf V}^T_{\spe}$ are defined by
\begin{equation}\label{eq:velocitiesPandT}
{\mathbf V}^p_\spe := \frac{1}{n_\spe B}\bun\times\nabla p_\spe, \quad
{\mathbf V}^T_\spe := \frac{1}{B}\bun\times\nabla T_\spe,
\end{equation}
and $p_\spe := n_\spe T_\spe$ is the pressure of species $\spe$.

One also gets an equation for $F_{\sigma 1}^\lw$ (recall from Section
\ref{sec:FokkerPlancklong0} that $F_{\spe 1}$ is
gyrophase-independent),
\begin{eqnarray}\label{eq:Vlasovorder1gyroav}
\fl
\left(u\bun\cdot\nabla_\bR - \mu\bun\cdot\nabla_\bR
B\partial_u
\right)
F_{\spe 1}^\lw
\nonumber\\[5pt]
\fl\hspace{1cm}
+
\left(
\frac{Z_\spe}{T_\spe}\partial_\psi\varphi_0
+\Upsilon_\spe
\right)
\bv_M\cdot\nabla_\bR\psi
\, F_{\spe 0}
\nonumber\\[5pt]
\fl\hspace{1cm}
+
\frac{Z_\spe\lambda_\spe}{T_\spe} u \bun\cdot\nabla_\bR\varphi_1^{\lw}
F_{\spe 0}
\nonumber\\[5pt]
\fl\hspace{1cm}
= \sum_{\sigma'}\cT_{\spe,0}^*
C_{\spe\spe'}\left[ \cT^{-1*}_{\spe,0}
F_{\spe
1}^\lw
,\cT^{-1*}_{\spe',0}F_{\spe' 0} \right]
\nonumber\\[5pt]
\fl\hspace{1cm} + \sum_{\sigma'}
\frac{\lambda_\spe}{\lambda_{\spe'}}\cT_{\spe,0}^*
C_{\spe\spe'} \Bigg[ \cT^{-1*}_{\spe,0}F_{\spe 0},
\cT^{-1*}_{\spe',0}
F_{\spe' 1}^\lw
\Bigg].
\end{eqnarray}
This is the well-known drift-kinetic equation for a general
stellarator. Here,
\begin{eqnarray}\label{eq:defUpsilon}
\Upsilon_\spe := 
\partial_\psi \ln n_\spe+
\left(\frac{u^2/2 +\mu B}{T_\spe}-\frac{3}{2}\right)
\partial_\psi \ln T_\spe\,
\end{eqnarray}
and
\begin{eqnarray}\label{eq:vmagnetic}
\bv_M&:=& \bv_\kappa + \bv_{\nabla B}
\end{eqnarray}
is the magnetic drift velocity, where
\begin{eqnarray}
 \bv_\kappa:=\frac{u^2}{B}\bun\times\kappabf,
\\[5pt]
 \bv_{\nabla B}:= \frac{\mu}{B}\bun\times\nabla_\bR B,
\end{eqnarray}
and
\begin{equation}
\kappabf:= \bun\cdot\nabla_\bR\bun
\end{equation}
is the magnetic field curvature. 

Finally, in terms of the non-adiabatic
part of the distribution function,
\begin{eqnarray}\label{eq:defGspe1}
G_{\spe 1}^\lw&:=& F_{\spe 1}^\lw
+
\frac{Z_\spe\lambda_\spe}{T_\spe}\varphi_1^\lw
F_{\spe 0},
\end{eqnarray}
the first-order Fokker-Planck equation reads
\begin{eqnarray}\label{eq:Vlasovorder1gyroav4}
\fl
\left(u\bun\cdot\nabla_\bR - \mu\bun\cdot\nabla_\bR
B\partial_u
\right)
G_{\spe 1}^\lw
\nonumber\\[5pt]
\fl\hspace{1cm}
+
\left(
\frac{Z_\spe}{T_\spe}\partial_\psi\varphi_0
+\Upsilon_\spe
\right)
\bv_M\cdot\nabla_\bR\psi
\, F_{\spe 0}
\nonumber\\[5pt]
\fl\hspace{1cm}
= \sum_{\sigma'}\cT_{\spe,0}^*
C_{\spe\spe'}\left[ \cT^{-1*}_{\spe,0}
G_{\spe
1}^\lw
,\cT^{-1*}_{\spe',0}F_{\spe' 0} \right]
\nonumber\\[5pt]
\fl\hspace{1cm} + \sum_{\sigma'}
\frac{\lambda_\spe}{\lambda_{\spe'}}\cT_{\spe,0}^*
C_{\spe\spe'} \left[ \cT^{-1*}_{\spe,0}F_{\spe 0},
\cT^{-1*}_{\spe',0}
G_{\spe' 1}^\lw
\right].
\end{eqnarray}

The electrostatic potential $\varphi^\lw_1$ does not appear in the
collision operator due to property \eq{eq:Maxwelliansannihilatecoll_nondim}.

\subsubsection{Short-wavelength Fokker-Planck and quasineutrality
  equations to $O(\epsilon_\spe)$.}
\label{sec:FPandQuasineutSW}

We will learn in subsection \ref{sec:FokkerPlancklong2} that the
long-wavelength Fokker-Planck equation to $O(\epsilon_s^2)$ involves
terms containing $F_{\spe 1}^\sw$ and $\phi_{\spe 1}^\sw$. Next, we
give the equations determining these pieces. In order to write in a
compact and precise way the collision terms of the short-wavelength
equations, it is convenient to define a new operator $\modTinv$ acting
on phase-space functions $F(\bR,u,\mu,\theta)$. Namely,
\begin{eqnarray}
\fl\modTinv F(\boldr,\bv) :=
F\Bigg(\boldr-\epsilon_\spe\frac{1}{B(\boldr)}\bun(\boldr)\times\bv,
\bv\cdot\bun(\boldr), \frac{v_\bot^2}{2B(\boldr)},
\nonumber\\[5pt]
\fl\hspace{1cm}
\arctan
\left(
\frac{\bv\cdot\eun_2(\boldr)}{\bv\cdot\eun_1(\boldr)}
\right)\Bigg).
\end{eqnarray}
This operator is useful to write some expressions that contain the
short-wavelength pieces of the distribution function and the
potential, for which it is not possible to expand the dependence on
$\boldr - \epsilon_\spe B(\boldr)^{-1}\bun(\boldr)\times\bv$
around $\boldr$.

The short-wavelength, $O(\epsilon_s)$ terms of
\eq{eq:FokkerPlancknondimgyro} yield
\begin{eqnarray}\label{eq:sworder1distfunction}
\fl\frac{1}{\tau_\spe}\partial_t F_{\spe 1}^\sw
+\left(u\bun\cdot\nabla_\bR-\mu\bun\cdot\nabla_\bR B\partial_u\right)
F_{\spe 1}^\sw
\nonumber\\[5pt]
\fl\hspace{0.5cm}
+
\left[
\frac{Z_\spe\lambda_\spe}{B}
\left(\bun\times\nabla_{\bR_\perp/\epsilon_\spe}
\langle
\phi_{\spe 1}^\sw
\rangle
\right)
\cdot\nabla_{\bR_\perp/\epsilon_\spe}F_{\spe 1}^\sw
\right]^\sw
\nonumber\\[5pt]
\fl\hspace{0.5cm}
+\left(
\bv_M + \bv_{E,\spe}^{(0)}
\right)
\cdot\nabla_{\bR_\perp/\epsilon_\spe}F_{\spe 1}^\sw
\nonumber\\[5pt]
\fl\hspace{0.5cm}
+\frac{Z_\spe\lambda_\spe}{B}
\left(\bun\times\nabla_{\bR_\bot/\epsilon_\spe}
\langle
\phi_{\spe 1}^\sw
\rangle
\right)
\cdot\nabla_\bR F_{\spe 0}
\nonumber\\[5pt]
\fl\hspace{0.5cm}
-Z_\spe\lambda_\spe
\Big(
\bun\cdot\nabla_\bR\langle\phi_{\spe 1}^\sw\rangle
\nonumber\\[5pt]
\fl\hspace{0.5cm}
+\frac{u}{B}
(\bun\times\kappabf)
\cdot
\nabla_{\bR_\perp/\epsilon_\spe}
\langle\phi_{\spe 1}^\sw\rangle
\Big)
\partial_u F_{\spe 0}
\nonumber\\[5pt]
\fl\hspace{0.5cm}
=
\sum_{\spe^\prime} \Bigg\langle \cT_{NP, \spe}^* C_{\sigma \sigma^\prime}
\Bigg [\modTinv F_{\sigma 1}^\sw
\nonumber\\[5pt]
\fl\hspace{0.5cm}
 -
\frac{Z_\spe\lambda_\spe}{T_\spe}
\modTinv\tilde\phi_{\spe 1}^\sw \cT_{\spe,0}^{-1
*} F_{\spe 0}, \cT_{\spe',0}^{-1
*}F_{\spe' 0}
 \Bigg] \Bigg\rangle\nonumber\\[5pt]
\fl\hspace{0.5cm}
 + \
\sum_{\spe^\prime} \frac{\lambda_\spe}{\lambda_{\spe'}}
\Bigg\langle \cT_{NP, \spe}^*
C_{\sigma
\sigma^\prime} \Bigg[ \cT_{\spe,0}^{-1 *}F_{\spe 0} ,
\modTinvprime F_{\sigma' 1}^\sw 
\nonumber\\[5pt]
\fl\hspace{0.5cm}
-
\frac{Z_{\spe'}\lambda_{\spe'}}{T_{\spe'}}
\modTinvprime\tilde\phi_{\spe'
1}^\sw \cT_{\spe',0}^{-1 *}F_{\spe' 0}
 \Bigg] \Bigg\rangle,
\end{eqnarray}
where
\begin{eqnarray}
\fl \bv_{E,\spe}^{(0)}=
\frac{Z_\spe}{B}\bun\times\nabla_{\bR}\varphi_0
\end{eqnarray}
 and the transformation $(\boldr,\bv) = \cT_{NP,\spe}(\bR,u,\mu,\theta)$ is
defined by
\begin{eqnarray}
  \boldr = \bR+\epsilon_\spe\rhobf(\bR,\mu,\theta),\nonumber\\[5pt]
  \bv = u\bun(\bR) + \rhobf(\bR,\mu,\theta)\times\bB(\bR).
\end{eqnarray}
This coincides with the {\it non-perturbative transformation}
introduced in \cite{ParraCalvo2011} and that is why we have kept
the notation $\cT_{NP,\spe}$.

The short-wavelength, $O(\epsilon_s)$ terms of
\eq{eq:gyroQuasineutrality} give
\begin{eqnarray}\label{eq:quasinautralitySWorder1appendix}
\fl\sum_\spe \frac{Z_\spe}{\lambda_\spe}
&\int
B
\Bigg[-Z_\spe\lambda_\spe \phiwig_{\spe
      1}^\sw \left(\boldr
-\epsilon_\spe\rhobf(\boldr,\mu,\theta),\mu,\theta,t\right)
\nonumber\\[5pt]
\fl&
\times\frac{F_{\spe 0}(\boldr,u,\mu,t)}{T_{\spe}(\boldr,t)}\nonumber\\[5pt]
\fl&
 + F_{\spe 1}^\sw
\left(\boldr
-\epsilon_\spe\rhobf(\boldr,\mu,\theta),u,\mu,t
\right)
\Bigg]
  \dd u \dd \mu \dd \theta = 0.
\end{eqnarray}

Equations \eq{eq:sworder1distfunction} and
\eq{eq:quasinautralitySWorder1appendix} constitute what is usually
understood by ``the $\delta f$ gyrokinetic set of equations'' for
electrostatic turbulence.

\subsubsection{Long-wavelength Fokker-Planck equation to 
$O(\epsilon_\spe^2)$.}
\label{sec:FokkerPlancklong2}

The gyroaverage of the pieces of order $\epsilon_\spe^2$ in
\eq{eq:FokkerPlancknondimgyro} yields
\begin{eqnarray} \label{eq:FPsecondorder2}
\fl
\left(
u \bun \cdot \nabla_\bR   - \mu \bun \cdot
\nabla_\bR B \, \partial_u
\right)
\left\langle F^\lw_{\sigma 2}\right\rangle
  +
\frac{\lambda_\spe^2}{\tau_\spe}\partial_{\epsilon_s^2 t} F_{\spe 0}
\nonumber\\[5pt]
\fl\hspace{0.5cm} + 
\left(
\bv_M + \bv_{E,\spe}^{(0)}
\right)
 \cdot \nabla_\bR F^\lw_{\sigma 1} \nonumber\\[5pt]
\fl\hspace{0.5cm}
+\left[- Z_\spe \lambda_\spe
\bun \cdot \nabla_\bR \varphi_1^\lw
+u\, \kappabf\cdot\left(\bv_{\nabla B} + \bv_{E,\spe}^{(0)}\right)
\right]
\partial_u F^\lw_{\sigma 1}
\nonumber\\[5pt]
\fl\hspace{0.5cm} + \Bigg[
\bv_{E,\spe}^{(1)}-\frac{u}{B}
(\bun \cdot \nabla_\bR \times \bun)
\left(\bv_M + \bv_{E,\spe}^{(0)}\right)
 \nonumber\\[5pt]
\fl\hspace{0.5cm} 
 - \frac{u \mu}{B}
 (\nabla_\bR \times \bK)_\bot
+ Z_\spe \lambda_\spe \partial_u\Psi_{\phi
B,\spe}^\lw \bun +
\partial_u\Psi_{B,\spe} \bun \Bigg]\cdot\nabla_\bR F_{\spe 0}
\nonumber\\[5pt]
\fl\hspace{0.5cm} - \Bigg \{ Z_\spe\lambda_\spe^2 \bun \cdot
\nabla_\bR \left [ \varphi_2^\lw + \frac{\mu}{2\lambda_\spe^2 B}
(\matI - \bun \bun) : \nabla_\bR \nabla_\bR \varphi_0 \right ] 
  \nonumber\\[5pt]
\fl\hspace{0.5cm}
+\bun \cdot \nabla_\bR \Psi_{B,\spe} + Z_\spe\lambda_\spe
\bun \cdot \nabla_\bR \Psi_{\phi
B,\spe}^\lw
 + Z_\spe^2 \lambda_\spe^2 \bun \cdot
\nabla_\bR \Psi_{\phi,\spe}^\lw
\nonumber\\[5pt]
\fl\hspace{0.5cm}
-u\,\kappabf\cdot\bv_{E,\spe}^{(1)}
+\Bigg[\frac{u^2}{B}\left(\bun\cdot\nabla_\bR\times\bun\right)
\kappabf
\nonumber\\[5pt]
\fl\hspace{0.5cm}
+ \mu\left(\left(\nabla_\bR\times\bK\right)\times\bun\right)\Bigg]
\cdot
\left(\bv_{\nabla B}+\bv_{E,\spe}^{(0)}\right)
\Bigg \} \partial_u F_{\spe 0} 
 \nonumber\\[5pt]
\fl\hspace{0.5cm}
 + 
\frac{Z_\spe \lambda_\spe}{B} \left[
\nabla_\bR \cdot \left(\bun\times\nabla_{\bR_\perp/\epsilon_\spe}
 \langle \phi_{\spe 1}^\sw \rangle
F_{\spe 1}^\sw\right)\right]^\lw
\nonumber\\[5pt]
\fl\hspace{0.5cm} - Z_\spe \lambda_\spe \partial_u \Bigg[
\Bigg( \bun\cdot\nabla_\bR 
\langle \phi_{\spe 1}^\sw \rangle
\nonumber\\[5pt]
\fl\hspace{0.5cm}
 +  \frac{u}{B}
\left(\bun \times \kappabf\right)\cdot
\nabla_{\bR_\perp/\epsilon_\spe}\langle \phi_{\spe 1}^\sw \rangle
\Bigg)F_{\spe 1}^\sw \Bigg]^\lw
\nonumber\\[5pt]
\fl\hspace{0.5cm} = \sum_{\spe'}
\left\langle
\left[\cT_{\spe,1}^*C_{\spe\spe'}^{(1)}\right]^\lw
\right\rangle
 +\sum_{\spe'}
\left\langle
\cT_{\spe,0}^*C_{\spe\spe'}^{(2)\lw}
\right\rangle
.
\end{eqnarray}

Here,
\begin{eqnarray}
\fl\bv_{E,\spe}^{(1)} = \frac{Z_\spe\lambda_\spe}{B}
\bun\times\nabla_{\bR}\varphi_1^\lw,
\end{eqnarray}
\begin{eqnarray} \label{eq:Psincphi}
\fl \Psi^\lw_{\phi,\spe} = 
 - \frac{1}{2\lambda_\spe^2 B^2} |\nabla_{\bR}
\varphi_0|^2 - \frac{1}{2B}\partial_\mu
\left[\langle (\phiwig_{\spe 1}^\sw)^2 \rangle
\right]^\lw,
\end{eqnarray}
\begin{eqnarray} \label{eq:PsincphiB}
\fl \Psi^\lw_{\phi B,\spe} &= - \frac{3 \mu}{2 \lambda_\spe B^2}
\nabla_\bR B \cdot \nabla_{\bR} \varphi_0
\nonumber \\
\fl & 
- \frac{u^2}{\lambda_\spe B^2} (\bun \cdot \nabla_\bR \bun)
 \cdot \nabla_{\bR}
\varphi_0,
\end{eqnarray}
and
\begin{eqnarray} \label{Psi2_B_TEXT}
\fl \Psi_{B,\spe} &=  - \frac{3u^2 \mu}{2B^2} \bun \cdot \nabla_\bR
\bun \cdot \nabla_\bR B
\nonumber \\
\fl & 
 + \frac{\mu^2}{4B} (\matI - \bun \bun) :
\nabla_\bR \nabla_\bR \bB \cdot \bun
\nonumber \\
\fl &
 - \frac{3\mu^2}{4B^2}
|\nabla_{\bR_\bot} B|^2 
+ \frac{u^2 \mu}{2B}
\nabla_\bR \bun : \nabla_\bR \bun
\nonumber \\
\fl &  + \left(\frac{\mu^2}{8} -
\frac{u^2 \mu}{4B}\right) \nabla_\bR \bun : (\nabla_\bR
\bun)^\mathrm{T}
\nonumber \\
\fl & 
 - \left(\frac{3 u^2 \mu}{8B} +
\frac{\mu^2}{16}\right) (\nabla_\bR \cdot \bun)^2 \nonumber \\ \fl
& + \left(\frac{3 u^2 \mu}{2B}-\frac{u^4}{2B^2}\right) |\bun \cdot
\nabla_\bR \bun|^2 
\nonumber \\
\fl & 
+ \left(\frac{ u^2 \mu}{8B} -
\frac{\mu^2}{16}\right) (\bun \cdot \nabla_\bR \times \bun)^2.
\end{eqnarray}

The collisional terms on the right side of \eq{eq:FPsecondorder2}
are spelled out in \ref{sec:SomePiecesOfCollisionOperator}.

The fact that the time derivative of $F_{\spe 0}$ appears in
\eq{eq:FPsecondorder2}, and not in lower-order pieces of the
equations, is important. It means that in this theoretical framework
transport equations for density and energy are obtained from the
$O(\epsilon_s^2)$ pieces of the equation. Specifically, we show in
Section \ref{sec:transportEqs} that such transport equations emerge as
solvability conditions of \eq{eq:FPsecondorder2}.

Equation \eq{eq:FPsecondorder2} is recast in \ref{sec:FPequationG2}
into a form, \eq{eq:FPequationG2final}, that is especially well suited
to work out the density and energy transport equations in Section
\ref{sec:transportEqs}.

\subsection{Long-wavelength quasineutrality equation}
\label{sec:longwavePoisson}

Since the calculation of the quasineutrality equation
\eq{eq:gyroQuasineutrality} at long-wavelengths is identical for
tokamaks and stellarators, we simply state the result derived in
\cite{CalvoParra2012} for tokamaks.
 
To order $\epsilon_s^0$,
\begin{eqnarray}\label{eq:gyroPoissonlw2order0}
\fl\sum_\spe
Z_\spe n_\spe(\boldr,t) = 0.
\end{eqnarray}

To order $\epsilon_s$,
\begin{eqnarray}\label{eq:gyroPoissonlw2order1}
\fl\sum_\spe
\frac{Z_\spe}{\lambda_\spe}
\int B(\boldr)F_{\spe 1}^\lw(\boldr,u,\mu,t)
\dd u\dd\mu\dd\theta = 0.
\end{eqnarray}

To order $\epsilon_s^2$,
\begin{eqnarray}\label{eq:gyroPoissonlw4}
\fl\sum_\spe
\frac{Z_\spe}{\lambda_\spe^2}
\Bigg[
\int(BF_{\spe 2}^\lw
+u\bun\cdot(\nabla_\boldr \times\bun)F_{\spe 1}^\lw)\dd u\dd\mu\dd\theta
\nonumber\\[5pt]
\fl\hspace{1cm}
-\bun\cdot(\nabla_\boldr \times\bK)\frac{n_{\spe}T_{\spe}}{B^2}
+
\nabla_\boldr\cdot\left(\frac{3}{2}
\frac{\nabla_{\boldr_\bot} B}{B^3}n_{\spe}T_{\spe}\right)
\nonumber\\[5pt]
\fl\hspace{1cm}
+\frac{1}{2}\nabla_\boldr\nabla_\boldr:
\left(
\left({\matI}-\bun\bun
\right)
\frac{n_{\spe}T_{\spe}}{B^2}
\right)\nonumber\\[5pt]
\fl\hspace{1cm}
+
\nabla_\boldr\cdot\left(
\frac{n_{\spe}T_{\spe}}{B^2}
\kappabf
+
\frac{Z_\spe n_{\spe}}{B^2}\nabla_\boldr\varphi_0
\right)
\Bigg]
=0.
\end{eqnarray}
Here, everything is evaluated at
$\bR=\boldr$. In writing the arguments of some functions we have
stressed that they are evaluated at $\bR=\boldr$, e.g. $n_\spe(\boldr)$,
but we should not forget that $n_\spe$, for example, depends only on
$\psi$ in flux coordinates.

\section{The long-wavelength radial electric field cannot be directly
  determined from the  quasineutrality equation}
\label{sec:indeterminacyRadialElectricField}

At this point we have already derived the Fokker-Planck and
quasineutrality equations to $O(\epsilon_s^2)$. One might think that
the quasineutrality equation is enough to solve for the
long-wavelength radial electric field, but this is not true {\em for
  any stellarator}. The argument is the same as that given in
\cite{CalvoParra2012} for the tokamak, but it is useful to restate it
briefly here.

Define
\begin{eqnarray}\label{eq:kernel}
 h_{\spe j} &=&
\left[
\frac{n_{\spe  j}}{n_\spe}
+\left(\frac{\mu B+u^2/2}{T_\spe}-\frac{3}{2}\right)
\frac{T_{\spe j}}{T_\spe}
\right]
F_{\spe 0}, \quad j = 1,2,
\end{eqnarray}
for an arbitrary set of flux functions $\{n_{\spe j}(\psi,t), T_{\spe
  j}(\psi,t)\}_\spe$ with the only restriction $T_{\spe
  j}/\lambda_\spe^j =T_{\spe' j}/\lambda_{\spe'}^j$, for all
$\sigma,\sigma'$. If $F_{\spe 1}^\lw$ and $\langle F_{\spe
  2}^\lw\rangle$ are solutions of the first and second-order
Fokker-Planck equations, \eq{eq:Vlasovorder1gyroav} (equivalently,
\eq{eq:Vlasovorder1gyroav4}) and \eq{eq:FPsecondorder2}, then so are
$F_{\spe 1}^\lw + h_{\spe 1}$ and $\langle F_{\spe 2}^\lw\rangle +
h_{\spe 2}$. That is, \eq{eq:kernel} gives the kernel of the operator
acting on $F_{\spe 1}^\lw$ in \eq{eq:Vlasovorder1gyroav} and on
$\langle F_{\spe 2}^\lw\rangle$ in \eq{eq:FPsecondorder2}. The freedom
due to the existence of a non-zero kernel can be removed
by imposing conditions such as
\begin{eqnarray}\label{eq:gaugefixingExample}
\fl
\left\langle
\int B F_{\spe 1}^\lw\dd u\dd\mu\dd\theta
\right\rangle_\psi
 = 0 \quad\mbox{for every $\spe$,}
\nonumber\\[5pt]
\fl
\left\langle
\int B\langle
F_{\spe 2}^\lw\rangle \dd u\dd\mu\dd\theta
\right\rangle_\psi
 = 0 \quad\mbox{for every $\spe$,}
\nonumber\\[5pt]
\fl
\left\langle
\sum_\spe\frac{1}{\lambda_\spe} \int B \left(u^2/2+\mu B\right)
F_{\spe 1}^\lw\dd u\dd\mu\dd\theta
\right\rangle_\psi
 = 0
\nonumber\\[5pt]
\fl\mbox{and}
\nonumber\\[5pt]
\fl
\left\langle
\sum_\spe\frac{1}{\lambda_\spe^2} \int B \left(u^2/2+\mu B\right)
\langle
F_{\spe 2}^\lw\rangle
\dd u\dd\mu\dd\theta
\right\rangle_\psi
 = 0.
\end{eqnarray}
Here, we have used the definition of
the flux-surface average of a function $G(\psi,\Theta,\zeta)$, given
by~\cite{dhaeseleer}
\begin{equation}
\fl\hspace{0.5cm}\langle G \rangle_\psi :=
\frac{\int_0^{2\pi}\int_0^{2\pi}\sqrt{g}\,
 G(\psi,\Theta,\zeta) \dd\Theta\dd\zeta}
{\int_0^{2\pi}\int_0^{2\pi}
\sqrt{g}\,
 \dd\Theta\dd\zeta}\, ,
\end{equation}
where
\begin{equation}\label{eq:sqrtg}
\sqrt{g}:=\frac{1}{\nabla_\bR\psi\cdot\left(
\nabla_\bR\Theta\times\nabla_\bR\zeta
\right)}
\end{equation}
is the square root of the determinant of the metric tensor in
coordinates $\{\psi,\Theta,\zeta\}$. It will also be useful to define
the volume enclosed by the flux surface labeled by $\psi$,
\begin{equation}\label{eq:defvolume}
V(\psi) = 
\int_0^\psi\int_0^{2\pi}\int_0^{2\pi}\sqrt{g}\,
\dd\psi'\dd\Theta\dd\zeta.
\end{equation}

Let $\textsf{F}^\lw_{\spe 1}$ and $\langle\textsf{F}^\lw_{\spe
  2}\rangle$ (note the different font) be solutions of
\eq{eq:Vlasovorder1gyroav} and \eq{eq:FPsecondorder2} satisfying
\eq{eq:gaugefixingExample} or any other set of conditions that fix the
component that belongs to the kernel. Then, any solution of
\eq{eq:Vlasovorder1gyroav} and \eq{eq:FPsecondorder2} is of the form
$\textsf{F}_{\spe 1}^\lw + h_{\spe 1}$, $\langle\textsf{F}_{\spe
  2}^\lw\rangle + h_{\spe 2}$. When introduced into
\eq{eq:gyroPoissonlw2order1} and \eq{eq:gyroPoissonlw4} one finds
\begin{eqnarray}\label{eq:gyroPoissonlw2order1Fix}
\fl
\sum_\spe\frac{Z_\spe}{\lambda_\spe}n_{\spe 1}
\nonumber\\[5pt]
\fl\hspace{0.5cm}
+
\sum_\spe
\frac{Z_\spe}{\lambda_\spe}
\int B(\boldr)\textsf{F}_{\spe 1}^\lw(\boldr,u,\mu,t)
\dd u\dd\mu\dd\theta = 0
\end{eqnarray}
and
\begin{eqnarray}\label{eq:gyroPoissonlw4Fix}
\fl
\sum_\spe\frac{Z_\spe}{\lambda^2_\spe}n_{\spe 2}
\nonumber\\[5pt]
\fl\hspace{0.5cm}
+
\sum_\spe
\frac{Z_\spe}{\lambda_\spe^2}
\Bigg[
\int(B\langle\textsf{F}_{\spe
  2}^\lw\rangle
+u\bun\cdot(\nabla_\boldr \times\bun)F_{\spe 1}^\lw)\dd u\dd\mu\dd\theta
\nonumber\\[5pt]
\fl\hspace{0.5cm}
-\bun\cdot(\nabla_\boldr \times\bK)\frac{n_{\spe}T_{\spe}}{B^2}
+
\nabla_\boldr\cdot\left(\frac{3}{2}
\frac{\nabla_{\boldr_\bot} B}{B^3}n_{\spe}T_{\spe}\right)
\nonumber\\[5pt]
\fl\hspace{0.5cm}
+\frac{1}{2}\nabla_\boldr\nabla_\boldr:
\left(
\left({\matI}-\bun\bun
\right)
\frac{n_{\spe}T_{\spe}}{B^2}
\right)\nonumber\\[5pt]
\fl\hspace{0.5cm}
+
\nabla_\boldr\cdot\left(
\frac{n_{\spe}T_{\spe}}{B^2}
\kappabf
+
\frac{Z_\spe n_{\spe}}{B^2}\nabla_\boldr\varphi_0
\right)
\Bigg]
=0.
\end{eqnarray}
The electrostatic potential $\varphi_0$ enters equation
\eq{eq:gyroPoissonlw4Fix} but it can never be determined from it. The
first and second-order pieces of the long-wavelength quasineutrality
equation simply give constraints on the corrections $n_{\spe 1}$ and
$n_{\spe 2}$. And these corrections cannot be simply set equal to
zero: In subsection \ref{sec:transportEqdensity} a transport equation
determining $n_\spe$ will be derived as a solvability condition for
equation \eq{eq:FPsecondorder2}. Analogously, $n_{\spe j}$ would be
determined by a transport equation obtained as a solvability
conditions for a higher-order piece of the Fokker-Planck equation.

It is important to note that the lowest order radial electric field
cannot be determined from the quasineutrality equation but the lowest
order pieces of the electric field parallel to the flux surface are
determined by equations \eq{eq:gyroPoissonlw2order1Fix} and
\eq{eq:gyroPoissonlw4Fix}. This becomes obvious when one writes
\eq{eq:gyroPoissonlw2order1Fix} and \eq{eq:gyroPoissonlw4Fix} in terms
of the non-adiabatic pieces of the distribution function and then acts
with the operators $\partial_\Theta$ and $\partial_\zeta$. In this
way, $\partial_\Theta\varphi_1^\lw$, $\partial_\zeta\varphi_1^\lw$,
$\partial_\Theta\varphi_2^\lw$, and $\partial_\zeta\varphi_2^\lw$ are
obtained. The key is to recall that $n_{\spe j}$ depends only on
$\psi$. Of course, $\varphi_1^\lw$ and $\varphi_2^\lw$ are determined
up to an arbitrary, additive function of $\psi$, that can be absorbed
by redefining the corrections $n_{\spe 1}$ and $n_{\spe 2}$. Without
loss of generality, we take
\begin{equation}\label{eq:fixvarphi1}
\left\langle\varphi_1^\lw\right\rangle_\psi = 0
\end{equation}
and
\begin{equation}\label{eq:fixvarphi2}
\left\langle\varphi_2^\lw\right\rangle_\psi = 0,
\end{equation}
therefore fixing the ambiguity.

\section{Transport equations as solvability
  conditions of the Fokker-Planck equations}
\label{sec:transportEqs}

The understanding of transport equations as solvability conditions of
kinetic equations dates back to the works by Chapman and Enskog on
gases~\cite{ChapmanCowling}. In our context, when we speak about
solvability conditions of the Fokker-Planck equations we mean the
following: \eq{eq:Vlasovorder1gyroav} and \eq{eq:FPsecondorder2} are
equations for $F_{\spe 1}^\lw$ and $\langle F_{\spe 2}^\lw\rangle$
but, in general, they cannot be solved for arbitrary values of the
lower-order quantities in them. The constraints that the lower-order
quantities entering \eq{eq:Vlasovorder1gyroav} and
\eq{eq:FPsecondorder2} must satisfy are called solvability conditions
of the first and second-order Fokker-Planck equations.

In Appendix M of reference \cite{CalvoParra2012} an exhaustive and
general computation of the solvability conditions of the Fokker-Planck
equations for strongly magnetized plasmas was given. Let us apply the
general results of that appendix to the stellarator problem. Equation
\eq{eq:Vlasovorder1gyroav4} can be rewritten in the form
\begin{eqnarray}\label{eq:FPequationsSolvCond_Text1}
\fl\tau_\spe\lambda_\spe^{-1}&\left(
u\bun\cdot\nabla_\bR - 
\mu\bun\cdot\nabla_\bR B\partial_u
\right) G_{\spe 1}^\lw
\nonumber\\[5pt]
\fl &-
\tau_\spe\sum_{\spe'}
\Bigg(
\cT_{\spe,0}^*
C_{\spe\spe'}
\left[
\lambda_\spe^{-1}
\cT_{\spe,0}^{-1*}G_{\spe 1}^\lw,
\cT_{\spe',0}^{-1*}F_{\spe' 0}
\right]
\nonumber\\[5pt]
\fl &
+
\cT_{\spe,0}^*
 C_{\spe\spe'}
\left[
\cT_{\spe,0}^{-1*}F_{\spe 0},
\lambda_{\spe'}^{-1}
\cT_{\spe',0}^{-1*}G_{\spe' 1}^\lw
\right]
\Bigg)
=
\tau_\spe\lambda_\spe^{-1} R_{\spe 1},
\end{eqnarray}
where
\begin{eqnarray}\label{eq:Rsigma1}
\fl
R_{\spe 1} =
-
\left(
\frac{Z_\spe}{T_\spe}\partial_\psi\varphi_0
+\Upsilon_\spe
\right)
\bv_M\cdot\nabla_\bR\psi
\, F_{\spe 0}.
\end{eqnarray}
As for the second-order Fokker-Planck equation, we use the form given
in \eq{eq:FPequationG2final}, which can be expressed as
\begin{eqnarray}\label{eq:FPequationsSolvCond_Text2}
\fl\tau_\spe\lambda_\spe^{-2}&\left(
u\bun\cdot\nabla_\bR - 
\mu\bun\cdot\nabla_\bR B\partial_u
\right) G_{\spe 2}^\lw
\nonumber\\[5pt]
\fl &-
\tau_\spe\sum_{\spe'}
\Bigg(
\cT_{\spe,0}^*
C_{\spe\spe'}
\left[
\lambda_\spe^{-2}
\cT_{\spe,0}^{-1*} G_{\spe 2}^\lw,
\cT_{\spe',0}^{-1*}F_{\spe' 0}
\right]
\nonumber\\[5pt]
\fl &
+
\cT_{\spe,0}^*
 C_{\spe\spe'}
\left[
\cT_{\spe,0}^{-1*}F_{\spe 0},
\lambda_{\spe'}^{-2}
\cT_{\spe',0}^{-1*} G_{\spe' 2}^\lw
\right]
\Bigg)
=
\tau_\spe\lambda_\spe^{-2} R_{\spe 2},
\end{eqnarray}
with
\begin{eqnarray} \label{eq:Rsigma2}
\fl
R_{\spe 2}=
-
\frac{\lambda_\spe^2}{\tau_\spe}\partial_{\epsilon_s^2 t} F_{\spe 0}
-\frac{1}{B}\nabla_\bR
\cdot
\Bigg[
(\mu\bun\times\nabla_\bR B
\nonumber\\[5pt]
\fl\hspace{0.5cm}
+
u^2\nabla_\bR\times\bun
+
Z_\spe
\bun\times\nabla_\bR\varphi_0
) G_{\spe 1}^\lw
\nonumber\\[5pt]
\fl\hspace{0.5cm}
-
Z_\spe\lambda_\spe
\varphi_1^\lw\bun\times\nabla_\bR\psi\left(\Upsilon_\spe+\frac{Z_\spe}{T_\spe}
\partial_\psi \varphi_0\right)F_{\spe 0}
\Bigg]
\nonumber\\[5pt]
\fl\hspace{0.5cm}
+\frac{1}{B}\partial_u
\Bigg[
u (\nabla_\bR\times\bun)\cdot
(\mu \nabla_\bR B + Z_\spe
\nabla_\bR\varphi_0)
 G_{\spe 1}^\lw
\nonumber\\[5pt]
\fl\hspace{0.5cm}
-
\frac{Z_\spe\lambda_\spe
  \varphi_1^\lw}{u}
\mu(\bun\times\nabla_\bR\psi)\cdot\nabla_\bR B
\left(\Upsilon_\spe+\frac{Z_\spe}{T_\spe}
\partial_\psi \varphi_0\right)F_{\spe 0}
\Bigg]
\nonumber\\[5pt]
\fl\hspace{0.5cm}
 + \frac{u \mu}{B}
 (\nabla_\bR \times \bK)_\bot
\cdot\nabla_\bR F_{\spe 0}
\nonumber\\[5pt]
\fl\hspace{0.5cm} -
\mu
\left(\nabla_\bR\times\bK\right)_\perp
\cdot
\left(\mu\nabla_\bR B + Z_\spe\nabla_\bR\varphi_0
\right)
\partial_u F_{\spe 0} 
 \nonumber\\[5pt]
\fl\hspace{0.5cm}
 -
\frac{Z_\spe \lambda_\spe}{B} \left[
\nabla_\bR \cdot \left(\bun\times\nabla_{\bR_\perp/\epsilon_\spe}
 \langle \phi_{\spe 1}^\sw \rangle
F_{\spe 1}^\sw\right)\right]^\lw
\nonumber\\[5pt]
\fl\hspace{0.5cm} + Z_\spe \lambda_\spe \partial_u \Bigg[
\Bigg( \bun\cdot\nabla_\bR 
\langle \phi_{\spe 1}^\sw \rangle
\nonumber\\[5pt]
\fl\hspace{0.5cm}
+  \frac{u}{B}
\left(\bun \times \kappabf\Bigg)\cdot
\nabla_{\bR_\perp/\epsilon_\spe}\langle \phi_{\spe 1}^\sw \rangle
\right)F_{\spe 1}^\sw \Bigg]^\lw 
\nonumber\\[5pt]
\fl\hspace{0.5cm}
+
\partial_u
\left[
  -\frac{Z_\spe\lambda_\spe\varphi_1^\lw}{u}
  \sum_{\spe'}
  \left\langle\cT_{\spe,0}^*C_{\spe\spe'}^{(1)\lw}\right\rangle
\right]
\nonumber\\[5pt]
\fl\hspace{0.5cm}
+\frac{u}{B}
\bun\cdot\nabla_\bR\times\bun
\sum_{\spe'}
\left\langle \cT_{\spe,0}^*C_{\spe\spe'}^{(1)\lw} \right\rangle
\nonumber\\[5pt]
\fl\hspace{0.5cm}
+
\sum_{\spe'}
\left\langle
  \left[\cT_{\spe,1}^*C_{\spe\spe'}^{(1)}\right]^\lw
\right\rangle
+\sum_{\spe'}\left\langle
\cT_{\spe,0}^*{\mathbb C}_{\spe\spe'}\right\rangle
  .
\end{eqnarray}
The term ${\mathbb C}_{\spe\spe'}$ is defined in \eq{eq:mathbbC}. As
shown in Appendix M of reference \cite{CalvoParra2012}, given the form
of equations \eq{eq:Rsigma1} and \eq{eq:Rsigma2}, the solvability
conditions are
\begin{eqnarray}
  \fl
  \left\langle
    \tau_\spe\lambda_\spe^{-j} \int B R_{\spe j}\dd u\dd\mu\dd\theta
  \right\rangle_\psi = 0, \mbox{ for each $\spe$}
\label{eq:solvcond_Density}
\\[5pt]
\fl
\mbox{and}\nonumber
\\[5pt]
  \fl
  \left\langle
    \sum_\spe
    \tau_\spe\lambda_\spe^{-j}\int B \left(u^2/2 + \mu B\right)
 R_{\spe j}\dd u\dd\mu\dd\theta
  \right\rangle_\psi = 0,\label{eq:solvcond_Energy}
\end{eqnarray}
with $j=1,2$.

\subsection{Triviality of the solvability condition of the
  first-order Fokker-Planck equations}
\label{sec:trivialFirstOrderSolvCond}

We start by showing that the first-order Fokker-Planck equations do
not give any constraint on lower-order quantities, i.e. we prove that
\eq{eq:solvcond_Density} and \eq{eq:solvcond_Energy} turn out to yield
trivial conditions when $R_{\spe 1}$ is given by \eq{eq:Rsigma1}. In
order to conclude that \eq{eq:solvcond_Density} and
\eq{eq:solvcond_Energy} vanish identically for $j=1$ we need to prove
the identity
\begin{eqnarray}\label{eq:propertyMagDrift}
  \left\langle
     \int B\,  h(\psi, \varepsilon) \bv_M\cdot
\nabla_\bR\psi \, \dd u\dd\mu\dd\theta
  \right\rangle_\psi \equiv 0,
\end{eqnarray}
for any function $h$ of the kinetic energy $\varepsilon$ and the
flux-surface label $\psi$. Here, $\varepsilon = u^2/2 + \mu B$. The
radial component of the magnetic drift velocity can be written as
\begin{eqnarray}
\fl  \bv_M\cdot\nabla_\bR\psi 
  = \frac{u^2}{B}(\nabla_\bR\times\bun)\cdot\nabla_\bR \psi 
\nonumber\\[5pt]
\fl\hspace{0.5cm}
+ \frac{\mu}{B}(\bun\times\nabla_\bR B)\cdot\nabla_\bR \psi.
\end{eqnarray}
Then, we have
\begin{eqnarray}\label{eq:propertyMagDriftAux}
\fl
\left\langle
     \int B\,  h(\psi,\varepsilon) \bv_M\cdot
\nabla_\bR\psi \, \dd u\dd\mu\dd\theta
  \right\rangle_\psi =
\nonumber\\[5pt]
\fl\hspace{0.5cm}
\left\langle
     \nabla_\bR\cdot
\int u^2 h \bun\times
\nabla_\bR\psi \, \dd u\dd\mu\dd\theta
  \right\rangle_\psi
\nonumber\\[5pt]
\fl\hspace{0.5cm}
-\Bigg\langle
\int  \Bigg(\mu h
(\bun\times\nabla_\bR \psi)\cdot\nabla_\bR B
\nonumber\\[5pt]
\fl\hspace{0.5cm}
+ u^2  (\bun\times
\nabla_\bR\psi)\cdot\nabla_\bR h
\Bigg)
\dd u\dd\mu\dd\theta
  \Bigg\rangle_\psi,
\end{eqnarray}
where we have used that $\nabla_\bR\cdot(\bA_1\times\bA_2)
=(\nabla_\bR\times\bA_1)\cdot\bA_2 -
(\nabla_\bR\times\bA_2)\cdot\bA_1$ for any two vector fields $\bA_1$
and $\bA_2$, and that the curl of the gradient of a function is zero.
Noting that $u (\bun\times \nabla_\bR\psi)\cdot\nabla_\bR h = \mu
(\bun\times \nabla_\bR\psi)\cdot\nabla_\bR B\partial_u h$ and
integrating by parts in $u$ the last term of
\eq{eq:propertyMagDriftAux}, we find that
\begin{eqnarray}
\fl
\Bigg\langle
\int  \Bigg(\mu h
(\bun\times\nabla_\bR \psi)\cdot\nabla_\bR B
\nonumber\\[5pt]
\fl\hspace{0.5cm}
+ u^2  (\bun\times
\nabla_\bR\psi)\cdot\nabla_\bR h
\Bigg)
\dd u\dd\mu\dd\theta
  \Bigg\rangle_\psi = 0.
\end{eqnarray}
The first term on the right side of \eq{eq:propertyMagDriftAux} also
vanishes, as can be easily seen by employing the formula
\begin{eqnarray}\label{eq:formuladivergence}
\left\langle
\nabla_\bR\cdot \bA
\right\rangle_\psi = \frac{1}{V'(\psi)}
\partial_\psi
\left\langle
V'(\psi)
\bA\cdot\nabla_\bR\psi
\right\rangle_\psi
\end{eqnarray}
for the flux-surface average of the divergence of a vector field
$\bA$. Here, $V'(\psi)$ is the derivative with respect to $\psi$ of
the volume defined in \eq{eq:defvolume}.

Therefore, we have proven \eq{eq:propertyMagDrift} and learnt that the
first-order Fokker-Planck equations do not impose any constraints on
lower-order quantities. The solvability conditions
\eq{eq:solvcond_Density} and \eq{eq:solvcond_Energy} for $j=2$ are
non-trivial; actually they are transport equations for
the particle density of each species and for the total kinetic
energy, respectively. We work them out in the next subsections.

\subsection{Transport equation for  particle density}
\label{sec:transportEqdensity}

The transport equation for particle density of species $\spe$ is
obtained from \eq{eq:solvcond_Density} when $j=2$ and $R_{\spe 2}$ is
given by \eq{eq:Rsigma2}. In \ref{sec:manipulations_densityEqEnergy}
it is shown that the result is
\begin{eqnarray} \label{eq:densityevolution4}
\fl
\partial_{\epsilon_s^2 t}\, n_{\spe}(\psi,t)
\nonumber\\[5pt]
\fl\hspace{0.5cm}
+V'^{-1}\partial_\psi
\Bigg\langle V'
\frac{\tau_\spe}{\lambda_\spe^2}
\int B\,
\bv_M 
\cdot\nabla_\bR\psi\,
G_{\spe 1}^\lw
\dd u\dd\mu\dd\theta
\Bigg\rangle_\psi =
\nonumber\\[5pt]
\fl\hspace{0.5cm}
V'^{-1}\partial_\psi
\Bigg\langle V'
\int
\Bigg\{ 
-\left[
 \left(\bun\times\nabla_{\bR_\perp/\epsilon_\spe}
 \langle \phi_{\spe 1}^\sw \rangle
F_{\spe 1}^\sw\right)\right]^\lw\cdot\nabla_\bR\psi
 \nonumber\\[5pt]
\fl\hspace{0.5cm}
+
\frac{\tau_\spe}{\lambda_\spe^2}
 \sum_{\spe'}
 \left \langle B
\rhobf\cdot\nabla_\bR\psi \,
\cT_{\spe,0}^*C_{\sigma \sigma^\prime}^{(1)\lw}
 \right \rangle
\Bigg\}
\dd u\dd\mu\dd\theta
\Bigg\rangle_\psi.
\end{eqnarray}
Observe that only the non-adiabatic piece of $F_{\spe 1}^\lw$,
$G_{\spe 1}^\lw = F_{\spe 1}^\lw + (Z_\spe\lambda_\spe /
T_\spe)\varphi_1^\lw F_{\spe 0}$, contributes to particle
transport. The quantity $C_{\sigma \sigma^\prime}^{(1)\lw}$ is defined
in \eq{eq:C1nc}.

\subsection{Transport equation for energy}
\label{sec:transportEqEnergy}

The solvability condition \eq{eq:solvcond_Energy} for $j=2$ and
$R_{\spe 2}$ corresponding to \eq{eq:Rsigma2} gives a transport
equation for the total energy. Its computation requires some algebra,
and this is done in \ref{sec:manipulations_transportEqEnergy}. The
result is 
\begin{eqnarray} \label{eq:TransEqEner_TotalFinal}
\fl
\partial_{\epsilon_s^2 t}
 \sum_\spe\frac{3}{2}n_{\spe}T_\spe
\nonumber\\[5pt]
\fl\hspace{0.5cm}
+
\frac{1}{V^\prime} \partial_\psi \Bigg\langle
V^\prime
\int B(u^2/2+\mu B)
\Bigg\{
\nonumber\\[5pt]
\fl\hspace{0.5cm}
\sum_\spe
\frac{\tau_\spe}{\lambda_\spe^2}
\bv_{M}\cdot\nabla_\bR\psi\,
G^\lw_{\sigma 1}
\nonumber\\[5pt]
\fl\hspace{0.5cm}
+
\frac{1}{B}\sum_\spe
\left[
 \left(\bun\times\nabla_{\bR_\perp/\epsilon_\spe}
 \langle \phi_{\spe 1}^\sw \rangle\cdot\nabla_\bR\psi
F_{\spe 1}^\sw\right)\right]^\lw
\nonumber\\[5pt]
\fl\hspace{0.5cm}
-\sum_{\spe,\spe'}
\frac{\tau_\spe}{\lambda_\spe^2} 
\left\langle
\rhobf\cdot\nabla_\bR\psi {\cal T}_{\spe,0}^* C_{\spe\spe'}^{(1)\lw}
\right\rangle
\Bigg\}\,
\dd u\dd\mu\dd\theta
\Bigg\rangle_\psi
=0.
\end{eqnarray}
Again, only the non-adiabatic piece of $F_{\spe 1}^\lw$ contributes to
energy transport.

\subsection{Ambipolarity condition}
\label{sec:timeevolutionquasineutrality}

In subsection \ref{sec:transportEqdensity} we have computed a
transport equation, \eq{eq:densityevolution4}, for the particle
density of each species. This is a solvability condition of the
Fokker-Planck equation to $O(\epsilon_s^2)$. But we must recall that
the quasineutrality equation to lowest order,
\eq{eq:gyroPoissonlw2order0}, imposes the constraint
\begin{equation}\label{eq:gyroPoissonlw2order0again}
\sum_\spe Z_\spe n_{\spe}(\psi,t) = 0,
\end{equation}
for every $\psi$ and $t$. Hence,
\begin{equation}
\partial_{\epsilon_s^2 t}
\sum_\spe Z_\spe n_{\spe}(\psi,t) = 0
\end{equation}
should also hold for every $\psi$ and $t$. The question is whether the
density transport equations \eq{eq:densityevolution4} automatically
imply $\partial_{\epsilon_s^2 t} \sum_\spe Z_\spe n_{\spe}(\psi,t)
\equiv 0$ or, on the contrary, the vanishing of
$\partial_{\epsilon_s^2 t} \sum_\spe Z_\spe n_{\spe}(\psi,t)$, needed
for consistency, yields an equation that sets new constraints on
lowest-order quantities. To find the answer we proceed to multiply
equation \eq{eq:densityevolution4} by $Z_\spe$ and sum over
$\spe$. The last term vanishes due to momentum conservation by the
collision operator (see \eq{eq:propcollnondim}). The contribution of
the first term on the right side of \eq{eq:densityevolution4} to
$\partial_{\epsilon_s^2 t} \sum_\spe Z_\spe n_\spe$ also vanishes,
irrespective of the magnetic geometry, i.e.
\begin{eqnarray} \label{eq:turbulentintrinsicambipolarity}
\fl
\sum_\spe Z_\spe \Bigg\langle \int B
\left[ F_{\sigma 1}^\sw (\nabla_{\bR_\perp/\epsilon_\spe}
\langle \phi_{\spe 1}^\sw
\rangle \times \bun) \cdot \nabla_\bR \psi \right]^\lw 
\dd u \dd\mu\dd\theta \Bigg\rangle_\psi = 0.
\end{eqnarray}
This was proven in \cite{Sugama96,parra09b} and also in subsection 5.3
of reference \cite{CalvoParra2012}. As a consequence, we have
\begin{eqnarray} \label{eq:ChargeDensityEvolution}
\fl
\partial_{\epsilon_s^2 t}\sum_\spe Z_\spe n_{\spe}(\psi,t)
\nonumber\\[5pt]
\fl\hspace{0.5cm}
+
\sum_\spe
V'^{-1}\partial_\psi
\Bigg\langle V'
\frac{1}{\lambda_\spe}
\int B
\bv_M 
\cdot\nabla_\bR\psi\,
G_{\spe 1}^\lw
\dd u\dd\mu\dd\theta
\Bigg\rangle_\psi = 0.
\end{eqnarray}
Of course, consistency with \eq{eq:gyroPoissonlw2order0again} enforces
\begin{eqnarray} \label{eq:AmbipolarityCondition_preliminar}
\fl\sum_\spe
V'^{-1}\partial_\psi
\Bigg\langle V'
\frac{1}{\lambda_\spe}
\int B\,
\bv_M 
\cdot\nabla_\bR\psi\,
G_{\spe 1}^\lw
\dd u\dd\mu\dd\theta
\Bigg\rangle_\psi = 0.
\end{eqnarray}
Regularity at $\psi=0$ makes \eq{eq:AmbipolarityCondition_preliminar} equivalent
to
\begin{eqnarray} \label{eq:AmbipolarityCondition} \fl\sum_\spe
  \Bigg\langle \frac{1}{\lambda_\spe} \int B\, \bv_M
  \cdot\nabla_\bR\psi\, G_{\spe 1}^\lw \dd u\dd\mu\dd\theta
  \Bigg\rangle_\psi = 0,
\end{eqnarray}
which is the standard neoclassical ambipolarity
condition. This is the lowest-order piece of the equation
\begin{equation}
\langle\bJ\cdot\nabla_\bR\psi\rangle_\psi = 0,
\end{equation}
where $\bJ$ is the electric current density.

\section{When does the ambipolarity condition add any information?}
\label{sec:ambipolarityConditionInformation}

In this section we want to discuss the circumstances under which
\eq{eq:AmbipolarityCondition} is actually an equation that imposes new
conditions on lowest-order quantities. The general answer to the above
question has already been given by Helander and Simakov in
\cite{Helander08}. Here, we give our own version of the proof for
completeness.

Since the temperature profile is the same for all species, in this
section we remove the subindex $\spe$, and $T_\spe \equiv T$.  At
every point of phase space $(\psi,\Theta,\zeta,u,\mu,\theta)$, the
function $G_{\spe 1}^\lw$, determined by \eq{eq:Vlasovorder1gyroav4}, is
a linear combination of the gradients of the profiles at
$\psi$. Namely,
\begin{equation}\label{eq:GredundantGradients}
G_{\spe 1}^\lw =  \partial_\psi\varphi_0 \, g_\spe +
\sum_\gamma
\partial_\psi n_{\gamma} \, \check{h}_{\spe\gamma }+\partial_\psi T \,  l_\spe,
\end{equation}
for some phase-space functions $g_\spe$, $\check{h}_{\spe\gamma}$,
$l_\spe$ that do not depend on the profile gradients and whose
defining equations are given below, in equations
\eq{eq:EquationCoeffVarphi},
\eq{eq:EquationCoeffn}, and \eq{eq:EquationCoeffT}. In
\eq{eq:GredundantGradients} the index $\gamma$ runs over all
species. Denote by $N$ the number of different species. There are not
$N$ independent density gradients, but $N-1$ due to the constraint
$\sum_\spe Z_\spe n_\spe(\psi,t) = 0$. For example, we can eliminate
the gradient of the electron density, $n_e$, by using
\begin{equation}
\partial_\psi n_e(\psi,t) = \sum_{\spe\neq e} Z_\spe 
\partial_\psi n_\spe(\psi,t).
\end{equation}
It is conceptually clearer to express $G_{\spe 1}^\lw$ in terms of
the $N+1$ independent gradients,
\begin{equation}\label{eq:G1_IndepGradients}
\fl
G_{\spe 1}^\lw =  \partial_\psi\varphi_0 \, g_\spe +
\sum_{\gamma\neq e}
\partial_\psi n_{\gamma} \, h_{\spe\gamma}
+\partial_\psi T \,  l_\spe,
\end{equation}
where $h_{\spe\gamma} := \check{h}_{\spe\gamma}+Z_\gamma
\check{h}_{\spe e}$ is defined only for $\gamma\neq e$. From
\eq{eq:Vlasovorder1gyroav4} and \eq{eq:G1_IndepGradients} we can
easily find the equations that determine the coefficients of the
gradients in \eq{eq:G1_IndepGradients},
\begin{eqnarray}\label{eq:EquationCoeffVarphi}
\fl
\left(u\bun\cdot\nabla_\bR - \mu\bun\cdot\nabla_\bR
B\partial_u
\right)
g_{\spe}
\nonumber\\[5pt]
\fl\hspace{1cm}
+
\frac{Z_\spe}{T}
\bv_M\cdot\nabla_\bR\psi
\, F_{\spe 0}
\nonumber\\[5pt]
\fl\hspace{1cm}
= \sum_{\sigma'}\cT_{\spe,0}^*
C_{\spe\spe'}\left[ \cT^{-1*}_{\spe,0}
g_{\spe}
,\cT^{-1*}_{\spe',0}F_{\spe' 0} \right]
\nonumber\\[5pt]
\fl\hspace{1cm} + \sum_{\sigma'}
\frac{\lambda_\spe}{\lambda_{\spe'}}\cT_{\spe,0}^*
C_{\spe\spe'} \left[ \cT^{-1*}_{\spe,0}F_{\spe 0},
\cT^{-1*}_{\spe',0}
g_{\spe'}
\right],
\end{eqnarray}
\begin{eqnarray}\label{eq:EquationCoeffn}
\fl
\left(u\bun\cdot\nabla_\bR - \mu\bun\cdot\nabla_\bR
B\partial_u
\right)
h_{\spe\gamma}
\nonumber\\[5pt]
\fl\hspace{1cm}
+
\frac{1}{n_\spe}(\delta_{\spe\gamma}+ Z_\gamma \delta_{\sigma\, e}) \, 
\bv_M\cdot\nabla_\bR\psi
\, F_{\spe 0}
\nonumber\\[5pt]
\fl\hspace{1cm}
= \sum_{\sigma'}\cT_{\spe,0}^*
C_{\spe\spe'}\left[ \cT^{-1*}_{\spe,0}
h_{\spe\gamma}
,\cT^{-1*}_{\spe',0}F_{\spe' 0} \right]
\nonumber\\[5pt]
\fl\hspace{1cm} + \sum_{\sigma'}
\frac{\lambda_\spe}{\lambda_{\spe'}}\cT_{\spe,0}^*
C_{\spe\spe'} \left[ \cT^{-1*}_{\spe,0}F_{\spe 0},
\cT^{-1*}_{\spe',0}
h_{\spe'\gamma}
\right],
\end{eqnarray}
and
\begin{eqnarray}\label{eq:EquationCoeffT}
\fl
\left(u\bun\cdot\nabla_\bR - \mu\bun\cdot\nabla_\bR
B\partial_u
\right)
l_{\spe}
\nonumber\\[5pt]
\fl\hspace{1cm}
+
\left(\frac{u^2/2 +\mu B}{T_\spe}-\frac{3}{2}\right)
\frac{1}{T} 
\bv_M\cdot\nabla_\bR\psi
\, F_{\spe 0}
\nonumber\\[5pt]
\fl\hspace{1cm}
= \sum_{\sigma'}\cT_{\spe,0}^*
C_{\spe\spe'}\left[ \cT^{-1*}_{\spe,0}
l_{\spe}
,\cT^{-1*}_{\spe',0}F_{\spe' 0} \right]
\nonumber\\[5pt]
\fl\hspace{1cm} + \sum_{\sigma'}
\frac{\lambda_\spe}{\lambda_{\spe'}}\cT_{\spe,0}^*
C_{\spe\spe'} \left[ \cT^{-1*}_{\spe,0}F_{\spe 0},
\cT^{-1*}_{\spe',0}
l_{\spe'}
\right].
\end{eqnarray}
Equations \eq{eq:EquationCoeffVarphi}, \eq{eq:EquationCoeffn} and
\eq{eq:EquationCoeffT} have a kernel (see Section
\ref{sec:indeterminacyRadialElectricField}). In order to choose
particular solutions $g_\spe$, $h_{\spe\gamma}$ and $l_\spe$, we
impose
\begin{eqnarray}\label{eq:gaugefixing_g}
\fl
\left\langle
\int B g_{\spe}\dd u\dd\mu\dd\theta
\right\rangle_\psi
 = 0
\quad\mbox{for every $\spe$,}
\nonumber\\
\fl
\left\langle
\sum_\spe\frac{1}{\lambda_\spe} \int B \left(u^2/2+\mu B\right)
g_{\spe}\dd u\dd\mu\dd\theta
\right\rangle_\psi
 = 0;
\end{eqnarray}
\begin{eqnarray}\label{eq:gaugefixing_h}
\fl
\left\langle
\int B h_{\spe\gamma}\dd u\dd\mu\dd\theta
\right\rangle_\psi
 = 0
\quad\mbox{for every $\spe$, $\gamma$,}
\nonumber\\
\fl
\left\langle
\sum_\spe\frac{1}{\lambda_\spe} \int B \left(u^2/2+\mu B\right)
h_{\spe\gamma}\dd u\dd\mu\dd\theta
\right\rangle_\psi
 = 0 \quad\mbox{for every $\gamma$;}
\end{eqnarray}
\begin{eqnarray}\label{eq:gaugefixing_l}
\fl
\left\langle
\int B l_{\spe}\dd u\dd\mu\dd\theta
\right\rangle_\psi
 = 0
\quad\mbox{for every $\spe$,}
\nonumber\\[5pt]
\fl\mbox{and}
\nonumber\\[5pt]
\fl
\left\langle
\sum_\spe\frac{1}{\lambda_\spe} \int B \left(u^2/2+\mu B\right)
l_{\spe}\dd u\dd\mu\dd\theta
\right\rangle_\psi
 = 0.
\end{eqnarray}
Other conditions are possible, but the radial current
\eq{eq:AmbipolarityCondition} will not depend on the conditions chosen
to fix the solutions of \eq{eq:EquationCoeffVarphi},
\eq{eq:EquationCoeffn} and \eq{eq:EquationCoeffT}.

Now, let us write the ambipolarity condition
\eq{eq:AmbipolarityCondition} in terms of $g_\spe,h_{\spe\gamma}$, and
$l_\spe$:
\begin{eqnarray}\label{eq:AmbipolarityCondition_LinearCombination}
\fl
\partial_\psi\varphi_0
\sum_\spe
\Bigg\langle
\frac{1}{\lambda_\spe}
\int B\,
\bv_M 
\cdot\nabla_\bR\psi\,
g_{\spe}
\dd u\dd\mu\dd\theta
\Bigg\rangle_\psi
\nonumber\\[5pt]
\fl\hspace{1cm}
+
\sum_{\stackrel{\spe}{\gamma\neq e}}
\partial_\psi n_\gamma
\Bigg\langle
\frac{1}{\lambda_\spe}
\int B\,
\bv_M 
\cdot\nabla_\bR\psi\,
h_{\spe \gamma}
\dd u\dd\mu\dd\theta
\Bigg\rangle_\psi
\nonumber\\[5pt]
\fl\hspace{1cm}
+
\partial_\psi T
\sum_\spe
\Bigg\langle
\frac{1}{\lambda_\spe}
\int B\,
\bv_M 
\cdot\nabla_\bR\psi\,
l_{\spe}
\dd u\dd\mu\dd\theta
\Bigg\rangle_\psi
 = 0.
\end{eqnarray}

We call the system {\em intrinsically ambipolar} if all of the
coefficients of the $N+1$ independent gradients in
\eq{eq:AmbipolarityCondition_LinearCombination} vanish. That is, if
\begin{eqnarray} \label{eq:IntrinsicAmbipolarity_g}
\fl\sum_\spe
\Bigg\langle
\frac{1}{\lambda_\spe}
\int B\,
\bv_M 
\cdot\nabla_\bR\psi\,
g_{\spe}
\dd u\dd\mu\dd\theta
\Bigg\rangle_\psi = 0,
\end{eqnarray}
\begin{eqnarray} \label{eq:IntrinsicAmbipolarity_h}
\fl\sum_\spe
\Bigg\langle
\frac{1}{\lambda_\spe}
\int B\,
\bv_M 
\cdot\nabla_\bR\psi\,
h_{\spe \gamma}
\dd u\dd\mu\dd\theta
\Bigg\rangle_\psi = 0, \quad \mbox{for every $\gamma\neq e$},
\end{eqnarray}
and
\begin{eqnarray} \label{eq:IntrinsicAmbipolarity_l}
\fl\sum_\spe
\Bigg\langle
\frac{1}{\lambda_\spe}
\int B\,
\bv_M 
\cdot\nabla_\bR\psi\,
l_{\spe}
\dd u\dd\mu\dd\theta
\Bigg\rangle_\psi = 0.
\end{eqnarray}
This is another way of saying that an intrinsically ambipolar system
is one for which \eq{eq:AmbipolarityCondition} is an identity, i.e. it
is satisfied for every value of the electrostatic potential,
density, and temperature gradients. Now, we want to derive necessary
and sufficient conditions for intrinsically ambipolar magnetic
configurations. We assume that
$(\nabla_\bR\times\bB)\cdot\nabla_\bR\psi = 0$.

First, we try to find necessary conditions for intrinsic
ambipolarity. The functions $g_\spe$ satisfy
\eq{eq:EquationCoeffVarphi}. Define $\hat{g}_\spe := g_\spe/F_{\spe
  0}$. Multiply \eq{eq:EquationCoeffVarphi} by
$Z_\spe^{-1}\lambda_\spe^{-1} B \hat{g}_\spe$, integrate over
$u,\mu,\theta$, take flux-surface average, and sum over $\spe$ to find
\begin{eqnarray}\label{eq:AlmostEntropyProduction}
\fl
\frac{1}{T}
\left\langle
\sum_\spe
\frac{1}{\lambda_\spe}
\int B 
\bv_M\cdot\nabla_\bR\psi
\, g_\spe
\dd u\dd\mu \dd\theta\right\rangle_\psi
\nonumber\\[5pt]
\fl\hspace{1cm}
= 
\left\langle\sum_\spe
\tau_\spe\int B \lambda_\spe^{-1}\hat{g}_\spe
\sum_{\sigma'}
\hat{C}_{\spe\spe'}\left[ \hat{g}_\spe,\hat{g}_{\spe'} \right]
\dd u\dd\mu \dd\theta\right\rangle_\psi,
\end{eqnarray}
where
\begin{eqnarray}
\fl
\hat{C}_{\spe\spe'}\left[ \hat{g}_\spe,\hat{g}_{\spe'} \right]=
\cT_{\spe,0}^*C_{\spe\spe'}\left[ \lambda_\spe^{-1}\cT^{-1*}_{\spe,0}
g_{\spe}
,\cT^{-1*}_{\spe',0}F_{\spe' 0} \right]
\nonumber\\[5pt]
\fl\hspace{1cm} + 
\cT_{\spe,0}^*C_{\spe\spe'} \left[ \cT^{-1*}_{\spe,0}F_{\spe 0},
\lambda_{\spe'}^{-1}\cT^{-1*}_{\spe',0}
g_{\spe'}
\right].
\end{eqnarray}
By hypothesis the system is instrinsically ambipolar, so
\eq{eq:IntrinsicAmbipolarity_g} holds. Thus,
\begin{eqnarray}\label{eq:EntropyProduction}
\left\langle\sum_\spe
\tau_\spe\int B \lambda_\spe^{-1}\hat{g}_\spe
\sum_{\sigma'}
\hat{C}_{\spe\spe'}\left[ \hat{g}_\spe,\hat{g}_{\spe'} \right]
\dd u\dd\mu \dd\theta\right\rangle_\psi
=0.
\end{eqnarray}
Recasting \eq{eq:linEntProdZero} and \eq{eq:linEntProdZeroResult} into
non-dimensionalized variables, one readily confirms that
\eq{eq:EntropyProduction} implies
\begin{equation}\label{eq:generalForm_g_preliminar}
 g_\spe = \alpha_{0,\spe} F_{\spe 0} 
+\alpha_1 Z_\spe u
F_{\spe 0}
+
\alpha_2\lambda_\spe \left(u^2/2+\mu B\right)F_{\spe 0},
\end{equation}
where $\alpha_{0,\spe}$, $\alpha_1$, and $\alpha_2$ are, in principle,
arbitrary functions of $\bR$. Conditions \eq{eq:gaugefixing_g} only
give $\langle\alpha_{0,\spe}\rangle_\psi \equiv 0$ for every $\spe$,
and $\langle\alpha_2\rangle_\psi \equiv 0$.

Imposing that $g_\spe$, i.e. the right-side of
\eq{eq:generalForm_g_preliminar}, satisfies
\eq{eq:EquationCoeffVarphi}, we get
\begin{eqnarray}\label{eq:Find_alphas}
\fl
u\bun\cdot\nabla_\bR\alpha_{0,\spe}
+ Z_\spe
\left(u\bun\cdot\nabla_\bR - \mu\bun\cdot\nabla_\bR
B\partial_u
\right)
(\alpha_1  u
)
\nonumber\\[5pt]
\fl\hspace{1cm}
+
\lambda_\spe u
\left(u^2/2 + \mu B\right)
\bun\cdot\nabla_\bR\alpha_2
+
\frac{Z_\spe}{T}\bv_M\cdot\nabla_\bR\psi = 0.
\end{eqnarray}
This equation must hold for every $u$ and $\mu$. Setting $\mu = 0$,
the terms linear in $u$ give $\bun\cdot\nabla_\bR \alpha_{0,\spe} \equiv 0$
and the terms cubic in $u$ give $\bun\cdot\nabla_\bR \alpha_{2} \equiv
0$. Therefore, we are left with
\begin{eqnarray}\label{eq:Find_alpha1}
\fl
\left(u\bun\cdot\nabla_\bR - \mu\bun\cdot\nabla_\bR
B\partial_u
\right)
(\alpha_1  u
)
\nonumber\\[5pt]
\fl\hspace{1cm}
+
\frac{1}{T}
\bv_M\cdot\nabla_\bR\psi
= 0.
\end{eqnarray}

Observe that if $\bB$ satisfies $(\nabla_\bR\times\bB)\cdot\nabla_\bR\psi
= 0$, then
\begin{eqnarray}
\bv_M\cdot\nabla_\bR\psi
=
\frac{u^2 + \mu B}{B^2}(\bun\times\nabla_\bR B)\cdot\nabla_\bR\psi.
\end{eqnarray}
Hence, \eq{eq:Find_alpha1} is recast into
\begin{eqnarray}\label{eq:Find_alpha1_2}
\fl
u^2\bun\cdot\nabla_\bR\alpha_1 - \alpha_1\mu\bun\cdot\nabla_\bR
B 
\nonumber\\[5pt]
\fl\hspace{1cm}
+
\frac{1}{T}
\frac{u^2 + \mu B}{B^2}(\bun\times\nabla_\bR B)\cdot\nabla_\bR\psi
= 0.
\end{eqnarray}
Again, this identity must hold for every value of $u$ and $\mu$. If
$\mu = 0$, then
\begin{eqnarray}\label{eq:Find_alpha1_muzero}
\fl
\frac{1}{T}
\frac{1}{B}(\bun\times\nabla_\bR B)\cdot\nabla_\bR\psi
= - B\bun\cdot\nabla_\bR\alpha_1.
\end{eqnarray}
If $u=0$, then
\begin{eqnarray}\label{eq:Find_alpha1_uzero}
\fl
\frac{1}{T}
\frac{1}{B}(\bun\times\nabla_\bR B)\cdot\nabla_\bR\psi
=  \alpha_1\bun\cdot\nabla_\bR B .
\end{eqnarray}
Subtracting the last two equations one finds $\bun\cdot\nabla_\bR
(B\alpha_1) = 0$, and consequently we can write
\begin{equation}\label{eq:alpha1_solution}
B\alpha_1 = -\frac{\chi}{T}
\end{equation}
for some flux function $\chi(\psi)$. Employing \eq{eq:alpha1_solution}
in \eq{eq:Find_alpha1_uzero} we get a condition that is expressed
solely in terms of the magnetic field:
\begin{eqnarray}\label{def:quasisymmetry}
(\bun\times\nabla_\bR \psi)\cdot\nabla_\bR B
=  \chi(\psi)\bun\cdot\nabla_\bR B.
\end{eqnarray}
A magnetic field satisfying \eq{def:quasisymmetry} for some flux
function $\chi$ is called {\em quasisymmetric}. Note that in order to
get this necessary condition we have not employed either
\eq{eq:IntrinsicAmbipolarity_h} or \eq{eq:IntrinsicAmbipolarity_l}.

Now, let us show that quasisymmetry, \eq{def:quasisymmetry}, is a
sufficient condition for intrinsic ambipolarity. The essential point
is to note that in a quasisymmetric system with
$(\nabla_\bR\times\bB)\cdot\nabla_\bR\psi = 0$, the radial magnetic drift
can be written as
\begin{equation}\label{eq:property_vdrift_QS}
\bv_M\cdot\nabla_\bR\psi
=
\chi(\psi)\left(u\bun\cdot\nabla_\bR - \mu\bun\cdot\nabla_\bR
B\partial_u
\right)\frac{u}{B}\, .
\end{equation}

We have to prove that \eq{eq:IntrinsicAmbipolarity_g},
\eq{eq:IntrinsicAmbipolarity_h}, and \eq{eq:IntrinsicAmbipolarity_l}
are satisfied. Take \eq{eq:IntrinsicAmbipolarity_g}. Using
\eq{eq:property_vdrift_QS} and integration by parts,
\begin{eqnarray} \label{eq:ProofSufficientCondition1}
\fl\sum_\spe
\Bigg\langle
\frac{1}{\lambda_\spe}
\int B\,
\bv_M 
\cdot\nabla_\bR\psi\,
g_{\spe}
\dd u\dd\mu\dd\theta
\Bigg\rangle_\psi =
\nonumber\\[5pt]
\fl\hspace{1cm}
-\chi(\psi)\sum_\spe
\Bigg\langle
\frac{1}{\lambda_\spe}
\int u\,
\left(u\bun\cdot\nabla_\bR - \mu\bun\cdot\nabla_\bR
B\partial_u
\right)g_{\spe}
\,
\dd u\dd\mu\dd\theta
\Bigg\rangle_\psi.
\end{eqnarray}
Recall that $g_\spe$ satisfies \eq{eq:EquationCoeffVarphi}. Using
equation \eq{eq:EquationCoeffVarphi} in
\eq{eq:ProofSufficientCondition1}, one sees that the last term on the
left side of \eq{eq:EquationCoeffVarphi} does not contribute because
it is even in $u$. The contribution of the right side of
\eq{eq:EquationCoeffVarphi} gives
\begin{eqnarray} \label{eq:ProofSufficientCondition2}
\fl\sum_\spe
\Bigg\langle
\frac{1}{\lambda_\spe}
\int B\,
\bv_M 
\cdot\nabla_\bR\psi\,
g_{\spe}
\dd u\dd\mu\dd\theta
\Bigg\rangle_\psi =
\nonumber\\[5pt]
\fl\hspace{1cm}
-\chi(\psi)\sum_\spe
\Bigg\langle
\frac{1}{\lambda_\spe}
\int u\,
\Bigg(
\sum_{\sigma'}\cT_{\spe,0}^*
C_{\spe\spe'}\left[ \cT^{-1*}_{\spe,0}
g_{\spe}
,\cT^{-1*}_{\spe',0}F_{\spe' 0} \right]
\nonumber\\[5pt]
\fl\hspace{1cm} + \sum_{\sigma'}
\frac{\lambda_\spe}{\lambda_{\spe'}}\cT_{\spe,0}^*
C_{\spe\spe'} \left[ \cT^{-1*}_{0,\spe}F_{\spe 0},
\cT^{-1*}_{0,\spe'}
g_{\spe'}
\right]
\Bigg)
\dd u\dd\mu\dd\theta
\Bigg\rangle_\psi,
\end{eqnarray}
which is zero due to the fact that the collision operator conserves
total momentum (see \eq{eq:propcollnondim}). Analogous steps lead to
\eq{eq:IntrinsicAmbipolarity_h} and \eq{eq:IntrinsicAmbipolarity_l}.

To summarize, a magnetic configuration with
$(\nabla_\bR\times\bB)\cdot\nabla_\bR\psi = 0$ is intrinsically
ambipolar if and only if it is quasisymmetric. If the stellarator is
not quasisymmetric, $\varphi_0$ can be found from neoclassical
theory. That is, from \eq{eq:AmbipolarityCondition} and
\eq{eq:Vlasovorder1gyroav4}. However, if the stellarator is
quasisymmetric, the ambipolarity condition
\eq{eq:AmbipolarityCondition} is identically satisfied to lowest
order. The determination of $\varphi_0$ is not a neoclassical problem
anymore, and one has to include both turbulent and neoclassical
terms. As explained in \cite{parra10a} for the tokamak, a moment
approach gives $\varphi_0$ if the second-order pieces of the
distribution function and turbulent electrostatic potential are
known. This is why the full second-order gyrokinetic equations are
required. The long-wavelength component of the second-order
Fokker-Planck and quasineutrality equations have to be solved to
calculate $\varphi_0$. The short-wavelength component of the equations
to second-order are also needed, and these will be given elsewhere.

\section{Violations of intrinsic ambipolarity due to unavoidable
  deviations from quasisymmetry}
\label{sec:deviationQS}

We have seen that for a non-quasisymmetric stellarator the
long-wavelength radial electric field is neoclassical, and therefore
determined by equations \eq{eq:Vlasovorder1gyroav4} and
\eq{eq:AmbipolarityCondition}.  For a quasisymmetric stellarator the
long-wavelength radial electric field is undetermined to this order,
and the higher-order pieces given by equations \eq{eq:FPsecondorder2}
and \eq{eq:gyroPoissonlw4} will be necessary to calculate it. Our
final objective is to give in the future a model valid for both types
of stellarators.  Garren and Boozer proved in reference
\cite{Garren1991} that, except for the axisymmetric case, no toroidal
quasisymmetric magnetic field exists. Therefore, it is important to
understand how certain important quantities such as
$\langle\bJ\cdot\nabla_\bR\psi\rangle_\psi$ respond to the unavoidable
deviation of the magnetic field from quasisymmetry.

Having finished our asymptotic expansions, we restore dimensionful
variables. We find it appropriate for this section, and especially for
the discussion in subsection \ref{sec:dependencecollisionality}. We
omit the gyrophase because we only deal with gyrophase-independent
functions. Then, we write the flux-surface average of the radial
electric current as
\begin{eqnarray}\label{eq:FSAofradialJdimensionful}
\fl
\left\langle
\bJ\cdot\nabla_\bR\psi
\right\rangle_\psi
=
2\pi \left\langle
\sum_{\spe}
Z_\spe e
\int B
\bv_{M,\spe}\cdot\nabla_\bR\psi
\, G_{\spe 1}
\dd u \dd\mu\right\rangle_\psi,
\end{eqnarray}
where we have simplified the notation by dropping the integration
limits, the superindex $\lw$ on $G_{\spe 1}$, and some arguments of
the functions in the integrand. The function $G_{\spe
  1}$ is the non-adiabatic piece of the distribution function,
$G_{\spe 1}= F_{\spe 1} + (Z_\spe e\varphi_1/T_\spe) F_{\spe 0}$, and
satisfies
\begin{eqnarray}\label{eq:FPG1dimensionful}
\fl
\left(u\bun\cdot\nabla_\bR - \mu\bun\cdot\nabla_\bR
B\partial_u
\right)
G_{\spe 1}
+
\nonumber\\[5pt]
\fl\hspace{1cm}
\left(
\frac{Z_\spe e }{T_\spe}\partial_\psi\varphi_0
+\Upsilon_\spe
\right)
\bv_{M,\spe}\cdot\nabla_\bR\psi
\, F_{\spe 0}
= C^\ell_\spe[G_1],
\end{eqnarray}
with $C^\ell_\spe[G_1]$ the linearized collision operator,
\begin{eqnarray}
\fl
C^\ell_\spe[G_1]:=
\sum_{\spe'}
\Big(\cT_{\spe,0}^*
C_{\spe\spe'}[\cT_{\spe,0}^{*-1}G_{\spe 1},\cT_{\spe',0}^{*-1}F_{\spe' 0}]
\nonumber\\[5pt]
\fl\hspace{1cm}
+
\cT_{\spe,0}^*C_{\spe\spe'}[\cT_{\spe,0}^{*-1}F_{\spe 0},\cT_{\spe',0}^{*-1}G_{\spe' 1}]
\Big),
\end{eqnarray}
\begin{eqnarray}
\fl\Upsilon_\spe := 
\frac{1}{n_\spe}\partial_\psi n_\spe+
\left(\frac{m_\spe(u^2/2 +\mu B)}{T_\spe}-\frac{3}{2}\right)
\frac{1}{T_\spe}\partial_\psi T_\spe,
\end{eqnarray}
and
\begin{eqnarray}
&& \fl F_{\spe 0}(\bR,u,\mu)=
n_{\spe}
\left(\frac{m_\spe}{2\pi T_{\spe}}\right)^{3/2}
\exp\left(-\frac{m_\spe(u^2/2 + \mu B)}{T_{\spe}}\right).
\end{eqnarray}
Recall that $n_\spe$ and $T_\spe$ depend only on $\psi$
and that $T_\spe = T_{\spe'}$ for every pair $\spe$, $\spe'$. We
stress that in dimensionful coordinates $F_{\spe} = F_{\spe 0} +
F_{\spe 1}+O(\epsilon_\spe^2 F_{\spe 0})$. Finally, the magnetic drift
velocity can be written as
\begin{eqnarray}\label{eq:vmDimensionful}
  \bv_{M,\spe} = \frac{1}{\Omega_\spe}\bun\times\big( 
u^2 \kappabf + \mu\nabla_\bR B\big).
\end{eqnarray}

Let us be more concrete about our aim. Assume that our
magnetic field can be written as $\bB = \bB_{0} + \alpha \bB_1$, where
$\bB_0$ is quasisymmetric and $\alpha$ is small. We want to expand
\eq{eq:FSAofradialJdimensionful} in powers of $\alpha$ and show that
\begin{eqnarray}\label{eq:FSAofradialJdimensionful2}
\fl
\left\langle
\bJ\cdot\nabla_\bR\psi
\right\rangle_\psi
= O(\alpha^2).
\end{eqnarray}
We will see that as soon as we employ the appropriate mathematical
language, the proof will become very easy.

For an arbitrary stellarator magnetic field that satisfies
$(\nabla_\bR\times\bB)\cdot\nabla_\bR\psi \equiv 0$ (a
magnetohydrodynamic equilibrium is a particular case), there exist
Boozer coordinates~\cite{Boozer81}, $\{\psi,\Theta,\zeta\}$, in
which
\begin{eqnarray}
\fl
\bB = -\tilde\eta\nabla_\bR\psi + \frac{I(\psi)}{2\pi}\nabla_\bR\Theta +
\frac{J(\psi)}{2\pi}\nabla_\bR\zeta
\end{eqnarray}
and
\begin{eqnarray}
\fl
\bB = \frac{\Psi'_p(\psi)}{2\pi}\nabla_\bR\zeta\times\nabla_\bR\psi +
\frac{\Psi'_t(\psi)}{2\pi}\nabla_\bR\psi\times\nabla_\bR\Theta.
\end{eqnarray}
Here, ${}^{'}$ denotes differentiation with respect to $\psi$,
$\Psi_t$ is the toroidal flux, $\Psi_p$ the poloidal flux, and
$\tilde\eta(\psi,\Theta,\zeta)$ is a singly-valued function. The
metric determinant can be expressed in terms of the magnitude of the
magnetic field,
\begin{equation}\label{eq:sqrtgBoozer}
\sqrt{g} = \frac{V' \langle B^2 \rangle_\psi}{4\pi^2 B^2}
\end{equation}
 and the derivative along the magnetic field reads
\begin{eqnarray}\label{eq:gradparBoozer}
\bun\cdot\nabla_\bR = \frac{2\pi\Psi'_t B}{\langle B^2 \rangle_\psi V'}
(\iotabar\partial_\Theta + \partial_\zeta),
\end{eqnarray}
where $\iotabar(\psi) = \Psi'_p(\psi)/\Psi'_t(\psi)$ is the rotational
transform. As for the radial component of the magnetic drift,
\begin{equation}\label{eq:vpsiBoozer2}
\fl
v_{\psi, \spe}
:= \bv_{M,\spe}\cdot\nabla_\bR\psi
=
\frac{2\pi m_\spe c (u^2 + \mu B)}
{  Z_\spe e V'\langle B^2 \rangle_\psi  B}
\big(
I
\partial_\zeta B
-J \partial_\Theta B
\big),
\end{equation}
where $(\nabla_\bR\times\bB)\cdot\nabla_\bR\psi \equiv
0$ has been used. Therefore,
\begin{eqnarray}\label{eq:FSAofradialJdimensionfulBoozer}
\fl
\left\langle
\bJ\cdot\nabla_\bR\psi
\right\rangle_\psi
=
\nonumber\\[5pt]
\fl\hspace{0.5cm}
\sum_\spe \frac{m_\spe c}{V'}
\oint
\dd\Theta\dd\zeta
\int
\frac{u^2 + \mu B}{B^2}
(I\partial_\zeta B - J \partial_\Theta B)
\, G_{\spe 1}
\dd u \dd\mu,
\end{eqnarray}
where $\oint$ denotes an integral over $[0,2\pi]$ in the angular
variables. Equations \eq{eq:FPG1dimensionful}, \eq{eq:gradparBoozer},
\eq{eq:vpsiBoozer2}, \eq{eq:FSAofradialJdimensionfulBoozer}, and the
fact that the kernel of the collision operator in drift-kinetic
coordinates depends on the magnetic field only through $B$ (see
\ref{sec:CollOp_DKcoor}) means that the function
$B(\psi,\Theta,\zeta)$ contains all the magnetic geometry information
that is needed for the drift kinetic equation and for $\left\langle
  \bJ\cdot\nabla_\bR\psi \right\rangle_\psi$. Put in different words,
take two different stellarators and find Boozer coordinates for each
of them. Then, formally, \eq{eq:FPG1dimensionful} and
\eq{eq:FSAofradialJdimensionfulBoozer} differ for these devices, at
most, in the function $B(\psi,\Theta,\zeta)$.

Now, we need to recall another way of characterizing quasisymmetric
systems. In reference \cite{Helander08} it was shown that a magnetic
field satisfies \eq{def:quasisymmetry} if and only if, in Boozer
coordinates $(\psi,\Theta,\zeta)$, the modulus of the magnetic field
depends only on a single helicity, say $B\equiv
B(\psi,M\Theta-N\zeta)$ for some pair $(M,N)$. Actually, equilibrium
conditions near the magnetic axis enforce $M=1$~\cite{Cary97}.

Thus, the result by Garren and Boozer~\cite{Garren1991} means that our
stellarator magnetic field, once recast in Boozer coordinates, has a
modulus that at best can be written as $B(\psi,\Theta,\zeta)=
B_0(\psi,\Theta-N\zeta) +\alpha B_1(\psi,\Theta,\zeta)$, with small
$\alpha$. The perturbation $B_1$ does not contain the
helicity $\Theta-N\zeta$. Without loss of generality, we can take
$B_0$ quasi-axisymmetric, $\partial_\zeta B_0 \equiv 0$, which
corresponds to $N=0$\footnote{If $\bB$ is helically symmetric, $N\neq
  0$, then one can define $\overline{\Theta} := \Theta -N\zeta$. The
  coordinates $\overline{\Theta},\zeta$ are Boozer angles and in terms
  of them the problem reduces to the quasi-axisymmetric case.}. Since $B_1$
does not contain the helicity of $B_0$, then
\begin{equation}\label{eq:conditionB1}
\int B_1(\psi,\Theta,\zeta)\dd\zeta = 0.
\end{equation}

We are ready to prove \eq{eq:FSAofradialJdimensionful2}. The
$O(\alpha^0)$ terms of \eq{eq:FSAofradialJdimensionfulBoozer} vanish
due to quasisymmetry. The $O(\alpha)$ terms, $\left\langle
  \bJ\cdot\nabla_\bR\psi \right\rangle_\psi^{(1)}$, are
\begin{eqnarray}\label{eq:J1}
\fl
\left\langle
\bJ\cdot\nabla_\bR\psi
\right\rangle_\psi^{(1)}
=
\nonumber\\[5pt]
\fl\hspace{1cm}
-\sum_\spe \frac{m_\spe c}{V'}
\oint
\dd\Theta\dd\zeta
\int
\frac{u^2 + \mu B_0}{B_0^2}
J \partial_\Theta B_0
\, G_{\spe }^{(1)}
\dd u \dd\mu
\nonumber\\[5pt]
\fl\hspace{1cm}
+\sum_\spe \frac{m_\spe c}{V'}
\oint
\dd\Theta\dd\zeta
\int
\Bigg[\frac{2u^2 + \mu B_0}{B_0^3}B_1 J\partial_\Theta B_0
\nonumber\\[5pt]
\fl\hspace{1cm}
+
\frac{u^2 + \mu B_0}{B_0^2}
(I\partial_\zeta B_1 - J \partial_\Theta B_1)
\Bigg] G_{\spe }^{(0)}
\dd u \dd\mu.
\end{eqnarray}
Here,
\begin{equation}
G_{\spe 1}:= G_{\spe}^{(0)}+\alpha G_{\spe}^{(1)} +
O(\alpha^2),
\end{equation}
 where the coefficients are determined by the equations
\begin{eqnarray}\label{eq:FPG1dimensionfulzero}
\fl
\left(u\bun\cdot\nabla_\bR - \mu\bun\cdot\nabla_\bR
B\partial_u
\right)^{(0)}
G_{\spe}^{(0)}
+
\nonumber\\[5pt]
\fl\hspace{1cm}
\left(
\frac{Z_\spe e }{T_\spe}\partial_\psi\varphi_0
+\Upsilon_\spe
\right)
(\bv_{M,\spe}\cdot\nabla_\bR\psi)^{(0)}
\, F_{\spe 0}^{(0)}
\nonumber\\[5pt]
\fl\hspace{1cm}
= C^{\ell (0)}_\spe[G^{(0)}],
\end{eqnarray}

\begin{eqnarray}\label{eq:FPG1dimensionfulone}
\fl
\left(u\bun\cdot\nabla_\bR - \mu\bun\cdot\nabla_\bR
B\partial_u
\right)^{(0)}
G_{\spe}^{(1)}
\nonumber\\[5pt]
\fl\hspace{1cm}
+ 
\left(u\bun\cdot\nabla_\bR - \mu\bun\cdot\nabla_\bR
B\partial_u
\right)^{(1)}
G_{\spe}^{(0)}
\nonumber\\[5pt]
\fl\hspace{1cm}
+\left[\left(
\frac{Z_\spe e }{T_\spe}\partial_\psi\varphi_0
+\Upsilon_\spe
\right)
\left(\bv_{M,\spe}\cdot\nabla_\bR\psi
\, F_{\spe 0}\right)\right]^{(1)}
\nonumber\\[5pt]
\fl\hspace{1cm}
= C^{\ell(1)}_\spe[G^{(0)}] + C^{\ell(0)}_\spe[G^{(1)}].
\end{eqnarray}
Obviously,
\begin{eqnarray}\label{eq:gradparBoozerzero}
\fl  (\bun\cdot\nabla_\bR)^{(0)} = 
\frac{2\pi\Psi'_t B_0}{\langle B^2 \rangle_\psi V'}
  (\iotabar\partial_\Theta + \partial_\zeta),
\end{eqnarray}
\begin{eqnarray}\label{eq:gradparBoozerone}
\fl   (\bun\cdot\nabla_\bR)^{(1)} = 
\frac{2\pi\Psi'_t B_1}{\langle B^2 \rangle_\psi V'}
  (\iotabar\partial_\Theta + \partial_\zeta),
\end{eqnarray}
\begin{eqnarray}\label{eq:gradparBBoozerzero}
\fl   (\bun\cdot\nabla_\bR B)^{(0)} = 
\frac{2\pi\Psi'_t B_0}{\langle B^2 \rangle_\psi V'}
  \iotabar\partial_\Theta B_0,
\end{eqnarray}
\begin{eqnarray}\label{eq:gradparBBoozerone}
\fl   (\bun\cdot\nabla_\bR B)^{(1)} = 
\frac{2\pi\Psi'_t}{\langle B^2 \rangle_\psi V'}
\left[
B_1
  \iotabar\partial_\Theta B_0
+
 B_0
  (\iotabar\partial_\Theta + \partial_\zeta) B_1
\right],
\end{eqnarray}
etc. In \ref{sec:CollOp_DKcoor} we have computed explicitly the form
of the Fokker-Planck collision operator in coordinates $(\bR,u,\mu) =
\cT_{\spe, 0}^{-1}(\boldr,\bv)$, and we make it clear that the kernel
depends on the magnetic geometry only via the magnitude of $\bB$.  We
will not give fully explicit expressions for the pieces of the
linearized operator, $C^{\ell (1)}_\spe[G^{(0)}]$ and $C^{\ell
  (0)}_\spe[G^{(1)}]$, because they are cumbersome and we will not
exploit them at all. Equations \eq{eq:collisionOpDKCoor},
\eq{eq:GammaComponents}, and \eq{eq:RosenbluthPotEqDKcoor} are enough
to realize that $C^{\ell (1)}_\spe[G^{(0)}]$ is linear in $B_1$ and
that the kernel defining the piece $C^{\ell (0)}_\spe[G^{(1)}]$ does
not depend on $\zeta$. Finally, from \eq{eq:FPG1dimensionfulzero} and
\eq{eq:FPG1dimensionfulone}, one immediately notes that
\begin{equation}
\partial_\zeta G_\spe^{(0)} = 0
\end{equation}
and
\begin{equation}
\int_0^{2\pi}
G^{(1)}_\spe\dd\zeta = 0.
\end{equation}
 Hence, we infer that every term on
the right-hand side of \eq{eq:J1} is of the form
\begin{eqnarray}
\oint
f(\psi,\Theta) g(\psi,\Theta,\zeta)
\dd\Theta\dd\zeta,
\end{eqnarray}
where
\begin{equation}
\int_0^{2\pi}
g(\psi,\Theta,\zeta)\dd\zeta = 0.
\end{equation}
Then, the result $\left\langle \bJ\cdot\nabla_\bR\psi
\right\rangle_\psi^{(1)}\equiv 0$ follows. The terms $O(\alpha^2)$ do
not vanish, in general, so $\left\langle \bJ\cdot\nabla_\bR\psi
\right\rangle_\psi = O(\alpha^2)$ as far as the lowest order
neoclassical terms are concerned. This result was obtained in a short
mean-free path plasma, employing fluid equations, in reference
\cite{Simakov2009}. Unsurprisingly, flow damping due to small magnetic
ripple in tokamaks also follows this scaling (see, for example,
\cite{Connor1973}).

Therefore, for small deviations from quasisymmetry,
the flux-surface-averaged radial current can be written as
\begin{eqnarray}\label{eq:schematicJalpha2}
\fl
\langle\bJ\cdot\nabla_\bR\psi\rangle_\psi
=
(
\epsilon_s^2\alpha^2 A + \epsilon_s^3C
)
e n_{e0}c_s|\nabla_\bR\psi|_0
\nonumber\\[5pt]
\fl\hspace{1cm}
+ O(\epsilon_s^4 e n_{e0}c_s|\nabla_\bR\psi|_0),
\end{eqnarray}
where $A$ and $C$ are $O(1)$, and $|\nabla_\bR\psi|_0$ is a
characteristic value of $|\nabla_\bR\psi|$. The coefficient $A$ is
determined from neoclassical theory, whereas $C$ has neoclassical and
turbulent contributions. Note that the dependence of
\eq{eq:schematicJalpha2} on $\alpha$ has been inferred in this section
and the dependence on $\epsilon_s$ is an immediate consequence of the
results of previous ones: the fact that the lowest-order neoclassical
contribution is $O(\epsilon_s^2)$ may be understood from inspection of
\eq{eq:FSAofradialJdimensionful}. In addition, when we derived the
lowest-order flux-surface-averaged radial current, we deduced that
turbulence does not contribute to it; that is, we proved that to
$O(\epsilon_s^2)$ the quantity $\left\langle \bJ\cdot\nabla_\bR\psi
\right\rangle_\psi$ is exactly given by the right-hand side of
\eq{eq:FSAofradialJdimensionful}. Hence, turbulent contributions can
only enter to $O(\epsilon_s^3)$ or higher.

It is clear that if $\alpha
> \epsilon_s^{1/2}$, then the long-wavelength radial electric field is
set neoclassically. Finally, if the magnetic field possesses
stellarator symmetry, $\epsilon_s^3C$ in \eq{eq:schematicJalpha2} has
to be replaced by $\epsilon_s^4C$~\cite{Sugama11b}, and therefore
$\varphi_0$ is neoclassical when $\alpha > \epsilon_s$.

The criterion for rotation, however, does not depend on the existence
or not of stellarator symmetry: the stellarator is able to rotate to
high speeds if $\alpha < \epsilon_s^{1/2}$. Consider the total
momentum conservation equation,
\begin{eqnarray}\label{eq:TotalMomentumConservEq}
\fl\partial_t(n_im_i\bV_i)
&=
-\nabla_\boldr\cdot\left[\matPi_i + p_\perp(\matI - \bun\bun)
+ p_{||}\bun\bun
\right]
\nonumber\\[5pt]
\fl&
+
\frac{1}{c}\bJ\times\bB,
\end{eqnarray}
where $\bV_i$ is the ion flow, $p_{\perp}$ and $p_{||}$ are the total
perpendicular and parallel pressures, and $\matPi_i$ contains the
turbulent contributions to the ion stress tensor. Electron inertial
terms and turbulent terms corresponding to electrons are small in a
$\sqrt{m_e/m_i} \ll 1$ expansion and have been neglected. As in reference
\cite{Helander08}, we will employ a solenoidal vector field
$\mathbf{S}$ of the form
\begin{equation}
\mathbf{S} = -\frac{c}{B}\bun\times\nabla_\boldr\psi + \xi\bB.
\end{equation}
The function $\xi(\psi,\Theta,\zeta)$ is determined by imposing
$\nabla_\boldr\cdot\mathbf{S} = 0$. This equation can be solved if
$(\nabla_\boldr\times\bB)\cdot\nabla_\boldr\psi \equiv 0$. Taking the
scalar product of \eq{eq:TotalMomentumConservEq} with $\mathbf{S}$
gives, in steady state,
\begin{eqnarray}
\fl\left\langle
\bJ\cdot\nabla_\bR\psi
\right\rangle_\psi
= \Lambda^{\rm neo} + \Lambda^{\rm tb},
\end{eqnarray}
where
\begin{eqnarray}
\fl
\Lambda^{\rm neo}
=
\left\langle
(p_{||} - p_\perp)(\matI/3 - \bun\bun):
\nabla_\boldr
\mathbf{S}
\right\rangle_\psi
\end{eqnarray}
and
\begin{eqnarray}
\fl
\Lambda^{\rm tb}
=
\frac{1}{V'}\partial_\psi
\left\langle
V'\mathbf{S}\cdot \matPi_i\cdot\nabla_\boldr\psi
\right\rangle_\psi
\nonumber\\[5pt]
\fl\hspace{1cm}
-
\left\langle
\matPi_i:
\nabla_\boldr
\mathbf{S}
\right\rangle_\psi.
\end{eqnarray}
If $\alpha$ is sufficiently large, $\left\langle
  \bJ\cdot\nabla_\bR\psi \right\rangle_\psi \sim \Lambda^{\rm neo}$
and $\Lambda^{\rm tb}$ can be neglected. Then, the radial electric
field is set by the neoclassical ambipolarity condition and the
resulting flow is subsonic. Hence, in order to have sonic flows we
need $\left\langle \bJ\cdot\nabla_\bR\psi \right\rangle_\psi \sim
\Lambda^{\rm tb}$. Recall the transport time scale $\tau_E =
\epsilon_s^{-2}L/c_s$ defined in subsection
\ref{sec:orderingANDseparationOFscales}. Assuming that the transport
of momentum satisfies
\begin{equation}
\Lambda^{\rm tb}
\sim
\frac{c n_im_iV_i|\nabla_\boldr\psi|}{\tau_E B}\,, 
\end{equation}
we have that $V_i\sim v_{ti}$ requires $\left\langle
  \bJ\cdot\nabla_\bR\psi \right\rangle_\psi \sim \epsilon_s^3
en_{e0}c_s|\nabla_\boldr\psi|_0$. Since
\begin{equation}
\Lambda^{\rm
  neo}\sim\epsilon_s^2\alpha^2 en_{e0}c_s|\nabla_\boldr\psi|_0,
\end{equation}
we need $\alpha < \epsilon_s^{1/2}$. It is true that $\Lambda^{\rm tb}
\sim \epsilon_s^4 e n_{e0}c_s|\nabla_\bR\psi|_0$ in a stellarator
symmetric device, and the consequence of this is that the intrinsic
rotation will be subsonic. To get sonic rotation, we need external
sources of momentum such as neutral beams or radio frequency momentum
injection.

\subsection{Breakdown of the expansion}
\label{sec:dependencecollisionality}

One might expect that our expansion in integer powers of $\alpha$, and
hence the scaling of the radial current with $\alpha^2$, break down in
certain circumstances (even if $\alpha\ll 1)$ because of the usual
notion that helically trapped particles give transport that scales
with a fractional power of the depth of the magnetic wells (see, for
example, reference \cite{Helander08}). In what follows, we discuss the
validity of the expansion in integer powers of $\alpha$.

\begin{figure}
\includegraphics[angle=0,width=0.8\columnwidth]{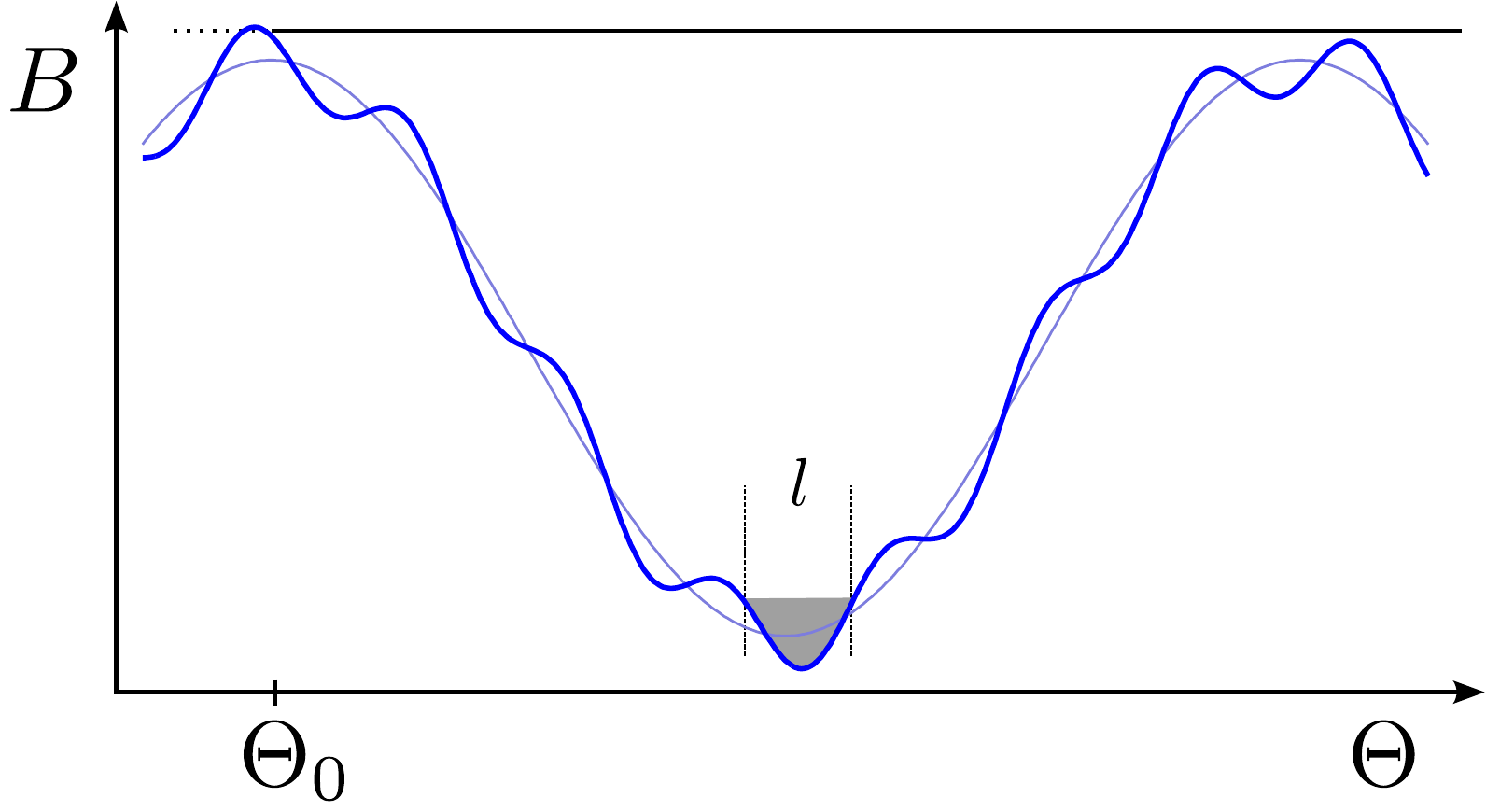}
\caption{(Color online) The dependence of the magnitude of the
  magnetic field on the distance along the field line in a
  quasisymmetric configuration (thin curve), and the result after a
  perturbation that makes it non-quasisymmetric has been added (thick
  curve). Trapped particles in a magnetic well of size $l$ created by
  the perturbation are confined to the shaded region. The trajectory
  of a particle that was passing and becomes barely trapped after
  including the perturbation is also represented.}
\label{fig:B_vs_Theta}
\end{figure}

In equation \eq{eq:FPG1dimensionful}, consider the factor
$\bun\cdot\nabla_\bR B$ in the parallel streaming operator. As long as
$\alpha(\bun\cdot\nabla_\bR)^{(0)} B_1\ll
(\bun\cdot\nabla_\bR)^{(0)} B_0$, our expansion can be carried
out for any value of the collisionality. However, if
\begin{equation}\label{eq:comparable_parallelgradients}
\alpha(\bun\cdot\nabla_\bR)^{(0)} B_1 
\sim (\bun\cdot\nabla_\bR)^{(0)} B_0
\end{equation}
somewhere, then we cannot guarantee that the expansion correctly
describes the behavior of particles in the whole phase space. The
gradients in \eq{eq:comparable_parallelgradients} can be comparable in
two qualitatively different situations that we treat separately (see
Figure \ref{fig:B_vs_Theta}). First, equation
\eq{eq:comparable_parallelgradients} holds near points where
$(\bun\cdot\nabla_\bR)^{(0)} B_0 = 0$, and this is related to almost
trapped, barely trapped, and deeply trapped particles in the magnetic
field $B_0$. There always exist points like these ones, and we deal
with them later on.

Second, in generic points where $(\bun\cdot\nabla_\bR)^{(0)} B_0 \sim B_0
L^{-1}$, condition \eq{eq:comparable_parallelgradients} is met if and
only if
\begin{equation}\label{eq:condition_helicity}
\frac{l}{L} \sim \alpha,
\end{equation}
where $L$ is the characteristic length of variation of $B_0$ and $l$
that of $B_1$. Then, the perturbation $\alpha B_1$ can create new
magnetic wells of typical width $l$. Given $\alpha$,
\eq{eq:condition_helicity} is satisfied only if the helicity of the
perturbation $B_1$ is large enough. Particles trapped in the new
helical wells may seem the main contribution to the breaking of
quasisymmetry (see reference \cite{Helander08}), but in most cases the
large gradients $\alpha\nabla_\bR B_1 \sim \nabla_\bR B_0$ will not be
exclusively parallel to the magnetic field. Typically, $\nabla_\bR
B_1$ will modify $v_{\psi,\spe}$ enough to make $\alpha
v_{\psi,\spe}^{(1)}\sim v_{\psi,\spe}^{(0)}$. The problem is that, if
this is the case, one cannot derive a scaling with $\alpha$ for the
distribution function outside the wells. Consequently, one cannot find
out whether helically trapped particles dominate transport, and no
scaling for the radial electric current can be derived either. As a
result, our expansion is only valid for
\begin{equation} \label{eq:validitycondition}
\frac{l}{L} \gg \alpha.
\end{equation}

\begin{figure}
\includegraphics[angle=0,width=0.8\columnwidth]{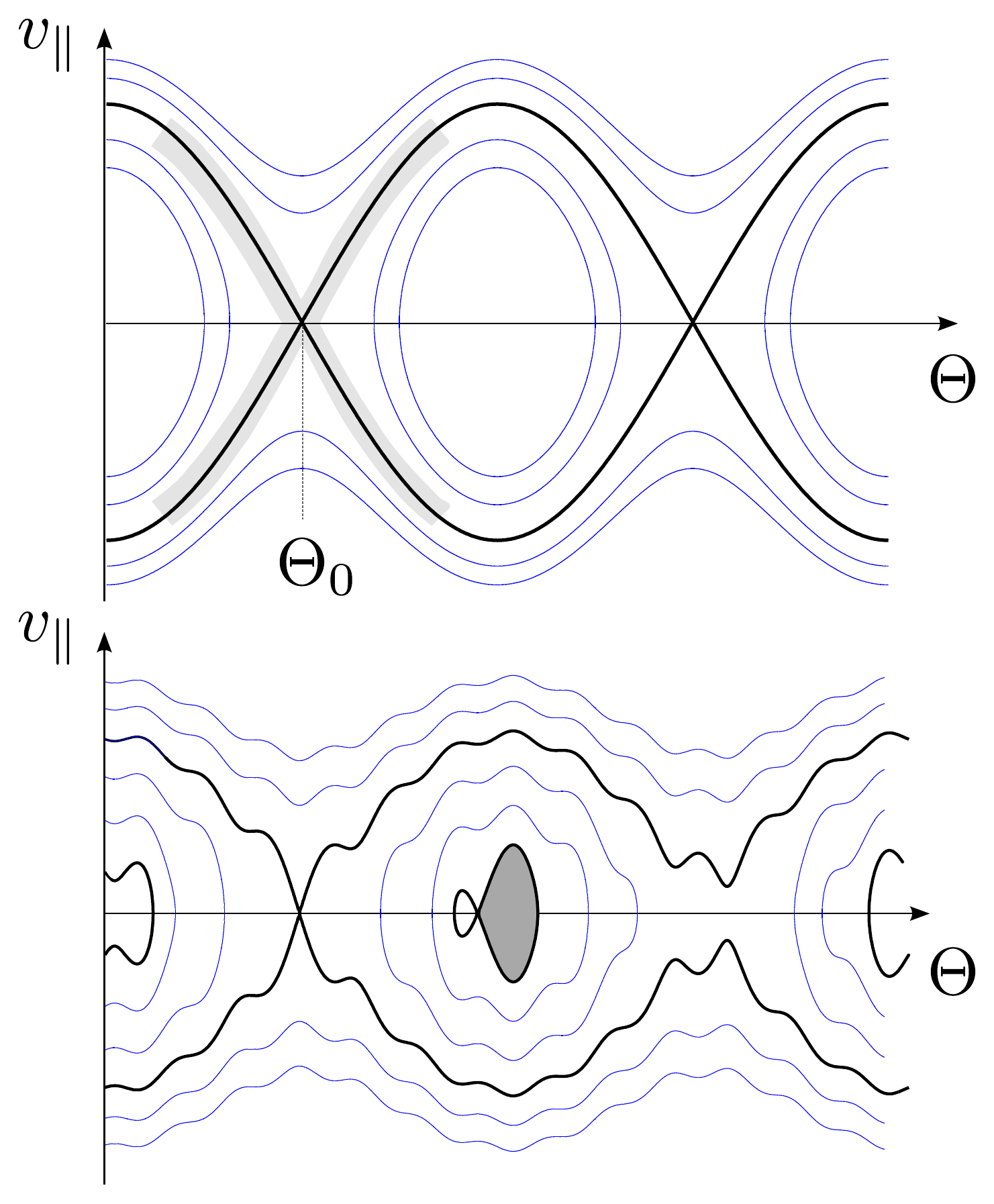}
\caption{(Color online) Some contours of constant kinetic energy for
  the quasisymmetric magnetic field (top) and perturbed magnetic field
  (bottom) of Figure \ref{fig:B_vs_Theta}.}
\label{fig:vpar_vs_Theta}
\end{figure}

If large helicity perturbations are unavoidable, then to get close to
quasisymmetry the design should make the large gradients
$\alpha\nabla_\bR B_1$ aligned in such a way that $\alpha
v_{\psi,\spe}^{(1)}\ll v_{\psi,\spe}^{(0)}$. Let us assume that
\eq{eq:validitycondition} is satisfied and hence no new magnetic wells
are created. Then, $\alpha(\bun\cdot\nabla_\bR)^{(0)} B_1 \sim
(\bun\cdot\nabla_\bR)^{(0)} B_0$ can only be satisfied in the vicinity
of points where $(\bun\cdot\nabla_\bR)^{(0)} B_0= 0$; points like
$\Theta_0$ in Figures \ref{fig:B_vs_Theta} and
\ref{fig:vpar_vs_Theta}. The trajectories that may qualitatively
change with the perturbation $\alpha B_1$ are the ones close to the
black, thick line in Figure \ref{fig:vpar_vs_Theta} (top). Those are
the trajectories of particles that are almost trapped or barely
trapped in $B_0$. For example, as is clearly seen in Figure
\ref{fig:B_vs_Theta} and in Figure \ref{fig:vpar_vs_Theta} (bottom), a
particle that was passing in the quasisymmetric field can become
trapped in a large magnetic well because $B = B_0 + \alpha B_1$ at
$\Theta_0$ is larger than $B_0$. These particles are not necessarily
trapped in the wells created by $B_0$, and in general will be trapped
in large wells that extend several of the wells due to $B_0$. The
trajectory shown in Figure \ref{fig:B_vs_Theta}, for example, spans at
least two of the wells created by $B_0$. We proceed to find the
scaling of the radial current due to these barely trapped or almost
trapped particles.

First, let us determine the size (in parallel velocity) of the region
of phase space that we are interested in. In the vicinity of $\Theta_0$,
\begin{eqnarray}
B_0 =
B_0(\Theta_0) +
\frac{1}{2} 
\partial_\Theta^2 B_0 (\Theta_0)(\Theta-\Theta_0)^2.
\end{eqnarray}
After adding the perturbation $\alpha
B_1$ we get, in that region,
\begin{eqnarray}
\fl
B = B_0(\Theta_0) +\frac{1}{2} \partial_\Theta^2 B_0 (\Theta_0)(\Theta-\Theta_0)^2
\nonumber\\[5pt]
  \fl\hspace{1cm}
 +
  \alpha B_1(\Theta_0)
+ \alpha\partial_\Theta B_1(\Theta_0) (\Theta-\Theta_0)
  =
  \nonumber\\[5pt]
  \fl\hspace{1cm}
  B_0(\Theta_0)  +
   \frac{1}{2} \partial_\Theta^2 B_0 (\Theta_0)
  \left(
    \Theta - \Theta_0 + 
    \alpha\frac{\partial_\Theta B_1(\Theta_0)}{\partial_\Theta^2 B_0 (\Theta_0)}
  \right)^2 
 \nonumber\\[5pt]
  \fl\hspace{1cm}
+ \alpha B_1(\Theta_0)
+ O(\alpha^2).
\end{eqnarray}
The perturbation $\alpha B_1$ moves the maximum of $B$ a distance
$O(\alpha)$ in $\Theta$ and it modifies the maximum value of $B$, $B_0
(\Theta_0)$, by a quantity of $O(\alpha)$. Denote by $v_{||}$ the
parallel velocity viewed as a function of the kinetic energy
$\varepsilon = u^2/2 + \mu B(\bR)$, the magnetic moment $\mu$, and the
position in space $\bR$, $v_{||} = \sqrt{2(\varepsilon - \mu
  B(\bR))}$. Due to these changes, the parallel velocity $v_{||}$ of a
barely trapped or almost trapped particle differs by a quantity of
$O(v_{||})$ from the velocity $v_{||0}$ obtained at the same location
$\Theta$ for the quasisymmetric magnetic field $B_0$. The expansion
for small $\alpha$ in \eq{eq:FPG1dimensionfulzero} and
\eq{eq:FPG1dimensionfulone} does not work when the perturbation to the
parallel velocity is not small in $\alpha$. We can estimate the number
of these particles and which part of their trajectory is significantly
modified.  In the vicinity of $\Theta_0$,
\begin{eqnarray}
  \fl v_{||} = \Bigg[v_{||0}^2 - 2 \mu \alpha B_1(\Theta_0)
\nonumber\\[5pt]
\fl\hspace{0.5cm}
 - 
\frac{\mu}{2} \partial_\Theta^2 B_0 (\Theta_0)
    \left(
      \Theta - \Theta_0 + 
      \alpha\frac{\partial_\Theta B_1(\Theta_0)}{\partial_\Theta^2 B_0 (\Theta_0)}
    \right)^2\Bigg]^{1/2},
\end{eqnarray}
where $v_{||0} = \sqrt{2(\varepsilon - \mu B_0 (\Theta_0))}$. It is
clear that the perturbation $\alpha B_1$ will significantly modify the
parallel velocity of particles with $v_{||} \sim \sqrt{\alpha} c_s$ in
a vicinity $|\Theta - \Theta_0| \sim \sqrt{\alpha}$ of $\Theta_0$. In
Figure \ref{fig:vpar_vs_Theta} (top), the region in phase space that
these particles occupy is a band with thickness $\Delta v_{||}/c_s
\sim \sqrt{\alpha}$ around the orbit plotted with a black, thick
line. Only in the part of that band that satisfies $|\Theta -
\Theta_0| \sim \sqrt{\alpha}$ is the orbit significantly modified by
the perturbation $\alpha B_1$. This part is lightly shaded in Figure
\ref{fig:vpar_vs_Theta} (top).

Now, we are ready to derive the scaling of the distribution function
and, then, the scaling of the radial electric current. The relevant
collisionality regime is the $1/\nu$ regime. In highly collisional
plasmas, it was proven in reference \cite{Simakov2009}, with fluid
equations, that an $\alpha^2$ scaling holds. Let $\nu_{\spe}$ be the
maximum of the collision frequencies $\nu_{\spe\spe'}$ when $\spe'$
runs over all the species, and $\nu_{*\spe} = \nu_\spe L/c_s$ the
corresponding collisionality. In the $1/\nu$ regime the parallel
streaming terms of \eq{eq:FPG1dimensionful} are the largest
ones. Define an expansion of $G_{\spe 1}$ in $\nu_{*\spe} \ll 1$
by
\begin{eqnarray}
G_{\spe 1} = G_{\spe}^{[-1]} + G_{\spe}^{[0]} + \dots,
\end{eqnarray}
where $G_{\spe}^{[j]} = O(\nu_{*\spe}^{j}\epsilon_\spe F_{\spe
  0})$. To lowest order in this small collisionality
expansion, equation
\eq{eq:FPG1dimensionful} reads
\begin{eqnarray}\label{eq:smallcolllowestorder}
\fl
\left(u\bun\cdot\nabla_\bR - \mu\bun\cdot\nabla_\bR
B\partial_u
\right)
G_{\spe}^{[-1]}=0,
\end{eqnarray}
so $G_{\spe}^{[-1]}$ is constant over orbits, $\overline{G_{\spe}^{[-1]}}=
G_{\spe}^{[-1]}$. Here,
\begin{eqnarray}
\fl\overline{h(\psi, \Theta, \zeta, \varepsilon, \mu)} = 
\nonumber\\[5pt]
\fl\hspace{1cm}
\frac{\int h(\psi, \Theta(l), \zeta(l), \varepsilon, \mu) v_{||}^{-1}(\psi, \Theta(l), \zeta(l), \varepsilon, \mu) \dd l}{\int v_{||}^{-1}(\psi, \Theta(l), \zeta(l), \varepsilon, \mu) \dd l}
\end{eqnarray}
is the transit average of the function $h(\psi, \Theta, \zeta,
\varepsilon, \mu)$, which is a time average over the particle
trajectory. To lowest order the trajectory is the magnetic field line,
$(\psi, \Theta(l), \zeta(l))$, with $l$ the arc length. The transit
average is taken holding the kinetic energy $\varepsilon$ and the
magnetic momentum $\mu$ fixed. An order higher in $\nu_{*\spe}$ than
\eq{eq:smallcolllowestorder}, the transit average of the
Fokker-Planck equation \eq{eq:FPG1dimensionful} gives
\begin{eqnarray}
\left(
\frac{Z_\spe e }{T_\spe}\partial_\psi\varphi_0
+\Upsilon_\spe
\right)
\overline{v_{\psi,\spe}}
\, F_{\spe 0}
= \overline{C^\ell_\spe[G^{[-1]}]}.
\end{eqnarray}
Clearly,
\begin{eqnarray}
G_{\spe}^{[-1]}\sim \frac{1}{\nu_{\rm  eff}B_0L^2}\overline{v_{\psi,\spe}}F_{\spe 0},
\end{eqnarray}
where $\nu_{\rm eff}:=\nu_{\spe}/\alpha$ is the effective collision
frequency. This effective collision frequency takes
  into account that the region of phase space where the perturbation
  $\alpha B_1$ modifies the trajectories significantly is a band with
  thickness $\Delta v_{||}/c_s \sim \sqrt{\alpha}$, and it takes very
  few collisions ($\sim \alpha$) for particles to diffuse out of this
  region into the part of phase space in which the trajectory is only
  modified by a correction of $O(\alpha)$. To obtain a final estimate
  for $G_{\spe}^{[-1]}$, we need to find a bound for
  $\overline{v_{\psi,\spe}}$. Away from the region $|\Theta -
  \Theta_0| \sim \sqrt{\alpha}$ where the orbits are modified
  significantly, the expansion in equations
  \eq{eq:FPG1dimensionfulzero} and \eq{eq:FPG1dimensionfulone} is
  valid. When $|\Theta - \Theta_0| \sim \sqrt{\alpha}$, the magnetic
  drift is composed of the piece due to $B_0$, $v_{\psi, \spe}^{(0)}$,
  and the piece due to $\alpha B_1$, $\alpha v_{\psi,
    \spe}^{(1)}$. For the quasisymmetric magnetic field, the radial
  magnetic drift has the form
\begin{equation}
\fl  v_{\psi,\spe}^{(0)} = \frac{m_\spe c}{Z_\spe e} 
  (u\bun\cdot\nabla_\bR - \mu\bun\cdot\nabla_\bR B\partial_u)^{(0)}
  \left(\chi(\psi)\frac{u}{B_0}
  \right).
\end{equation}
Thus, $v_{\psi,\spe}^{(0)}$ is zero at $\Theta_0$, where $(\bun \cdot
\nabla_\bR)^{(0)} B_0$, and it satisfies $v_{\psi,\spe}^{(0)} \sim
(\Theta - \Theta_0) \rho_i c_s B_0$ around $\Theta = \Theta_0$. In the
region of interest, $|\Theta - \Theta_0| \sim \sqrt{\alpha}$, giving
$v_{\psi,\spe}^{(0)} \sim \sqrt{\alpha} \rho_i c_s B_0 \gg \alpha v_{\psi,
  \spe}^{(1)}$. We need to find how much this $v_{\psi, \spe}^{(0)}$
contributes to the average $\overline{v_{\psi, \spe}}$. The particle
spends a time interval of $O(L/c_s)$ in the region $|\Theta -
\Theta_0| \gg \sqrt{\alpha}$, and $O(\sqrt{\alpha} L/\sqrt{\alpha}
c_s) = O(L/c_s)$ in the region $|\Theta - \Theta_0| \sim
\sqrt{\alpha}$, where the trajectory has a length of only
$\sqrt{\alpha} L$ and a parallel velocity $v_{||} \sim \sqrt{\alpha}
c_s$. Thus, the time spent in the much smaller region $|\Theta -
\Theta_0| \sim \sqrt{\alpha}$ is equivalent to the time spent in the
rest of the trajectory. Since the expansion in equations
\eq{eq:FPG1dimensionfulzero} and \eq{eq:FPG1dimensionfulone} works for
$|\Theta - \Theta_0| \gg \sqrt{\alpha}$, giving a contribution of
$O(\alpha)$ to $\overline{v_{\psi, \spe}}$ (only the vanishing
contribution of the quasisymmetric magnetic field enters to
$O(\alpha^0)$), $\overline{v_{\psi, \spe}}$ is dominated by the region
$|\Theta - \Theta_0| \sim \sqrt{\alpha}$, giving $\overline{v_{\psi,
    \spe}} \sim \sqrt{\alpha} \rho_i c_s B_0$. Then,
$G_{\spe}^{[-1]}\propto \alpha^{3/2}$ in the phase space region where
the expansion in \eq{eq:FPG1dimensionfulzero} and
\eq{eq:FPG1dimensionfulone} breaks. By employing
\eq{eq:FSAofradialJdimensionful}, one gets the scaling of the
contribution of these particles to the radial current, given by $e
\sqrt{\alpha} c_s^3 G_{\spe}^{[-1]} \overline{v_{\psi, \spe}} \propto
\alpha^{5/2}$, that is subdominant with respect to the $\alpha^2$
scaling. Thus, these particles do not dominate transport, and the
scaling $\langle\bJ\cdot\nabla_\bR\psi\rangle_\psi \sim \alpha^2$ will
be observed.

In addition to barely trapped and almost trapped particles, deeply
trapped particles can also be affected by perturbations that satisfy
\eq{eq:validitycondition} because they move in a region close to a
point with $(\bun\cdot\nabla_\bR)^{(0)} B_0 = 0$ (see the grey wells
in Figures \ref{fig:B_vs_Theta} and \ref{fig:vpar_vs_Theta}). Even if
new wells are created in this region, and the trajectories of deeply
trapped particles are modified, their contribution to the radial
electric current scales also as $\alpha^{5/2}$ and hence is
negligible.

\section{Conclusions}
\label{sec:conclusions}

Quasisymmetric stellarators can rotate freely in the symmetry
direction whereas non-quasisymmetric ones cannot. In the latter,
parallel viscosity damps the flow and sets the long-wavelength radial
electric field. In the former, the component of the flow in the
symmetry direction remains neoclassically undamped and the
long-wavelength radial electric field undetermined in calculations to
lowest-order in $\epsilon$, the Larmor radius over the characteristic
macroscopic scale. One of the objectives of this paper is to take the
first step towards a unified treatment of the long-wavelength radial
electric field and the rotation that is valid for quasisymmetric and
non-quasisymmetric stellarators. We have derived the gyrokinetic
Fokker-Planck and quasineutrality equations to $O(\epsilon^2)$ for an
arbitrary stellarator. Then, we have taken their long-wavelength
limit. To $O(\epsilon)$ one gets the well-known drift-kinetic equation
\eq{eq:Vlasovorder1gyroav4} and the first-order quasineutrality
equation \eq{eq:gyroPoissonlw2order1}. By themselves, these
first-order equations do not give the long-wavelength radial electric
field. In a generic stellarator one expects to be able to solve for it
by adding the neoclassical ambipolarity condition.  How does it show
up in the typical framework of gyrokinetic theory, i.e. a hierarchy of
Fokker-Planck and quasineutrality equations derived order by order in
$\epsilon$?  It comes from the solvability conditions of the
$O(\epsilon^2)$ piece of the Fokker-Planck equation, as explained in
subsection \ref{sec:timeevolutionquasineutrality}. Then, one can ask
whether the Fokker-Planck and quasineutrality equations to
$O(\epsilon)$, together with the neoclassical ambipolarity condition,
determine the long-wavelength radial electric field. The answer is
affirmative if and only if the stellarator is not quasisymmetric.

When the stellarator is quasisymmetric, the neoclassical radial
current does not give an equation for the long-wavelength radial
electric field (or equivalently, for the rotation along the symmetry
direction) and the full gyrokinetic computation to order
$O(\epsilon^2)$ is necessary. In particular, the long-wavelength
equations derived here must be implemented in a code to
obtain the second-order correction $\langle F_{\spe 2}^\lw\rangle$
(see \eq{eq:FPsecondorder2}). This piece of the distribution function
will eventually enter the equation for the long-wavelength radial
electric field obtained from higher-order solvability
conditions. Equivalently, this piece is necessary to get the correct
radial flux of momentum in a quasisymmetric stellarator. This is the
reason that we have worked out all the expressions to second order
explicitly. In order to have a complete set of equations that make it
possible to obtain the long-wavelength radial electric field, the
short-wavelength gyrokinetic equations to $O(\epsilon^2)$ must be
derived. This will be done elsewhere.

As mentioned above, a stellarator is able to sustain large rotation,
beneficial for confinement, if and only if it is
quasisymmetric~\cite{Helander08}. Since in reference \cite{Garren1991}
it was proven that, apart from the axisymmetric case, quasisymmetric
stellarators do not exist, it is pertinent to ask when a stellarator
can be considered quasisymmetric in practice. All the above results
allow us to pose this question in precise terms. It is the absence of
intrinsic ambipolarity that makes the computation of the radial
electric field radically different, and prevents the stellarator from
freely rotating, so it seems relevant to investigate how a
sufficiently small deviation $\alpha\bB_1$ from a quasisymmetric
magnetic field $\bB_0$ translates into a violation of intrinsic
ambipolarity. We have found that, generally, the stellarator can be
considered quasisymmetric in practice if
\begin{equation}\label{eq:ourScaling_conclusions}
\alpha < \epsilon^{1/2}.
\end{equation}


\ack We acknowledge Matt Landreman for very interesting discussions
and for suggesting some modifications to the first version of this
paper. This research was supported in part by grant ENE2012-30832,
Ministerio de Econom\'{\i}a y Competitividad, and by US DoE grant
DE-SC008435.

\appendix

\section{Some basic properties of the collision operator}
\label{sec:collisionoperator}

We recall (see, for example, reference \cite{Lenard1960}) that the
collision operator \eq{eq:collisionoperator} satisfies, for every
$\spe,\spe'$, the particle number, momentum, and energy conservation
properties
\begin{eqnarray}\label{eq:propcoll}
\int C_{\spe\spe'}\dd^3 v = 0,
\nonumber\\[5pt]
\int m_\spe\bv C_{\spe\spe'}\dd^3 v = -\int m_{\spe'}\bv C_{\spe'\spe}\dd^3 v,
\nonumber\\[5pt]
\int \frac{1}{2}m_\spe\bv^2 C_{\spe\spe'}\dd^3 v
= -\int \frac{1}{2} m_{\spe'}\bv^2 C_{\spe'\spe}\dd^3 v.
\end{eqnarray}

The entropy production term is non-negative. That is,
\begin{eqnarray}\label{eq:entropyProduction}
s = -\sum_{\spe,\spe'}\int \ln f_\spe C_{\spe\spe'}[f_\spe,f_{\spe'}]\dd^3 v
\end{eqnarray}
always satisfies $s\ge 0$, and $s=0$ if and only if all of the
distribution functions are Maxwellians with the same temperature and
flow; i.e. $f_\spe = f_{M\spe}$, where
\begin{eqnarray}\label{eq:maxwelliancoll}
{f_{M\spe}}({\boldr},{\bv})
=n_{\spe}(\boldr)
\left(
\frac{m_\spe}{2\pi T_{\spe}(\boldr)}
\right)^{3/2}
\exp\left(-\frac{m_\spe ({\bv}-\bV(\boldr))^2}{2T_{\spe}
(\boldr)}\right),
\end{eqnarray}
with the restriction $T_\spe = T_{\spe'}$ for every pair
$\spe,\spe'$. This implies that, given $\spe$ and $\spe'$,
\begin{equation}\label{Maxwellianannihilatecoll}
C_{\spe\spe'}[f_{\spe},f_{\spe'}] = 0
\end{equation}
if and only if $f_{\spe}$ and $f_{\spe'}$ are of the form
\eq{eq:maxwelliancoll} with $T_\spe = T_{\spe'}$.

Other well-known property, derived from
\eq{Maxwellianannihilatecoll}, is
\begin{equation}\label{eq:flowcoll}
C_{\spe\spe'}\left[\frac{m_\spe}{T_\spe}\bv f_{M\sigma},f_{M\sigma'}\right]
+C_{\spe\spe'}\left[f_{M\sigma},\frac{m_{\spe'}}{T_{\spe'}}\bv f_{M\sigma'}\right]
= 0.
\end{equation}

 Let us take $f_\spe = f_{M\spe}
+\delta f_{\spe 1} +O(\delta^2)$ for a small
parameter $\delta$ and write the lowest-order contribution to
\eq{eq:entropyProduction},
\begin{eqnarray}
\fl  s = - \delta^2
  \sum_{\spe,\spe'}\int \frac{f_{\spe 1}}{f_{M\spe}}
\Big(C_{\spe\spe'}[f_{M\spe},f_{\spe' 1}]+
C_{\spe\spe'}[f_{\spe 1},f_{M \spe'}]
\Big)\dd^3 v
\nonumber\\[5pt]
\fl\hspace{1cm} + O(\delta^3).
\end{eqnarray}
Since $s=0$ if and only if all the distribution functions are
Maxwellians with the same temperature and flow, then
\begin{eqnarray}\label{eq:linEntProdZero}
  \sum_{\spe,\spe'}\int \frac{f_{\spe 1}}{f_{M\spe}}
\Big(C_{\spe\spe'}[f_{M\spe},f_{\spe' 1}]+
C_{\spe\spe'}[f_{\spe 1},f_{M \spe'}]
\Big)\dd^3 v = 0
\end{eqnarray}
if and only if, for some $a_{\spe,0}(\boldr)$, $\mathbf{a}_1(\boldr)$,
$a_2(\boldr)$, the distribution functions $f_{\spe 1}$ are of the form
\begin{equation}\label{eq:linEntProdZeroResult}
\fl
f_{\spe 1} = \left(
a_{\spe,0}(\boldr)+m_\spe \mathbf{a}_1(\boldr)\cdot\bv+m_\spe a_2(\boldr)\bv^2
\right)
f_{M\spe}(\boldr,\bv), 
\end{equation}
where the temperature and flow of $f_{M \spe}$ are the same for
all species.

It is useful to have the explicit translation of some of the
properties of the collision operator into our non-dimensional
variables. With the definition \eq{eq:collisionoperatornondim} we have
\begin{eqnarray}\label{eq:propcollnondim}
\int \underline{C_{\spe\spe'}}\dd^3 \underline{v} = 0
\nonumber\\[5pt]
\int  \underline{\bv}\, \underline{C_{\spe\spe'}}\,\dd^3
\underline{v} = -\int \underline{\bv} \,
\underline{C_{\spe'\spe}}\,\dd^3 \underline{v}
\nonumber\\[5pt]
\int\frac{1}{2}\,\tau_\spe \, \underline{\bv}^2 \,
\underline{C_{\spe\spe'}} \, \dd^3 \underline{v} =
-\int \frac{1}{2}\, \tau_{\spe'} \, \underline{\bv}^2 \,
\underline{C_{\spe'\spe}} \, \dd^3 \underline{v}.
\end{eqnarray}
Also,
\begin{equation}\label{eq:Maxwelliansannihilatecoll_nondim}
\underline{C_{\spe\spe'}} [\underline{f_{M\spe}},\underline{f_{M\spe'}}]
=
0
\end{equation}
when
\begin{eqnarray}\label{eq:maxwelliancollnondim}
&&\underline{f_{M\spe}}(\underline{\boldr},\underline{\bv})
=\frac{\underline{n_{\spe}}(\underline{\boldr})}
{(2\pi \underline{T_{\spe}}(\underline{\boldr}))^{3/2}}
\exp\left(-\frac{(\underline{\bv}
-\underline{\bV}(\underline{\boldr})/\tau_\spe)^2}
{2\underline{T_{\spe}}
(\underline{\boldr})}\right),\nonumber\\[5pt]
&&\underline{f_{M\spe'}}(\underline{\boldr},\underline{\bv})
=\frac{\underline{n_{\spe'}}(\underline{\boldr})}
{(2\pi \underline{T_{\spe'}}(\underline{\boldr}))^{3/2}}
\exp\left(-\frac{(\underline{\bv}
-\underline{\bV}(\underline{\boldr})/\tau_{\spe'})^2}{2\underline{T_{\spe'}}
(\underline{\boldr})}\right),
\end{eqnarray}
and $\underline{T_{\spe}}(\underline{\boldr})= 
\underline{T_{\spe'}} (\underline{\boldr})$ at every
point. Finally,
\begin{equation}\label{eq:flowcollnondim}
\underline{C_{\spe\spe'}}
\left[\frac{1}{\tau_{\spe}\underline{T_\spe}}\underline{\bv}\
 \underline{f_{M\sigma}},
\underline{f_{M\sigma'}}\right]
+\underline{C_{\spe\spe'}}\left[\underline{f_{M\sigma}},
\frac{1}{\tau_{\spe'}\underline{T_{\spe'}}}\underline{\bv}\
 \underline{f_{M\sigma'}}\right]
=
0.
\end{equation}

\section{Explicit expressions for some pieces of the collision term}
\label{sec:SomePiecesOfCollisionOperator}

Reference \cite{CalvoParra2012} defines
\begin{eqnarray} \label{eq:C1nc}
\fl  C_{\sigma \sigma^\prime}^{(1)\lw}
=
 C_{\sigma \sigma^\prime} \Bigg[
\frac{1}{T_\spe} \bv\cdot \left (\bV_{\sigma}^p
+ \left ( \frac{v^2}{2T_\sigma} - \frac{5}{2} \right )
\bV^T_{\sigma} \right ) \cT_{\spe,0}^{-1 *}F_{\spe 0}
\nonumber\\[5pt]
\fl\hspace{0.5cm} + \cT_{\spe,0}^{-1 *}
G_{\spe 1}
,
\cT_{\spe',0}^{-1 *}F_{\spe' 0} \Bigg]
\nonumber\\[5pt]
\fl\hspace{0.5cm} + \
\frac{\lambda_\spe}{\lambda_{\spe'}}C_{\sigma
\sigma^\prime} \Bigg[ \cT_{\spe,0}^{-1 *}F_{\spe 0} ,
\frac{1}{T_{\spe'}} \bv\cdot \Bigg(\bV_{
\sigma'}^p
\nonumber\\[5pt]
\fl\hspace{0.5cm}
 + \left( \frac{v^2}{2T_{\sigma'}}
 - \frac{5}{2} \Bigg) \bV^T_{ \sigma'} \right ) \cT_{\spe',0}^{-1 *}F_{\spe' 0}
+ \cT_{\spe',0}^{-1*}
G_{\spe' 1}
\Bigg]
\end{eqnarray}
and
\begin{eqnarray}
\fl  C_{\sigma \sigma^\prime}^{(1)\sw} = C_{\sigma \sigma^\prime}
\Bigg[\modTinv F_{\sigma 1}^\sw
\nonumber\\[5pt]
\fl\hspace{0.5cm}
 -
\frac{Z_\spe\lambda_\spe}{T_\spe}
\modTinv\tilde\phi_{\spe 1}^\sw \cT_{\spe,0}^{-1
*} F_{\spe 0}, \cT_{\spe',0}^{-1
*}F_{\spe' 0}
 \Bigg]\nonumber\\[5pt]
\fl\hspace{0.5cm} + \
\frac{\lambda_\spe}{\lambda_{\spe'}}C_{\sigma
\sigma^\prime} \Bigg[ \cT_{\spe,0}^{-1 *}F_{\spe 0} ,
\modTinvprime F_{\sigma' 1}^\sw
\nonumber\\[5pt]
\fl\hspace{0.5cm}
 -
\frac{Z_{\spe'}\lambda_{\spe'}}{T_{\spe'}}
\modTinvprime\tilde\phi_{\spe'
1}^\sw \cT_{\spe',0}^{-1 *}F_{\spe' 0}
 \Bigg].
\end{eqnarray}

Therefore,
\begin{eqnarray}\label{eq:T1C1lw}
\fl\left[\cT_{\spe,1}^* C_{\spe\spe'}^{(1)}\right]^\lw &=
\left(
\rhobf\cdot\nabla_\bR+
\hat{u}_1\partial_u
+
\hat{\mu}_1^\lw\partial_\mu +
\hat{\theta}_1^\lw
 \partial_\theta\right)
\cT_{\spe,0}^{*}C_{\spe\spe'}^{(1)\lw}\nonumber\\[5pt]
\fl
&+
\left[
\cT_{\spe,1}^*C_{\spe\spe'}^{(1)\sw}\right]^\lw.
\end{eqnarray}
The first-order coordinate transformation that enters expression
\eq{eq:T1C1lw} is given in \ref{sec:pullback}; in particular,
$\hat{u}_1,\hat{\mu}_1^\lw, \hat{\theta}_1^\lw$ are defined in
\eq{eq:totalchangecoorfirstorderCorrections} and
\eq{eq:totalchangecoorfirstorderCorrectionsLW}. We use those results
and
\begin{eqnarray}
\fl
B^{-1}\nabla_\bR\cdot(B\rhobf)
+\partial_u\hat{u}_1
+\partial_\mu\hat{\mu}_1
+\partial_\theta\hat{\theta}_1
\nonumber\\[5pt]
\fl
\hspace{1cm}
=
\frac{u}{B}\bun\cdot\nabla_\bR\times\bun
\end{eqnarray}
to write
\begin{eqnarray}\label{eq:T1C1lwGyroaveraged}
\fl
\left\langle
\left[
\cT_{\sigma,1}^\ast C_{\sigma
\sigma^\prime}^{(1)}
\right]^\lw
\right\rangle = \nonumber\\
\fl\hspace{0.5cm}\partial_u 
\Bigg(
\left\langle u \bun \cdot \nabla_\bR \bun \cdot \rhobf\,
\cT_{\spe,0}^*C_{\sigma \sigma^\prime}^{(1)\lw}
\right \rangle
\nonumber\\[5pt]
\fl
\hspace{0.5cm}
-\mu
 \bun \cdot \nabla_\bR \times \bun \left \langle
\cT_{\spe,0}^*C_{\sigma \sigma^\prime}^{(1)\lw}
\right\rangle
\Bigg)
\nonumber\\
\fl\hspace{0.5cm}
+ \partial_\mu
\Bigg\{
\frac{u \mu}{B} \bun \cdot \nabla_\bR \times \bun
\left \langle
\cT_{\spe,0}^*C_{\sigma \sigma^\prime}^{(1)\lw}
\right \rangle
\nonumber\\
\fl\hspace{0.5cm}
-
\Bigg\langle \Bigg(
\frac{Z_\spe}{B} \rhobf \cdot \nabla_\bR
\varphi_0 + \frac{\mu}{B} \rhobf \cdot \nabla_\bR B
\nonumber\\
\fl\hspace{0.5cm}
+
\frac{u^2}{B} \bun \cdot \nabla \bun \cdot \rhobf \Bigg)
\cT_{\spe,0}^*C_{\sigma \sigma^\prime}^{(1)\lw}
\Bigg\rangle
\Bigg\}
 \nonumber\\
\fl\hspace{0.5cm}
- \frac{u}{B}
\bun \cdot \nabla_\bR \times \bun
\left \langle
\cT_{\spe,0}^*C_{\sigma \sigma^\prime}^{(1)\lw}
\right \rangle
 \nonumber\\
\fl\hspace{0.5cm}
+ \frac{1}{B} \nabla_\bR \cdot \left \langle B
\rhobf \,
\cT_{\spe,0}^*C_{\sigma \sigma^\prime}^{(1)\lw}
 \right \rangle  \nonumber\\
\fl\hspace{0.5cm} +
\left\langle\left[
\cT_{\spe,1}^\ast C_{\sigma
\sigma^\prime}^{(1)\sw}
\right]^\lw\right\rangle.
\end{eqnarray}

The turbulent piece $[ \cT_{\spe,1}^*C_{\spe\spe'}^\sw]^\lw$ appearing
in \eq{eq:T1C1lw} was computed in Appendix E of
\cite{CalvoParra2012}. We collect here the result,
\begin{eqnarray}\label{eq:turbpiececoll2}
\fl\left[ \cT_{\spe,1}^*C_{\spe\spe'}^\sw
\right]^\lw=
\Bigg[
\frac{Z_\spe\lambda_\spe}{B}
\Bigg(
-\tilde\phi_{\spe 1}^\sw\partial_\mu
+\partial_\mu\tilde\Phi_{\spe 1}^\sw\partial_\theta
\Bigg)
\nonumber\\[5pt]
\fl\hspace{0.5cm}
\Bigg\{
\cT_{NP, \spe}^* C_{\sigma \sigma^\prime}
\left [\modTinv F_{\sigma 1}^\sw -
\frac{Z_\spe\lambda_\spe}{T_\spe}
\modTinv\tilde\phi_{\spe 1}^\sw \cT_{\spe,0}^{-1
*} F_{\spe 0}, \cT_{\spe',0}^{-1
*}F_{\spe' 0}
 \right ]\nonumber\\[5pt]
\fl\hspace{0.5cm} + \
\frac{\lambda_\spe}{\lambda_{\spe'}}
\cT_{NP, \spe}^*
C_{\sigma
\sigma^\prime} \Bigg[ \cT_{\spe,0}^{-1 *}F_{\spe 0} ,
\modTinvprime F_{\sigma' 1}^\sw
\nonumber\\[5pt]
\fl\hspace{0.5cm}
 -
\frac{Z_{\spe'}\lambda_{\spe'}}{T_{\spe'}}
\modTinvprime\tilde\phi_{\spe'
1}^\sw \cT_{\spe',0}^{-1 *}F_{\spe' 0}
 \Bigg]
\Bigg\}
\Bigg]^\lw.
\end{eqnarray}
As for its gyroaverage,
\begin{eqnarray}\label{eq:turbpiececollgyroave}
\fl
\Big\langle
\Big[ &\cT_{\spe,1}^*C_{\spe\spe'}^\sw
\Big]^\lw
\Big\rangle
=-
\partial_\mu
\Bigg\langle\Bigg[
\frac{Z_\spe\lambda_\spe}{B}
\tilde\phi_{\spe 1}^\sw
\nonumber\\[5pt]
\fl&
\Bigg\{
\cT_{NP, \spe}^* C_{\sigma \sigma^\prime}
\left [\modTinv F_{\sigma 1}^\sw -
\frac{Z_\spe\lambda_\spe}{T_\spe}
\modTinv\tilde\phi_{\spe 1}^\sw \cT_{\spe,0}^{-1
*} F_{\spe 0}, \cT_{\spe',0}^{-1
*}F_{\spe' 0}
 \right ]
\nonumber\\[5pt]
\fl& + \
\frac{\lambda_\spe}{\lambda_{\spe'}}
\cT_{NP, \spe}^*
C_{\sigma
\sigma^\prime} \Bigg[ \cT_{\spe,0}^{-1 *}F_{\spe 0} ,
\modTinvprime F_{\sigma' 1}^\sw
\nonumber\\[5pt]
\fl&
 -
\frac{Z_{\spe'}\lambda_{\spe'}}{T_{\spe'}}
\modTinvprime\tilde\phi_{\spe'
1}^\sw \cT_{\spe',0}^{-1 *}F_{\spe' 0}
 \Bigg]
\Bigg\}
\Bigg]^\lw
\Bigg\rangle.
\end{eqnarray}
In order to get the last expression we have integrated by parts in
$\theta$ and $\mu$.

The full computation of the last term in equation
\eq{eq:FPsecondorder2} is tedious and was performed in Appendix F of
\cite{CalvoParra2012}. The necessary result is
\begin{eqnarray}\label{eq:C2lwgyro}
\fl\Big\langle \cT_{\spe, 0}^{*}C_{\spe \spe'}^{(2)\lw}\Big\rangle =
\cT_{\spe, 0}^{*} C_{\sigma \sigma^\prime}
 \Big[\cT_{\spe, 0}^{-1*}\left\langle F_{\sigma 2}^\lw \right\rangle +
\cT_{\spe, 0}^{-1*}\left\langle 
\cT_{\spe, 0}^{*} [\cT_{\sigma,1}^{-1 *} F_{\sigma 1}^\lw]^\lw\right\rangle 
\nonumber\\\hspace{0.5cm}
\fl
 +
\cT_{\spe, 0}^{-1*}\left\langle \cT_{\spe, 0}^{*}
[\cT_{\sigma,2}^{-1 *} F_{\spe 0}]^\lw\right\rangle ,
\cT^{-1*}_{\spe',0}F_{\spe' 0} \Big] 
\nonumber\\
\fl
\hspace{0.5cm}
+\left(\frac{\lambda_\spe}{\lambda_{\spe'}}\right)^2
\cT_{\spe, 0}^{*}C_{\sigma \sigma^\prime}
\Big[
\cT^{-1*}_{\spe,0}F_{\spe 0},
\nonumber\\\hspace{0.5cm}
\fl
\cT_{\spe^\prime, 0}^{-1*}
 \left\langle F_{\sigma' 2}^\lw\right\rangle
+
\cT_{\spe^\prime, 0}^{-1*}
\left\langle
\cT_{\spe^\prime, 0}^{*}[\cT_{\sigma',1}^{-1 *} F_{\sigma' 1}^\lw]^\lw\right\rangle
\nonumber\\
\fl
\hspace{0.5cm}
+
\cT_{\spe^\prime, 0}^{-1*}\left\langle\cT_{\spe^\prime, 0}^{*}
[\cT_{\sigma',2}^{-1 *} F_{\spe' 0}]^\lw\right\rangle
\Big] \nonumber\\\hspace{0.5cm}
\fl +
\frac{\lambda_\spe}{\lambda_{\spe'}}
\Bigg\langle
\cT_{\spe, 0}^{*}C_{\sigma\sigma^\prime}
\Bigg[\cT^{-1*}_{\spe,0} F_{\sigma 1}^\lw
+ [\cT_{\sigma,1}^{-1 *} F_{\sigma 0}]^\lw,
\nonumber\\
\fl
\hspace{0.5cm}
\cT^{-1*}_{\spe',0}F_{\sigma^\prime 1}^\lw +
[\cT_{\sigma',1}^{-1 *} F_{\sigma' 0}]^\lw
\Bigg]
\Bigg\rangle
\nonumber\\\hspace{0.5cm}
\fl +
\frac{\lambda_\spe}{\lambda_{\spe'}}
\Bigg
\langle
\cT_{\spe, 0}^{*} \Bigg[C_{\sigma\sigma^\prime} \Bigg[ 
\modTinv F_{\spe 1}^\sw
\nonumber\\
\fl
\hspace{0.5cm}
-
\frac{Z_\spe\lambda_\spe
}{T_\spe}
\modTinv \phiwig_{\spe 1}^\sw\cT_{\spe,0}^{-1 *}F_{\spe 0}
, 
\modTinvprime F_{\spe' 1}^\sw \nonumber\\\hspace{0.5cm}
\fl
-
\frac{Z_{\spe'}\lambda_{\spe'} 
}{T_{\spe'}}
\modTinvprime\phiwig_{\spe' 1}^\sw\cT_{\spe',0}^{-1 *}F_{\spe' 0}
\Bigg] \Bigg]^\lw
\Bigg
\rangle.
\end{eqnarray}
Here,
\begin{eqnarray}\label{eq:T1F1lwAveraged}
\fl \left\langle
\cT_{\spe, 0}^{*}[\cT_{\sigma,1}^{-1 *} F_{\spe 1}^\lw]^\lw
\right\rangle = 
\mu\bun\cdot\nabla_\bR\times\bun
\left(
\partial_u
-\frac{u}{B} 
\partial_\mu
\right)
F_{\spe 1}^\lw,
\end{eqnarray}
\begin{eqnarray}\label{gyroaverageT2F0}
\fl\left\langle\cT_{\spe, 0}^{*} [\cT_{\sigma,2}^{-1 *} F_{\spe 0}]^\lw\right\rangle
 =\nonumber\\
\fl\hspace{0.5cm} \frac{\mu}{2B} (\matI - \bun \bun) : \Bigg [
\nabla_\bR \nabla_\bR \ln n_\spe
\nonumber\\
\fl
\hspace{0.5cm}
+ \left ( \frac{u^2/2 + \mu B}{T_\spe} -
\frac{3}{2} \right ) \nabla_\bR \nabla_\bR \ln T_\spe \Bigg ] F_{\spe 0}
\nonumber\\
\fl\hspace{0.5cm} 
- \frac{\mu}{B} \frac{Z_\spe}{T_\spe^2}
\nabla_\bR \varphi_0 \cdot \nabla_\bR T_\spe F_{\spe 0}
\nonumber\\
\fl
\hspace{0.5cm}
 -
\frac{\mu}{2B} \frac{u^2/2 + \mu B}{T_\spe^3} |\nabla_\bR T_\spe|^2
F_{\spe 0} \nonumber\\
\fl\hspace{0.5cm} 
+ \frac{\mu}{2B} \Bigg |
\frac{\nabla_\bR n_\spe}{n_\spe} + \frac{Z_\spe \nabla_\bR
\varphi_0}{T_\spe}
\nonumber\\
\fl
\hspace{0.5cm}
+ \left ( \frac{u^2/2 + \mu B}{T_\spe} - \frac{3}{2}
\right ) \frac{\nabla_\bR T_\spe}{T_\spe} \Bigg |^2 F_{\spe 0}
\nonumber\\
\fl\hspace{0.5cm} 
- \frac{\mu}{2B^2} \nabla_{\bR_\bot} B \cdot \Bigg(
\frac{\nabla_\bR n_\spe}{n_\spe} + \frac{Z_\spe \nabla_\bR
\varphi_0}{T_\spe}
\nonumber\\
\fl
\hspace{0.5cm}
+ \left ( \frac{u^2/2 + \mu B}{T_\spe} - \frac{3}{2}
\right ) \frac{\nabla_\bR T_\spe}{T_\spe} \Bigg) F_{\spe 0}
\nonumber\\
\fl\hspace{0.5cm} 
+ \frac{Z_\spe^2\lambda_\spe^2}{2T_\spe^2}
\left[ \left \langle
(\phiwig_{\spe 1}^\sw)^2 \right \rangle \right]^\lw F_{\spe
0} + \frac{1}{T_\spe} \Bigg [ - \frac{Z_\spe^2}{2B^2}
|\nabla_\bR \varphi_0|^2 \nonumber\\
\fl\hspace{0.5cm} 
- \frac{Z_\spe^2\lambda_\spe^2}{2B}
\partial_\mu \left[ \left
\langle (\phiwig_{\spe 1}^\sw)^2 \right \rangle \right]^\lw
- \frac{3 Z_\spe \mu}{2 B^2} \nabla_{\bR_\bot} B \cdot
\nabla_\bR \varphi_0
\nonumber\\
\fl\hspace{0.5cm} 
 - \frac{Z_\spe u^2}{B^2} \bun
\cdot \nabla_\bR \bun \cdot \nabla_\bR \varphi_0  + \Psi_{B, \spe}
\nonumber\\
\fl\hspace{0.5cm} 
+ 
\frac{Z_\spe
\mu}{B} (\matI - \bun \bun): \nabla_\bR \nabla_\bR \varphi_0 \Bigg
] F_{\spe 0},
\end{eqnarray}
and $\cT_{\spe,1}^{-1*}$ is obtained from the results of
\ref{sec:pullback}.

Finally, the last term in equation
\eq{eq:Rsigma2} is given by
\begin{eqnarray}
\fl\left\langle\cT_{\spe, 0}^{*}{\mathbb C}_{\spe\spe'}\right\rangle
=
\left\langle\cT_{\spe, 0}^{*}C_{\spe \spe'}^{(2)\lw}\right\rangle -
\cT_{\spe, 0}^{*} C_{\sigma \sigma^\prime}
 \left[\cT_{\spe, 0}^{-1*} G_{\sigma 2}^\lw
 ,
\cT^{-1*}_{\spe',0}F_{\spe' 0} \right]
\nonumber\\\hspace{0.5cm}
\fl
-
 \left(\frac{\lambda_\spe}{\lambda_\spe'}\right)^2
 C_{\sigma \sigma^\prime}
 \left[\cT^{-1*}_{\spe,0}F_{\spe 0},
\cT_{\spe', 0}^{-1*} G_{\sigma' 2}^\lw
\right].
\end{eqnarray}
Explicitly,
\begin{eqnarray}\label{eq:mathbbC}
\fl
\left\langle\cT_{\spe, 0}^{*}{\mathbb C}_{\spe\spe'}\right\rangle
=
\cT_{\spe, 0}^{*} C_{\sigma \sigma^\prime}
 \Bigg[
\cT_{\spe, 0}^{-1*}\left\langle 
\cT_{\spe, 0}^{*} [\cT_{\sigma,1}^{-1 *} F_{\sigma 1}^\lw]^\lw\right\rangle 
\nonumber\\[5pt]\hspace{0.5cm}
\fl
 +
\cT_{\spe, 0}^{-1*}\left\langle \cT_{\spe, 0}^{*}
[\cT_{\sigma,2}^{-1 *} F_{\spe 0}]^\lw\right\rangle 
\nonumber\\[5pt]\hspace{0.5cm}
\fl
+\cT_{\spe, 0}^{-1*}\Bigg\{- \frac{1}{T_\spe}
\Bigg(
 \frac{Z_\spe\mu}{2 B}
(\matI - \bun \bun) : \nabla_\bR \nabla_\bR \varphi_0
\nonumber\\[5pt]
\fl\hspace{0.5cm}
+ \Psi_{B,\spe} + Z_\spe\lambda_\spe \Psi_{\phi
B,\spe}^\lw
 + Z_\spe^2 \lambda_\spe^2
 \Psi_{\phi,\spe}^\lw
\Bigg)
F_{\spe 0}
\nonumber\\[5pt]
\fl\hspace{0.5cm}
+
\frac{Z_\spe\lambda_\spe\varphi_1^\lw}{u}
\partial_u G_{\spe 1}^\lw\Bigg\}
,
\cT^{-1*}_{\spe',0}F_{\spe' 0} \Bigg] 
\nonumber\\
\fl
\hspace{0.5cm}
+\left(\frac{\lambda_\spe}{\lambda_{\spe'}}\right)^2
\cT_{\spe, 0}^{*}C_{\sigma \sigma^\prime}
\Bigg[
\cT^{-1*}_{\spe,0}F_{\spe 0},
\nonumber\\[5pt]
\hspace{0.5cm}
\fl
\cT_{\spe^\prime, 0}^{-1*}
\left\langle
\cT_{\spe^\prime, 0}^{*}[\cT_{\sigma',1}^{-1 *} F_{\sigma' 1}^\lw]^\lw\right\rangle +
\cT_{\spe^\prime, 0}^{-1*}\left\langle\cT_{\spe^\prime, 0}^{*}
[\cT_{\sigma',2}^{-1 *} F_{\spe' 0}]^\lw\right\rangle
\nonumber\\[5pt]
\fl\hspace{0.5cm}
+\cT_{\spe', 0}^{-1*}\Bigg\{
- \frac{1}{T_{\spe'}}
\Bigg(
 \frac{Z_{\spe'}\mu}{2 B}
(\matI - \bun \bun) : \nabla_\bR \nabla_\bR \varphi_0
\nonumber\\[5pt]
\fl\hspace{0.5cm}
+ \Psi_{B,\spe'} + Z_{\spe'}\lambda_{\spe'} \Psi_{\phi
B,\spe'}^\lw
 + Z_{\spe'}^2 \lambda_{\spe'}^2
 \Psi_{\phi,\spe'}^\lw
\Bigg)
F_{\spe' 0}
\nonumber\\[5pt]
\fl\hspace{0.5cm}
+
\frac{Z_\spe'\lambda_\spe'\varphi_1^\lw}{u}
\partial_u G_{\spe' 1}^\lw
\Bigg\}
\Bigg] \nonumber\\[5pt]
\hspace{0.5cm}
\fl +
\frac{\lambda_\spe}{\lambda_{\spe'}}
\Big\langle
\cT_{\spe, 0}^{*}C_{\sigma\sigma^\prime}
[\cT^{-1*}_{\spe,0} F_{\sigma 1}^\lw
+ [\cT_{\sigma,1}^{-1 *} F_{\sigma 0}]^\lw,
 \nonumber\\[5pt]
\hspace{0.5cm}
\fl
\cT^{-1*}_{\spe',0}F_{\sigma^\prime 1}^\lw +
[\cT_{\sigma',1}^{-1 *} F_{\sigma' 0}]^\lw
]
\Big\rangle
\nonumber\\\hspace{0.5cm}
\fl +
\frac{\lambda_\spe}{\lambda_{\spe'}}
\Bigg
\langle
\cT_{\spe, 0}^{*} \Bigg[C_{\sigma\sigma^\prime} \Big[ 
\modTinv F_{\spe 1}^\sw -
\frac{Z_\spe\lambda_\spe
}{T_\spe}
\modTinv \phiwig_{\spe 1}^\sw\cT_{\spe,0}^{-1 *}F_{\spe 0}
, 
 \nonumber\\[5pt]
\hspace{0.5cm}
\fl
\modTinvprime F_{\spe' 1}^\sw
-
\frac{Z_{\spe'}\lambda_{\spe'} 
}{T_{\spe'}}
\modTinvprime\phiwig_{\spe' 1}^\sw\cT_{\spe',0}^{-1 *}F_{\spe' 0}
\Big] \Bigg]^\lw
\Bigg
\rangle.
\end{eqnarray}

\section{Gyrokinetic transformation to first order}
\label{sec:pullback}

The explicit expressions for the gyrokinetic transformation
$(\boldr,\bv)=\cT_\spe(\bR,u,\mu,\theta,t)$ to order $\epsilon_\spe$
were derived in detail in Appendix C of \cite{CalvoParra2012}. Here,
we only quote the results. Namely,
\begin{eqnarray}\label{eq:totalchangecoorfirstorder}
\fl \boldr &= \bR + \epsilon_\spe\rhobf
+ O(\epsilon_\spe^2)
,
\nonumber\\[5pt]
\fl v_{||} &= u
+ \epsilon_\spe\hat{u}_1
+ O(\epsilon_\spe^2),\nonumber\\[5pt]
\fl \mu_0 &= \mu + \epsilon_\spe\hat{\mu}_1
+ O(\epsilon_\spe^2),\nonumber\\[5pt]
\fl \theta_0 &= \theta
+\epsilon_\spe\hat{\theta}_1
+ O(\epsilon_\spe^2),
\end{eqnarray}
where
\begin{eqnarray}\label{eq:totalchangecoorfirstorderCorrections}
\fl \hat{u}_1 &= 
u\bun\cdot\nabla_\bR\bun\cdot\rhobf+
\frac{B}{4}[\rhobf(\rhobf\times\bun)
+(\rhobf\times\bun)\rhobf]
:\nabla_{\bR}\bun
\nonumber\\[5pt]
\fl&-\mu\bun\cdot\nabla_\bR\times\bun,
\nonumber\\[5pt]
\fl\hat{\mu}_1 &=
-\frac{\mu}{B}\rhobf\cdot\nabla_{\bR}B
-\frac{u}{4}
\left(
\rhobf(\rhobf\times\bun)
+(\rhobf\times\bun)\rhobf
\right):\nabla_{\bR}\bun
\nonumber\\[5pt]
\fl&+\frac{u\mu}{B}\bun\cdot\nabla_\bR\times\bun
-\frac{u^2}{B}\bun\cdot\nabla_\bR\bun\cdot\rhobf
-\frac{Z_\spe\lambda_\spe}{B}\tilde\phi_{\spe 1},
\nonumber\\[5pt]
\fl\hat{\theta}_1&= 
(\rhobf\times\bun)
\cdot
\Bigg(
\nabla_\bR\ln B + \frac{u^2}{2\mu B}\bun\cdot\nabla_\bR\bun
\nonumber\\[5pt]
\fl&-\bun\times\nabla_\bR\eun_2\cdot\eun_1
\Bigg)
-\frac{u}{8\mu}
\left(
\rhobf\rhobf - (\rhobf\times\bun)(\rhobf\times\bun)
\right):\nabla_\bR\bun
\nonumber\\[5pt]
\fl&
+\frac{u}{2B^2}\bun\cdot\nabla_\bR B
+\frac{Z_\spe\lambda_\spe}{B}\partial_\mu\tilde\Phi_{\spe 1}
.
\end{eqnarray}
In the main text of this paper we need the long-wavelength limit of the previous
expressions. Using
\eq{eq:potentiallw2} and \eq{eq:potentiallw3} we get
\begin{eqnarray}\label{eq:totalchangecoorfirstorderCorrectionsLW}
\fl \hat{u}_1^\lw &= \hat{u}_1
\nonumber\\[5pt]
\fl\hat{\mu}_1^\lw &=
-\frac{\mu}{B}\rhobf\cdot\nabla_{\bR}B
-\frac{u}{4}
\left(
\rhobf(\rhobf\times\bun)
+(\rhobf\times\bun)\rhobf
\right):\nabla_{\bR}\bun
\nonumber\\[5pt]
\fl&+\frac{u\mu}{B}\bun\cdot\nabla_\bR\times\bun
-\frac{u^2}{B}\bun\cdot\nabla_\bR\bun\cdot\rhobf
-\frac{Z_\spe}{B}\rhobf\cdot\nabla_\bR\varphi_0,
\nonumber\\[5pt]
\fl\hat{\theta}_1^\lw&= 
(\rhobf\times\bun)
\cdot
\Bigg(
\nabla_\bR\ln B + \frac{u^2}{2\mu B}\bun\cdot\nabla_\bR\bun
\nonumber\\[5pt]
\fl&-\bun\times\nabla_\bR\eun_2\cdot\eun_1
\Bigg)
-\frac{u}{8\mu}
\left(
\rhobf\rhobf - (\rhobf\times\bun)(\rhobf\times\bun)
\right):\nabla_\bR\bun
\nonumber\\[5pt]
\fl&
+\frac{u}{2B^2}\bun\cdot\nabla_\bR B
+\frac{Z_\spe}{2\mu B}\left(\rhobf\times\bun\right)\cdot\nabla_\bR\varphi_0
.
\end{eqnarray}

Next, we proceed to calculate the long-wavelength limit of
$\cT_\spe^{-1*}F_{\spe 0}$ to first order in $\epsilon_\spe$, employed
in Section \ref{sec:FokkerPlancklong1}. Inverting
\eq{eq:totalchangecoorfirstorder} to first order, and recalling
\eq{eq:totalchangecoorfirstorderCorrectionsLW} and the relations
$\partial_u F_{\spe0 } = -(u/T_\spe) F_{\spe0 }$, $\partial_\mu
F_{\spe0 } = -(B/T_\spe) F_{\spe0 }$, one finds that
\begin{eqnarray}\label{eq:pullback_order1_Maxwellian}
\fl \left[\cT^{-1*}_{\spe,1} F_{\spe 0}\right]^\lw &=&
\frac{1}{ T_{\spe}}
\Bigg[
\bv\cdot{\mathbf V}^p_\spe +
\left(
\frac{v^2}{2T_{\spe}}-\frac{5}{2}
\right)\bv\cdot{\mathbf V}^T_\spe
\nonumber\\[5pt]
\fl&+&
\frac{Z_\spe}{B}
\bv\cdot(\bun\times\nabla_\boldr\varphi_0)
\Bigg]\cT^{-1*}_{\spe,0}F_{\spe 0},
\end{eqnarray}
with ${\mathbf V}^p_\spe$ and ${\mathbf V}^T_\spe$ defined in
\eq{eq:velocitiesPandT}.

\section{A convenient way to write the second-order Fokker-Planck equation}
\label{sec:FPequationG2}

In this appendix we rewrite equation \eq{eq:FPsecondorder2} in a way
that is very useful to compute the transport equations for density and
energy. Defining $G_{\spe 2}^\lw$ by
\begin{eqnarray}
\fl\left\langle F^\lw_{\sigma 2}\right\rangle
=
G_{\spe 2}^\lw - \frac{1}{T_\spe}
\Bigg[
Z_\spe\lambda_\spe^2
\Bigg(
\varphi_2^\lw
\nonumber\\[5pt]
\fl\hspace{0.5cm}
+ \frac{\mu}{2\lambda_\spe^2 B}
(\matI - \bun \bun) : \nabla_\bR \nabla_\bR \varphi_0
\Bigg)
\nonumber\\[5pt]
\fl\hspace{0.5cm}
+ \Psi_{B,\spe} + Z_\spe\lambda_\spe \Psi_{\phi
B,\spe}^\lw
 + Z_\spe^2 \lambda_\spe^2
 \Psi_{\phi,\spe}^\lw
\Bigg]
F_{\spe 0}
\nonumber\\[5pt]
\fl\hspace{0.5cm}
+
\frac{1}{2}
\left(\frac{Z_\spe\lambda_\spe\varphi_1^\lw}{T_\spe}\right)^2
F_{\spe 0}
+
\frac{Z_\spe\lambda_\spe\varphi_1^\lw}{u}
\partial_u G_{\spe 1}^\lw
\end{eqnarray}
and using \eq{eq:defGspe1}, equation \eq{eq:FPsecondorder2} becomes
\begin{eqnarray} \label{eq:FPequationG2}
\fl
\left(
u \bun \cdot \nabla_\bR   - \mu \bun \cdot
\nabla_\bR B \, \partial_u
\right)
G^\lw_{\sigma 2}
  +
\frac{\lambda_\spe^2}{\tau_\spe}\partial_{\epsilon_s^2 t} F_{\spe 0}
\nonumber\\[5pt]
\fl\hspace{0.5cm} + 
\left(
\bv_M + \bv_{E,\spe}^{(0)}
\right)
 \cdot \nabla_\bR G^\lw_{\sigma 1} \nonumber\\[5pt]
\fl\hspace{0.5cm}
+
u\, \kappabf\cdot\left(\bv_{\nabla B} + \bv_{E,\spe}^{(0)}\right)
\partial_u G^\lw_{\sigma 1}
\nonumber\\[5pt]
\fl\hspace{0.5cm} + \Bigg[
-\frac{u}{B}
(\bun \cdot \nabla_\bR \times \bun)
\left(\bv_M + \bv_{E,\spe}^{(0)}\right)
 \nonumber\\[5pt]
\fl\hspace{0.5cm} 
 - \frac{u \mu}{B}
 (\nabla_\bR \times \bK)_\bot
\Bigg]\cdot\nabla_\bR F_{\spe 0}
\nonumber\\[5pt]
\fl\hspace{0.5cm} - 
\Bigg[\frac{u^2}{B}\left(\bun\cdot\nabla_\bR\times\bun\right)
\kappabf
\nonumber\\[5pt]
\fl\hspace{0.5cm}
+ \mu
\left(\nabla_\bR\times\bK\right)\times\bun
\Bigg]
\cdot
\left(\bv_{\nabla B}+\bv_{E,\spe}^{(0)}\right)
\partial_u F_{\spe 0} 
 \nonumber\\[5pt]
\fl\hspace{0.5cm}
 + 
\frac{Z_\spe \lambda_\spe}{B} \left[
\nabla_\bR \cdot \left(\bun\times\nabla_{\bR_\perp/\epsilon_\spe}
 \langle \phi_{\spe 1}^\sw \rangle
F_{\spe 1}^\sw\right)\right]^\lw
\nonumber\\[5pt]
\fl\hspace{0.5cm} - Z_\spe \lambda_\spe \partial_u \Bigg[
\Bigg( \bun\cdot\nabla_\bR 
\langle \phi_{\spe 1}^\sw \rangle
\nonumber\\[5pt]
\fl\hspace{0.5cm}
+  \frac{u}{B}
\left(\bun \times \kappabf\right)\cdot
\nabla_{\bR_\perp/\epsilon_\spe}\langle \phi_{\spe 1}^\sw \rangle
\Bigg)F_{\spe 1}^\sw \Bigg]^\lw
\nonumber\\[5pt]
\fl\hspace{0.5cm}
-\frac{1}{B}
\nabla_\bR\cdot
\Bigg[
Z_\spe\lambda_\spe
\varphi_1^\lw\bun\times\nabla_\bR\psi\left(\Upsilon_\spe+\frac{Z_\spe}{T_\spe}
\partial_\psi \varphi_0\right)F_{\spe 0}
\Bigg]
\nonumber\\[5pt]
\fl\hspace{0.5cm}
+
\frac{1}{B}
\partial_u
\Bigg[
\frac{Z_\spe\lambda_\spe
  \varphi_1^\lw}{u}
\mu(\bun\times\nabla_\bR\psi)\cdot\nabla_\bR B
\Bigg(\Upsilon_\spe
\nonumber\\[5pt]
\fl\hspace{0.5cm}
+\frac{Z_\spe}{T_\spe}
\partial_\psi \varphi_0\Bigg)F_{\spe 0}
\Bigg]
=
\partial_u
\left[
-\frac{Z_\spe\lambda_\spe\varphi_1^\lw}{u}
\sum_{\spe'}
\left\langle\cT_{\spe,0}^* C_{\spe\spe'}^{(1)\lw}\right\rangle
\right]
\nonumber\\[5pt]
\fl\hspace{0.5cm}
+
\sum_{\spe'}
\left\langle
\left[\cT_{\spe,1}^*C_{\spe\spe'}^{(1)}\right]^\lw
\right\rangle
 +\sum_{\spe'}
\left\langle
\cT_{\spe,0}^*C_{\spe\spe'}^{(2)\lw}
\right\rangle,
\end{eqnarray}
where we have used the relations
\begin{eqnarray}
\fl
Z_\spe\lambda_\spe\bun\cdot\nabla_\bR\varphi_1^\lw
\partial_u\left(
\frac{Z_\spe\lambda_\spe\varphi_1^\lw}{T_\spe}
F_{\spe 0}
\right)
=
\nonumber\\[5pt]
\fl\hspace{0.5cm}
\left(
u \bun \cdot \nabla_\bR   - \mu \bun \cdot
\nabla_\bR B \, \partial_u
\right)
\left[-
\frac{1}{2}
\left(\frac{Z_\spe\lambda_\spe\varphi_1^\lw}{T_\spe}\right)^2
F_{\spe 0}
\right],
\end{eqnarray}
\begin{eqnarray}
\fl
-Z_\spe\lambda_\spe\bun\cdot\nabla_\bR\varphi_1^\lw
\partial_u G_{\spe 1}^\lw
=
\nonumber\\[5pt]
\fl\hspace{0.5cm}
\left(
u \bun \cdot \nabla_\bR   - \mu \bun \cdot
\nabla_\bR B \, \partial_u
\right)
\left[-
\frac{Z_\spe\lambda_\spe\varphi_1^\lw}{u}
\partial_u G_{\spe 1}^\lw
\right]
\nonumber\\[5pt]
\fl\hspace{0.5cm}
+
\partial_u
\left[
\frac{Z_\spe\lambda_\spe\varphi_1^\lw}{u}
\left(
u \bun \cdot \nabla_\bR   - \mu \bun \cdot
\nabla_\bR B \, \partial_u
\right)
 G_{\spe 1}^\lw
\right],
\end{eqnarray}
\begin{eqnarray}\label{eq:relationRewriteFPorder2}
\fl
(\bv_M+\bv_{E,\spe}^{(0)})
\cdot
\nabla_\bR
\left(
-
\frac{Z_\spe\lambda_\spe\varphi_1^\lw}{T_\spe}
F_{\spe 0}
\right)
\nonumber\\[5pt]
\fl\hspace{0.5cm}
-\frac{u}{B}(\bun\times\kappabf)
\cdot
(\mu\nabla_\bR B + Z_\spe\nabla_\bR\varphi_0)
\partial_u\left(
-
\frac{Z_\spe\lambda_\spe\varphi_1^\lw}{T_\spe}
F_{\spe 0}
\right)
\nonumber\\[5pt]
\fl\hspace{0.5cm}
+\bv_{E,\spe}^{(1)}\cdot\nabla_\bR F_{\spe 0}
-Z_\spe\lambda_\spe
\frac{u}{B}(\bun\times\kappabf)
\cdot
\nabla_\bR \varphi_1^\lw\partial_u F_{\spe 0}
=
\nonumber\\[5pt]
\fl\hspace{0.5cm}
-\frac{1}{B}
\nabla_\bR\cdot
\Bigg[
Z_\spe\lambda_\spe
\varphi_1^\lw\bun\times\nabla_\bR\psi\left(\Upsilon_\spe+\frac{Z_\spe}{T_\spe}
\partial_\psi \varphi_0\right)F_{\spe 0}
\Bigg]
\nonumber\\[5pt]
\fl\hspace{0.5cm}
+
\frac{1}{B}
\partial_u
\Bigg[
\frac{Z_\spe\lambda_\spe
  \varphi_1^\lw}{u}
\mu(\bun\times\nabla_\bR\psi)\cdot\nabla_\bR B
\left(\Upsilon_\spe+\frac{Z_\spe}{T_\spe}
\partial_\psi \varphi_0\right)F_{\spe 0}
\Bigg]
\nonumber\\[5pt]
\fl\hspace{0.5cm}
+\partial_u
\left[
\frac{Z_\spe\lambda_\spe\varphi_1^\lw}{u}
\bv_M\cdot\nabla_\bR\psi
\left(\Upsilon_\spe+\frac{Z_\spe}{T_\spe}
\partial_\psi \varphi_0\right)F_{\spe 0}
\right],
\end{eqnarray}
and
\begin{eqnarray}\label{eq:relationRewriteFPorder2anotherone}
\fl
\partial_u
\Bigg[
\frac{Z_\spe\lambda_\spe\varphi_1^\lw}{u}
\Bigg\{\left(
u \bun \cdot \nabla_\bR   - \mu \bun \cdot
\nabla_\bR B \, \partial_u
\right)
 G_{\spe 1}^\lw
\nonumber\\[5pt]
\fl\hspace{0.5cm}
+
\bv_M\cdot\nabla_\bR\psi
\left(\Upsilon_\spe+\frac{Z_\spe}{T_\spe}
\partial_\psi \varphi_0\right)F_{\spe 0}\Bigg\}
\Bigg]
\nonumber\\[5pt]
\fl\hspace{0.5cm}
=
\partial_u
\Bigg[
\frac{Z_\spe\lambda_\spe\varphi_1^\lw}{u}
\sum_{\spe'}
\left\langle\cT_{\spe, 0}^* C^{(1)\lw}_{\spe\spe'}\right\rangle
\Bigg].
\end{eqnarray}
To obtain \eq{eq:relationRewriteFPorder2anotherone} we have employed
equation \eq{eq:Vlasovorder1gyroav4}.

Using \eq{eq:turbpiececoll2}, the relation
\begin{eqnarray}
\fl
(\bv_M + \bv_{E,\spe}^{(0)})\cdot
\nabla_\bR G_{\spe 1}^\lw
+
u\kappabf\cdot
(\bv_{\nabla B} + \bv_{E,\spe}^{(0)})\partial_u G_{\spe 1}^\lw
=
\nonumber\\[5pt]
\fl\hspace{0.5cm}
\frac{1}{B}\nabla_\bR
\cdot
\Bigg[
(\mu\bun\times\nabla_\bR B
+
u^2\nabla_\bR\times\bun
+
Z_\spe
\bun\times\nabla_\bR\varphi_0
) G_{\spe 1}^\lw
\Bigg]
\nonumber\\[5pt]
\fl\hspace{0.5cm}
-\frac{1}{B}\partial_u
\Bigg[
u (\nabla_\bR\times\bun)\cdot
(\mu \nabla_\bR B + Z_\spe
\nabla_\bR\varphi_0)
 G_{\spe 1}^\lw
\Bigg]
\nonumber\\[5pt]
\fl\hspace{0.5cm}
-\frac{u}{B}
\bun\cdot\nabla_\bR\times\bun\,
(u\bun\cdot\nabla_\bR-\mu\bun\cdot\nabla_\bR B \partial_u)G_{\spe 1}^\lw
\end{eqnarray}
and again \eq{eq:Vlasovorder1gyroav4} to write
\begin{eqnarray}
\fl
-\frac{u}{B}(\bun\cdot\nabla_\bR\times\bun)
\Bigg[
\left(
u \bun \cdot \nabla_\bR   - \mu \bun \cdot
\nabla_\bR B \, \partial_u
\right)
 G_{\spe 1}^\lw
\nonumber\\[5pt]
\fl\hspace{0.5cm}
+
(\bv_M+\bv_{E,\spe}^{(0)})\cdot\nabla_\bR F_{\spe 0}
+
u\kappabf\cdot(\bv_{\nabla B}+\bv_{E,\spe}^{(0)})\partial_u F_{\spe 0}
\Bigg]=
\nonumber\\[5pt]
\fl\hspace{0.5cm}
-\frac{u}{B}(\bun\cdot\nabla_\bR\times\bun)
\sum_{\spe'}
\left\langle\cT_{\spe, 0}^* C^{(1)\lw}_{\spe\spe'}\right\rangle
,
\end{eqnarray}
we find
\begin{eqnarray} \label{eq:FPequationG2final}
\fl
\left(
u \bun \cdot \nabla_\bR   - \mu \bun \cdot
\nabla_\bR B \, \partial_u
\right)
G^\lw_{\sigma 2}
  +
\frac{\lambda_\spe^2}{\tau_\spe}\partial_{\epsilon_s^2 t} F_{\spe 0}
\nonumber\\[5pt]
\fl\hspace{0.5cm}
\frac{1}{B}\nabla_\bR
\cdot
\Bigg[
(\mu\bun\times\nabla_\bR B
+
u^2\nabla_\bR\times\bun
+
Z_\spe
\bun\times\nabla_\bR\varphi_0
) G_{\spe 1}^\lw
\nonumber\\[5pt]
\fl\hspace{0.5cm}
-
Z_\spe\lambda_\spe
\varphi_1^\lw\bun\times\nabla_\bR\psi\left(\Upsilon_\spe+\frac{Z_\spe}{T_\spe}
\partial_\psi \varphi_0\right)F_{\spe 0}
\Bigg]
\nonumber\\[5pt]
\fl\hspace{0.5cm}
-\frac{1}{B}\partial_u
\Bigg[
u (\nabla_\bR\times\bun)\cdot
(\mu \nabla_\bR B + Z_\spe
\nabla_\bR\varphi_0)
 G_{\spe 1}^\lw
\nonumber\\[5pt]
\fl\hspace{0.5cm}
-
\frac{Z_\spe\lambda_\spe
  \varphi_1^\lw}{u}
\mu(\bun\times\nabla_\bR\psi)\cdot\nabla_\bR B
\left(\Upsilon_\spe+\frac{Z_\spe}{T_\spe}
\partial_\psi \varphi_0\right)F_{\spe 0}
\Bigg]
\nonumber\\[5pt]
\fl\hspace{0.5cm}
 - \frac{u \mu}{B}
 (\nabla_\bR \times \bK)_\bot
\cdot\nabla_\bR F_{\spe 0}
\nonumber\\[5pt]
\fl\hspace{0.5cm} +
\mu
\left(\nabla_\bR\times\bK\right)_\perp
\cdot
\left(\mu\nabla_\bR B + Z_\spe\nabla_\bR\varphi_0
\right)
\partial_u F_{\spe 0} 
 \nonumber\\[5pt]
\fl\hspace{0.5cm}
 + 
\frac{Z_\spe \lambda_\spe}{B} \left[
\nabla_\bR \cdot \left(\bun\times\nabla_{\bR_\perp/\epsilon_\spe}
 \langle \phi_{\spe 1}^\sw \rangle
F_{\spe 1}^\sw\right)\right]^\lw
\nonumber\\[5pt]
\fl\hspace{0.5cm} - Z_\spe \lambda_\spe \partial_u \left[
\left( \bun\cdot\nabla_\bR 
\langle \phi_{\spe 1}^\sw \rangle
 +  \frac{u}{B}
\left(\bun \times \kappabf\right)\cdot
\nabla_{\bR_\perp/\epsilon_\spe}\langle \phi_{\spe 1}^\sw \rangle
\right)F_{\spe 1}^\sw \right]^\lw =
\nonumber\\[5pt]
\fl\hspace{0.5cm}
\partial_u
\left[
-\frac{Z_\spe\lambda_\spe\varphi_1^\lw}{u}
\sum_{\spe'}
\left\langle \cT_{\spe,0}^* C_{\spe\spe'}^{(1)\lw}\right\rangle
\right]
\nonumber\\[5pt]
\fl\hspace{0.5cm}
+\partial_u 
\Bigg(
\left \langle u \bun \cdot \nabla_\bR \bun \cdot \rhobf\,
\cT_{\spe,0}^*C_{\sigma \sigma^\prime}^{(1)\lw}
\right \rangle
\nonumber\\
\fl\hspace{0.5cm}
-\mu
 \bun \cdot \nabla_\bR \times \bun \left \langle
\cT_{\spe,0}^*C_{\sigma \sigma^\prime}^{(1)\lw}
\right \rangle
\Bigg)
\nonumber\\
\fl\hspace{0.5cm}
+ \partial_\mu
\Bigg\{
\frac{u \mu}{B} \bun \cdot \nabla_\bR \times \bun
\left \langle
\cT_{\spe,0}^*C_{\sigma \sigma^\prime}^{(1)\lw}
\right \rangle
\nonumber\\
\fl\hspace{0.5cm}
-
\Bigg\langle \Bigg(
\frac{Z_\spe}{B} \rhobf \cdot \nabla_\bR
\varphi_0 + \frac{\mu}{B} \rhobf \cdot \nabla_\bR B +
\nonumber\\
\fl\hspace{0.5cm}
\frac{u^2}{B} \bun \cdot \nabla \bun \cdot \rhobf \Bigg)
\cT_{\spe,0}^*C_{\sigma \sigma^\prime}^{(1)\lw}
\Bigg\rangle
\Bigg\}
 \nonumber\\
\fl\hspace{0.5cm}
+ \frac{1}{B} \nabla_\bR \cdot \left \langle B
\rhobf \,
\cT_{\spe,0}^*C_{\sigma \sigma^\prime}^{(1)\lw}
 \right \rangle  \nonumber\\
\fl\hspace{0.5cm} +
\left\langle\left[
\cT_{\spe,1}^\ast C_{\sigma
\sigma^\prime}^{(1)\sw}
\right]^\lw\right\rangle
 +\sum_{\spe'}
\left\langle
\cT_{\spe,0}^*C_{\spe\spe'}^{(2)\lw}
\right\rangle.
\end{eqnarray}
Finally, we remind the reader that the term $\langle [
\cT_{\spe,1}^\ast C_{\sigma \sigma^\prime}^{(1)\sw} ]^\lw\rangle$ is a
total derivative with respect to $\mu$ (see
\eq{eq:turbpiececollgyroave}).

\section{Calculation of the particle density transport equation}
\label{sec:manipulations_densityEqEnergy}

The transport equation for the particle density of species $\spe$ is
obtained from \eq{eq:solvcond_Density}, with $j=2$ and $R_{\spe 2}$
given by \eq{eq:Rsigma2}. Easily, one gets
\begin{eqnarray} \label{eq:transEqDensityApp2}
\fl
\partial_{\epsilon_s^2 t}n_\spe=
\Bigg\langle
\frac{\tau_\spe}{\lambda_\spe^2}\int
\Bigg\{
-\nabla_\bR
\cdot
\Bigg[
\Big(\mu\bun\times\nabla_\bR B
\nonumber\\[5pt]
\fl\hspace{0.5cm}
+
u^2\nabla_\bR\times\bun
+
Z_\spe
\bun\times\nabla_\bR\varphi_0
\Big) G_{\spe 1}^\lw
\nonumber\\[5pt]
\fl\hspace{0.5cm}
-
Z_\spe\lambda_\spe
\varphi_1^\lw\bun\times\nabla_\bR\psi\left(\Upsilon_\spe+\frac{Z_\spe}{T_\spe}
\partial_\psi \varphi_0\right)F_{\spe 0}
\Bigg]
\nonumber\\[5pt]
\fl\hspace{0.5cm}
 -
Z_\spe \lambda_\spe \left[
\nabla_\bR \cdot \left(\bun\times\nabla_{\bR_\perp/\epsilon_\spe}
 \langle \phi_{\spe 1}^\sw \rangle
F_{\spe 1}^\sw\right)\right]^\lw
\nonumber\\[5pt]
\fl\hspace{0.5cm} 
+
\sum_{\spe'}
\nabla_\bR \cdot \left \langle B
\rhobf \,
\cT_{\spe,0}^*C_{\sigma \sigma^\prime}^{(1)\lw}
 \right \rangle 
\Bigg\}
\dd u\dd\mu\dd\theta
\Bigg\rangle_\psi,
\end{eqnarray}
where the particle conservation property of the collision operator,
integrations by parts in $u$ and arguments of parity in $u$ have been
used to drop many terms from $R_{\spe 2}$. Also, we have used the fact
that the term $\langle [ \cT_{\spe,1}^\ast C_{\sigma
  \sigma^\prime}^{(1)\sw} ]^\lw\rangle$ is an exact derivative in
$\mu$, as shown in \eq{eq:turbpiececollgyroave}.

Finally, we recall \eq{eq:formuladivergence} and the fact that
$\varphi_0$ is a flux function to obtain \eq{eq:densityevolution4}.

\section{Calculation of the energy
  transport equation}
\label{sec:manipulations_transportEqEnergy}

It is illustrative to show that the transport equation for the energy
density of species $\spe$ is not a solvability condition, because it
still involves $\langle F_{\spe 2}^\lw\rangle$. It is obtained by
multiplying \eq{eq:FPequationG2final} by $\tau_\spe
\lambda_\spe^{-2}B(u^2/2 +\mu B)$, integrating over $u$, $\mu$ and
$\theta$, and taking the flux-surface average, giving
\begin{eqnarray} \label{eq:TransEqEner2}
\fl
\partial_{\epsilon_s^2 t}
\left( \frac{3}{2}n_{\spe}T_\spe\right)
\nonumber\\[5pt]
\fl\hspace{0.5cm}
+
\Bigg\langle
\nabla_\bR\cdot
\int
\frac{\tau_\spe}{\lambda_\spe^2}
 B(u^2/2+\mu B)
\Big[
(\bv_{\nabla B}
+
\bv_{E,\spe}^{(0)}) G^\lw_{\sigma 1}
\nonumber\\[5pt]
\fl\hspace{0.5cm}
-
Z_\spe\lambda_\spe
\varphi_1^\lw\bun\times\nabla_\bR\psi\left(\Upsilon_\spe+\frac{Z_\spe}{T_\spe}
\partial_\psi \varphi_0\right)F_{\spe 0}
\Big]
\dd u\dd\mu\dd\theta
\Bigg\rangle_\psi 
\nonumber\\[5pt]
\fl\hspace{0.5cm}
+
\Bigg\langle
\nabla_\bR\cdot
\int
\frac{\tau_\spe}{\lambda_\spe^2}
 u^2(u^2/2 + \mu B) \nabla_\bR\times\bun
\, G_{\spe 1}^\lw
\dd u\dd\mu\dd\theta\Bigg\rangle_\psi
\nonumber\\[5pt]
\fl\hspace{0.5cm}
+
\Bigg\langle
\nabla_\bR\cdot
\int (u^2/2 +\mu B)\Bigg\{
\left[
 \left(\bun\times\nabla_{\bR_\perp/\epsilon_\spe}
 \langle \phi_{\spe 1}^\sw \rangle
F_{\spe 1}^\sw\right)\right]^\lw
\nonumber\\[5pt]
\fl\hspace{0.5cm}
-\frac{\tau_\spe}{\lambda_\spe^2}
\sum_{\spe'}
\left\langle
B\rhobf {\cal T}_{\spe,0}^* C_{\spe\spe'}^{(1)\lw}
\right\rangle
\Bigg\}
\dd u\dd\mu\dd\theta\Bigg\rangle_\psi
\nonumber\\[5pt]
\fl\hspace{0.5cm}
+
\Bigg\langle
\int B\bv_M
\cdot
\left[
\nabla_{\bR_\perp/\epsilon_\spe}
 \langle \phi_{\spe 1}^\sw \rangle
F_{\spe 1}^\sw
\right]^\lw
\dd u\dd\mu\dd\theta
\Bigg\rangle_\psi
 \nonumber\\[5pt]
\fl\hspace{0.5cm}
+\Bigg\langle
\int
 Bu \Big[ \bun\cdot\nabla_\bR 
\langle \phi_{\spe 1}^\sw \rangle
F_{\spe 1}^\sw \Big]^\lw
\dd u\dd\mu\dd\theta
\Bigg\rangle_\psi
 \nonumber\\[5pt]
\fl\hspace{0.5cm}
+\frac{1}{\lambda_\spe}
\Bigg\langle\int
B \bv_M\cdot\nabla_\bR\varphi_0 G_{\spe 1}^\lw\,
\dd u\dd\mu\dd\theta
\Bigg\rangle_\psi
 =
\nonumber\\[5pt]
\fl\hspace{0.5cm}
\Bigg\langle\frac{\tau_\spe}{\lambda_\spe^2}
\int B(u^2/2+\mu B)
\sum_{\spe'}
\left\langle
\left[
\cT_{\spe, 1}^{*} C_{\sigma \sigma^\prime}^{(1)\sw}
\right]^\lw
\right\rangle
\dd u\dd\mu\dd\theta
\Bigg\rangle_\psi
\nonumber\\[5pt]
\fl\hspace{0.5cm}
+
\Bigg\langle\frac{\tau_\spe}{\lambda_\spe^2}
\int B Z_\spe
\sum_{\spe'}
\left\langle
\rhobf\cdot\nabla_\bR\varphi_0
 {\cal T}_{\spe,0}^* C_{\spe\spe'}^{(1)\lw}
\right\rangle\dd u\dd\mu\dd\theta\Bigg\rangle_\psi
\nonumber\\[5pt]
\fl\hspace{0.5cm}
+\Bigg\langle\frac{\tau_\spe}{\lambda_\spe^2}
\int B(u^2/2+\mu B)\sum_{\spe'}
\left\langle
\cT_{\spe, 0}^{*} C_{\sigma \sigma^\prime}^{(2)\lw}
\right\rangle
\dd u\dd\mu\dd\theta
\Bigg\rangle_\psi
.
\end{eqnarray}
Here, conservation of particles by the collision operator,
integrations by parts in $\bR$, $u$ and $\mu$, and arguments of parity
in $u$ have been employed. The last term of \eq{eq:TransEqEner2} still
contains $\langle F_{\spe 2}^\lw\rangle$. Summing \eq{eq:TransEqEner2}
over $\spe$, one gets the solvability condition corresponding to
\eq{eq:solvcond_Energy} for $j=2$ and $R_{\spe 2}$ given in
\eq{eq:Rsigma2}, which is a transport equation for the total
energy. Hence, adding \eq{eq:TransEqEner2} over all the species:
\begin{eqnarray} \label{eq:TransEqEner_Total}
\fl
\partial_{\epsilon_s^2 t}
\sum_\spe\frac{3}{2}n_{\spe}T_\spe
\nonumber\\[5pt]
\fl\hspace{0.5cm}
+
\Bigg\langle
\nabla_\bR\cdot
\sum_\spe
\frac{\tau_\spe}{\lambda_\spe^2}
\int B(u^2/2+\mu B)
\Big[
(\bv_{\nabla B}
\nonumber\\[5pt]
\fl\hspace{0.5cm}
+\bv_{E,\spe}^{(0)} 
+u^2 \nabla_\bR\times\bun)G^\lw_{\sigma 1}
\nonumber\\[5pt]
\fl\hspace{0.5cm}
-
Z_\spe\lambda_\spe
\varphi_1^\lw\bun\times\nabla_\bR\psi\left(\Upsilon_\spe+\frac{Z_\spe}{T_\spe}
\partial_\psi \varphi_0\right)F_{\spe 0}
\Big]
\dd u\dd\mu\dd\theta
\Bigg\rangle_\psi 
\nonumber\\[5pt]
\fl\hspace{0.5cm}
+
\Bigg\langle
\nabla_\bR\cdot
\int (u^2/2 +\mu B)\Bigg\{
\sum_\spe
\left[
 \left(\bun\times\nabla_{\bR_\perp/\epsilon_\spe}
 \langle \phi_{\spe 1}^\sw \rangle
F_{\spe 1}^\sw\right)\right]^\lw
\nonumber\\[5pt]
\fl\hspace{0.5cm}
-
\sum_{\spe,\spe'}
\frac{\tau_\spe}{\lambda_\spe^2}
\left\langle
B
\rhobf {\cal T}_{\spe,0}^* C_{\spe\spe'}^{(1)\lw}
\right\rangle\Bigg\}
\dd u\dd\mu\dd\theta\Bigg\rangle_\psi
\nonumber\\[5pt]
\fl\hspace{0.5cm}
+
\sum_\spe
\Bigg\langle
\int B\bv_M
\cdot
\left[
\nabla_{\bR_\perp/\epsilon_\spe}
 \langle \phi_{\spe 1}^\sw \rangle
F_{\spe 1}^\sw
\right]^\lw
\dd u\dd\mu\dd\theta
\Bigg\rangle_\psi
 \nonumber\\[5pt]
\fl\hspace{0.5cm}
+
\sum_\spe
\Bigg\langle
\int Bu
 \Big[
\bun\cdot\nabla_\bR 
\langle \phi_{\spe 1}^\sw \rangle
 F_{\spe 1}^\sw \Bigg]^\lw
\dd u\dd\mu\dd\theta
\Big\rangle_\psi
 \nonumber\\[5pt]
\fl\hspace{0.5cm}
-\Bigg\langle
\int B(u^2/2+\mu B)
\sum_{\spe,\spe'} \frac{\tau_\spe}{\lambda_\spe^2}
\left\langle
\left[
\cT_{\spe, 1}^{*} C_{\sigma \sigma^\prime}^{(1)\sw}
\right]^\lw
\right\rangle
\dd u\dd\mu\dd\theta
\Bigg\rangle_\psi
 \nonumber\\[5pt]
\fl\hspace{0.5cm}
+\partial_\psi \varphi_0 \sum_\spe \frac{1}{\lambda_\spe}
\Bigg\langle\int
B \bv_M\cdot\nabla_\bR\psi G_{\spe 1}^\lw\,
\dd u\dd\mu\dd\theta
\Bigg\rangle_\psi
=0
.
\end{eqnarray}
We have employed the momentum and energy conserving properties of the
collision operator to deduce that
\begin{eqnarray}\label{eq:cancellationcollisions}
\fl
\sum_{\spe,\spe'}
\Bigg\langle\frac{1}{\lambda_\spe}
\int B 
\left\langle
\rhobf\cdot\nabla_\bR\psi
 {\cal T}_{\spe,0}^* C_{\spe\spe'}^{(1)\lw}
\right\rangle\dd u\dd\mu\dd\theta\Bigg\rangle_\psi = 0,
\nonumber\\[5pt]
\fl
\sum_{\spe,\spe'}
\Bigg\langle\frac{\tau_\spe}{\lambda_\spe^2}
\int B(u^2/2+\mu B)
\left\langle
\cT_{\spe, 0}^{*} C_{\sigma \sigma^\prime}^{(2)\lw}
\right\rangle
\dd u\dd\mu\dd\theta
\Bigg\rangle_\psi = 0.
\end{eqnarray}
The vanishing of these terms is checked easily after realizing that
the sums over $\spe$ and $\spe'$ can be split into terms of the form
\begin{eqnarray}
\fl \int B\, \rhobf \cdot \nabla_\bR \psi
( \cT_{\spe,0}^*C_{\sigma \sigma^\prime} [ \lambda_\spe^{-1} g_\spe, \cT_{\spe^\prime, 0}^{-1*} F_{\spe^\prime 0} ] \nonumber\\
\fl\hspace{1cm}+ \cT_{\spe^\prime,0}^*C_{\spe^\prime \spe} [  \cT_{\spe^\prime, 0}^{-1*} F_{\spe^\prime 0}, \lambda_{\spe}^{-1} g_\spe]  )
\dd u\dd\mu\dd\theta = 0,
\nonumber\\
\fl
\int B \left ( u^2/2 + \mu B \right )
( \tau_\spe \cT_{\spe,0}^*C_{\sigma \sigma^\prime} [ \lambda_\spe^{-1} g_\spe, \lambda_{\spe^\prime}^{-1} g_{\spe^\prime} ]
\nonumber\\
\fl \hspace{1cm}
+ \tau_{\spe^\prime} \cT_{\spe^\prime,0}^*C_{\spe^\prime \spe} [  \lambda_{\spe^\prime}^{-1} g_{\spe^\prime}, \lambda_\spe^{-1} g_\spe]  )
\dd u\dd\mu\dd\theta = 0,
\nonumber\\
\fl \int B \left ( u^2/2 + \mu B \right )
( \tau_\spe \cT_{\spe,0}^*C_{\sigma \sigma^\prime} [ \lambda_\spe^{-2} g_\spe, \cT_{\spe^\prime, 0}^{-1*} F_{\spe^\prime 0} ] \nonumber\\
\fl
\hspace{1cm}
+ \tau_{\spe^\prime} \cT_{\spe^\prime,0}^*C_{\spe^\prime \spe} [  \cT_{\spe^\prime, 0}^{-1*} F_{\spe^\prime 0}, \lambda_{\spe}^{-2} g_\spe]  )
\dd u\dd\mu\dd\theta = 0,
\end{eqnarray}
that are zero because of \eq{eq:propcollnondim} for any pair of
functions $g_{\spe} (\boldr, \bv, t)$ and $g_{\spe^\prime} (\boldr,
\bv, t)$.

In Appendix L of reference \cite{CalvoParra2012} it has been shown
that, independently of the magnetic geometry,
\begin{eqnarray}\label{eq:cancellationenergy}
\fl \Bigg\langle \sum_\spe
 \int B \Bigg [ F_{\sigma 1}^\sw
 \Big(  u \bun\cdot\nabla_\bR
\langle\phi_{\spe
 1}^\sw\rangle
\nonumber\\[5pt]
\fl \hspace{0.5cm}
+ \bv_M \cdot
\nabla_{\bR_\perp/\epsilon_\spe}\langle\phi_{\spe 1}^\sw\rangle 
 \Big)  \Bigg]^\lw \dd u \dd \mu \dd \theta \Bigg \rangle_\psi
\nonumber\\[5pt]
\fl \hspace{0.5cm} -
 \Bigg\langle
\sum_{\spe,\spe'} \frac{\tau_\spe}{\lambda_\spe^2}
\int
B
\left(u^2/2+\mu B\right)
 \left[\left\langle
\cT_{\spe, 1}^* C_{\sigma
\sigma^\prime}^{(1)\sw}
\right\rangle
 \right]^\lw
\dd u\dd\mu\dd\theta\Bigg\rangle_\psi
\nonumber\\[5pt]
\fl \hspace{0.5cm}
 = O(\epsilon_s).
\end{eqnarray}
Besides, since $\varphi_0$ is a flux function, the ambipolarity
condition \eq{eq:AmbipolarityCondition} makes the last term on the
left side of \eq{eq:TransEqEner_Total} vanish. Then, we can drop the
terms that are not written explicitly as a divergence in
\eq{eq:TransEqEner_Total}. Using formula \eq{eq:formuladivergence} to
rewrite the remaining ones, we reach \eq{eq:TransEqEner_TotalFinal}.

\section{Collision operator in drift-kinetic coordinates}
\label{sec:CollOp_DKcoor}

For the proof of Section \ref{sec:deviationQS} it is useful to exhibit
the form of the kernel of the collision operator in coordinates
$(\bR,u,\mu,\theta)=\cT_{\spe,0}^{-1}(\boldr,\bv)$. Recall that when
$\epsilon_\spe = 0$ the change of coordinates is simply given by
\begin{eqnarray}\label{eq:Tspe0appendix}
\bR = \boldr
,\nonumber\\[5pt]
u = \bv\cdot\bun(\boldr)
,\nonumber\\[5pt]
\mu = \frac{1}{2 B(\boldr)}
\left(\bv - \bv\cdot\bun(\boldr)\bun(\boldr)\right)^2
,\nonumber\\[5pt]
\theta = \arctan
\left(\frac{\bv\cdot\eun_2(\boldr)}{\bv\cdot\eun_1(\boldr)}
\right).
\end{eqnarray}
Since $\bR = \boldr$, to this order the change of coordinates only
affects the velocities and the spatial position enters as a
parameter. An additional simplification is provided by the fact that,
for the purposes of Section \ref{sec:deviationQS}, we can restrict
ourselves to gyrophase-independent distribution functions.

It is convenient to rewrite the collision operator
\eq{eq:collisionoperator} in terms of the Rosenbluth potentials
$\potH_{\spe'}(\boldr,\bv,t)$ and $\potL_{\spe'}(\boldr,\bv,t)$,
defined by
\begin{eqnarray}\label{eq:defpotH}
\potH_{\spe'} = \int_{{\mathbb R}^3}|\bv-\bv'|f_{\spe'}(\boldr,\bv',t)\dd^3v'
\end{eqnarray}
and
\begin{eqnarray}\label{eq:defpotL}
\potL_{\spe'} = \int_{{\mathbb R}^3}\frac{2}{|\bv-\bv'|}
f_{\spe'}(\boldr,\bv',t)\dd^3v'.
\end{eqnarray}
Namely,
\begin{eqnarray}\label{eq:collisionoperatorRosenbluthPotentials}
\fl
C_{\spe \spe'}[f_\spe,f_{\spe'}](\boldr,\bv) =
\nonumber\\[5pt]
\fl\hspace{1cm}
\frac{\gamma_{\spe\spe'}}{m_\sigma}
\nabla_\bv\cdot
\Bigg(
\frac{1}{m_\spe}\nabla_\bv\nabla_\bv \potH_{\spe'}(\boldr,\bv,t)
\cdot \nabla_\bv f_\spe(\boldr,\bv,t)
\nonumber\\[5pt]
\fl\hspace{1cm}
-
\frac{1}{m_{\spe'}}
f_{\spe}(\boldr,\bv,t)\nabla_\bv\potL_{\spe'}(\boldr,\bv,t)
\Bigg).
\end{eqnarray}
This can be immediately checked by using
\begin{equation}
\matW(\bw) = \nabla_\bw\nabla_\bw w
\end{equation}
and
\begin{equation}
\nabla_\bw\cdot\matW(\bw)= \nabla_\bw\left(\frac{2}{w}\right),
\end{equation}
where $\matW(\bw)$ has been defined in \eq{eq:defmatW}. An equivalent
definition of the Rosenbluth potentials is provided by saying that
they are the solutions of
\begin{eqnarray}\label{eq:potHequation}
\nabla_\bv^2\potH_{\spe'}(\boldr,\bv,t) = \potL_{\spe'}(\boldr,\bv,t)
\end{eqnarray}
and
\begin{eqnarray}\label{eq:potLequation}
\nabla_\bv^2\potL_{\spe'}(\boldr,\bv,t) = -8\pi f_{\spe'}(\boldr,\bv,t)
\end{eqnarray}
with the appropriate boundary conditions. In order to show that
\eq{eq:defpotH} and \eq{eq:defpotL} fulfill these equations, one has
to use
\begin{equation}
\nabla_\bw^2 w= \frac{2}{w}
\end{equation}
and
\begin{equation}
\nabla_\bw^2 \left(\frac{1}{w}\right)= -4\pi\delta(\bw).
\end{equation}
It is more convenient for us to take \eq{eq:potHequation}
and \eq{eq:potLequation} as the expressions defining the
potentials.

Denote by $F_\spe(\bR,u,\mu,t)$, $P_{\spe'}(\bR,u,\mu,t)$,
$Q_{\spe'}(\bR,u,\mu,t)$, the distribution function and
Rosenbluth potentials in the new coordinates. That is,
\begin{eqnarray}
F_\spe(\bR,u,\mu,t) = \cT_{\spe,0}^*f_\spe,\nonumber\\[5pt]
P_{\spe'}(\bR,u,\mu,t) = \cT_{\spe',0}^*\,\potH_{\spe'},\nonumber\\[5pt]
Q_{\spe'}(\bR,u,\mu,t) = \cT_{\spe',0}^*\,\potL_{\spe'}.
\end{eqnarray}
Since the distribution functions are gyrophase independent,
$P_{\spe'}$ and $Q_{\spe'}$ are also required to be so.  We have to
apply to \eq{eq:collisionoperatorRosenbluthPotentials} the divergence
formula for a transformation from euclidean coordinates $\bv$ to
arbitrary coordinates ${\mathbf Y}$:
\begin{equation}
\nabla_\bv\cdot\Gammabf_{\spe\spe'}
=\sum_{i=1}^3
\frac{1}{{\cal J}}\frac{\partial}{\partial Y^i}({\cal J}\Gammabf_{\spe\spe'}\cdot\nabla_\bv Y^i),
\end{equation}
where $\cal J = \det(\partial\bv/\partial{\mathbf Y})$.  In our case,
${\mathbf Y}\equiv\{u,\mu,\theta\}$ and the transformation is given in
\eq{eq:Tspe0appendix} (recall that the spatial coordinates do not
change, so we abuse a bit the notation and understand the
transformation as acting only on velocity coordinates). The vector
field $\Gammabf_{\spe\spe'}$ reads
\begin{eqnarray}\label{eq:defGammabf}
\fl
\Gammabf_{\spe\spe'} =
\frac{\gamma_{\spe\spe'}}{m_\sigma}
\Bigg(
\frac{1}{m_\spe}\nabla_\bv\nabla_\bv \potH_{\spe'}(\boldr,\bv,t)
\cdot \nabla_\bv f_\spe(\boldr,\bv,t)
\nonumber\\[5pt]
\fl\hspace{1cm}
-
\frac{1}{m_{\spe'}}
f_{\spe}(\boldr,\bv,t)\nabla_\bv\potL_{\spe'}(\boldr,\bv,t)
\Bigg),
\end{eqnarray}
${\cal J} = B$, $\nabla_\bv u = \bun$, $\nabla_\bv \mu =
B^{-1}\bv_\perp$, and $\nabla_\bv\theta =
(\bun\times\bv_\perp)/\bv_\perp^2$. Here, $\bv_\perp = \bv
-\bv\cdot\bun\bun$.

We proceed to write \eq{eq:collisionoperatorRosenbluthPotentials} in
the new coordinates, and denote it by $C^{\cT_{\spe, 0}}_{\spe
  \spe'}[F_\spe,F_{\spe'}]$,
\begin{eqnarray}\label{eq:collisionOpDKCoor}
C^{\cT_{\spe, 0}}_{\spe \spe'}[F_\spe,F_{\spe'}] =
\partial_u\Gamma_{\spe\spe'}^u+\partial_\mu\Gamma_{\spe\spe'}^\mu
+\partial_\theta\Gamma_{\spe\spe'}^\theta.
\end{eqnarray}
Here, $\Gamma_{\spe\spe'}^i := \Gammabf_{\spe\spe'}\cdot\nabla_\bv
Y^i$.  Noting that $\partial_u(\nabla_\bv\mu)=0$,
$\partial_\mu(\nabla_\bv\mu)=(2B\mu)^{-1}\bv_\perp$, it is easy to
reach explicit expressions for $\Gamma_{\spe\spe'}^u$,
$\Gamma_{\spe\spe'}^\mu$, and $\Gamma_{\spe\spe'}^\theta$:
\begin{eqnarray}\label{eq:GammaComponents}
\fl
\Gamma_{\spe\spe'}^u
=
\frac{\gamma_{\spe\spe'}}{m_\spe^2}
\Bigg(
\partial_u^2 P_{\spe'}\partial_u F_\spe
\nonumber\\[5pt]
\fl\hspace{0.5cm}
+
\frac{2\mu}{B}\partial_u\partial_\mu P_{\spe'}\partial_\mu F_\spe
-\frac{m_\spe}{m_{\spe'}}\partial_u Q_{\spe'}F_\spe
\Bigg),
\nonumber\\[5pt]
\fl
\Gamma_{\spe\spe'}^\mu
=
\frac{\gamma_{\spe\spe'}}{m_\spe^2}\frac{2\mu}{B}
\Bigg[
\partial_u\partial_\mu P_{\spe'}\partial_u F_\spe
\nonumber\\[5pt]
\fl\hspace{0.5cm}
+
\frac{1}{B}\Bigg(
\partial_\mu P_{\spe'} + 2\mu\partial_\mu^2P_{\spe'}
\Bigg)\partial_\mu F_\spe
-\frac{m_\spe}{m_{\spe'}}\partial_\mu Q_{\spe'}F_\spe
\Bigg],
\nonumber\\[5pt]
\fl
\Gamma_{\spe\spe'}^\theta
= 0.
\end{eqnarray}

Finally, we write the equations that determine the Rosenbluth
potentials, \eq{eq:potHequation} and \eq{eq:potLequation}, in the new
coordinates,
\begin{eqnarray}\label{eq:RosenbluthPotEqDKcoor}
\fl\Bigg(\partial_u^2+\frac{2\mu}{B}\partial_\mu^2+
\frac{1}{B}\partial_\mu
\Bigg)
P_{\spe'} = Q_{\spe'},
\nonumber\\[5pt]
\fl
\Bigg(\partial_u^2+\frac{2\mu}{B}\partial_\mu^2+
\frac{1}{B}\partial_\mu
\Bigg)
Q_{\spe'} = -8\pi F_{\spe'}.
\end{eqnarray}

\section*{References}

\end{document}